\documentclass[12pt]{elsarticle}

\usepackage{bm}
\usepackage{mathstyle00}
\usepackage{amsmath}
\usepackage{ulem}

\makeatletter
\def\ps@pprintTitle{%
  \let\@oddhead\@empty
  \let\@evenhead\@empty
  \let\@oddfoot\@empty
  \let\@evenfoot\@oddfoot
}
\makeatother
\usepackage[margin=1.0in]{geometry}    
\usepackage{caption}
\usepackage{tikz}
\usetikzlibrary{matrix,shapes,arrows,positioning}
\newcommand{\beq}{\begin{equation}}
\newcommand{\eeq}{\end{equation}}

\newcommand{\g}{{\gamma}}
\def\eps{\varepsilon}
\def\bx{\bm{x}}
\newcommand{\Om}{{\Omega}}
\begin{document}
\begin{frontmatter}
		\title{
		A Deep Learning Approach to Nonconvex Energy Minimization for Martensitic Phase Transitions}
		
		\author[nus]{Xiaoli Chen}
		\ead{xlchen@nus.edu.sg}
		\author[uoc,iacm]{Phoebus Rosakis\corref{cor1}}
		\ead{rosakis@uoc.gr}
		\author[hku]{Zhizhang Wu}
		\ead{wuzz@hku.hk}
		\author[hku]{Zhiwen Zhang\corref{cor1}}
		\ead{zhangzw@hku.hk}

	    \address[nus]{Institute for Functional Intelligent Materials \& Department of Mathematics, National University of Singapore, 119077, Singapore}
	    \address[uoc]{Department of Mathematics and Applied Mathematics, University of Crete, Greece}		
	    \address[iacm]{Institute of Applied and Computational Mathematics, Foundation for Research and Technology-Hellas, Heraklion, Greece}
	    \address[hku]{Department of Mathematics, The University of Hong Kong, Pokfulam Road, Hong Kong SAR, China.}
		\cortext[cor1]{Corresponding author}
\begin{abstract}
We propose a mesh-free method to solve nonconvex energy minimization problems for martensitic phase transitions and twinning in crystals, using the deep learning approach. These problems pose multiple challenges to both analysis and computation, as they involve multiwell gradient energies with large numbers of local minima, each involving a topologically complex microstructure of free boundaries with gradient jumps. We use the Deep Ritz method, whereby candidates for minimizers are represented by parameter-dependent deep neural networks, and the energy is minimized with respect to network parameters. The new essential ingredient is a novel activation function proposed here,  which is a smoothened rectified linear unit we call SmReLU; this captures the structure of minimizers where usual activation functions fail. The method is mesh-free and thus can approximate free boundaries essential to this problem without any special treatment, and is extremely simple to implement. We show the results of many numerical computations demonstrating the success of our method.



\end{abstract}
		
\end{frontmatter}
\vfill\eject
\section{Introduction}
\noindent
Physics-informed deep learning methods \cite{karniadakis2021physics,JinchaoXu:2018,QiangDU:2018,weinan2017deep,weinan2018deep,khoo2017solving,Karniadakis2018learning,raissi2019physics,shin2020convergence} have recently achieved great success in solving PDE problems arising in different fields of STEM. This entails  using deep neural networks to approximate the solution of PDEs, often in many dimensions, leading to a mesh-free numerical method in place of more established schemes, such as finite-element  or finite-difference methods.

In this approach \cite{Lagaris1998, raissi2019physics}, the loss function to be minimized is the mean square of the PDE residual. This method is rigorously known to converge for linear elliptic and parabolic problems \cite{shin2020convergence}, and is successful with a number of nonlinear PDEs \cite{chen2021learning,lou2021physics}. It is not clear how it would perform in situations where there are weak solutions of the PDE that suffer from reduced smoothness. Examples are hyperbolic conservation laws, as in nonlinear elastodynamics, and solid-solid phase transitions described by nonconvex gradient energy functions  in nonlinear elastostatics \cite{ericksen,knowles,ball}. In the latter case, the associated Euler-Lagrange equations change type and lose ellipticity at some values of the solution gradient  \cite{knowles}.  In these situations, there are weak solutions with discontinuous gradient across interfaces which are unknown \textit{a priori} but are part of the solution.

Another type of interface problem involves heterogeneous media whose properties have \textit{a priori} specified high-contrast or discontinuous features, such as PDEs whose coefficients are discontinuous across a fixed interface, where solutions suffer a loss of smoothness. A number of sophisticated finite element and finite difference  methods for interface problems have been developed \cite{babuvska1970finite,li2003new,gong2008immersed,chen1998finite,GrahamHou:10,peskin1977numerical, leveque1994immersed}. These  methods are quite accurate, but they often require special treatment of the interface and jump conditions across it in creating the finite element mesh and/or associated basis functions for elements that intersect it.

Recently,  deep learning  was used to solve interface problems involving linear elliptic PDE systems with discontinuous  coefficients and/or discontinuous forcing \cite{wang2020}, motivated from biomechanics \cite{kerato}.
%
The problems studied in \cite{wang2020} are solved by energy minimization. The energy functional is strictly convex, and a weak solution of the PDE (Euler-Lagrange equation) is its unique minimizer. The interface is fixed at the location of the jump in the coefficients of the PDE. The Deep Ritz method \cite{weinan2018deep} used in \cite{wang2020}  approximates solutions by a deep neural network (DNN) and minimizes the energy as a function of the DNN parameters (weights and biases). The DNN method is mesh-free and does not require any special structure to capture the location of the interface inside the domain spontaneously and accurately.


In  phase transition problems, interfaces separating parts of the body in different phases (solid and liquid, or austenite and martensite) are  free boundaries whose position or shape is not known \textit{a priori}. Rather, they are part of the solution and can evolve in time-dependent problems. Phase boundaries (domain walls, twin boundaries, etc. in different physical settings) are characterized by a jump discontinuity or narrow transition layer of the unknown scalar or vector field (e.g., order parameter, deformation, polarization) or  its spatial derivatives.  To identify the unknown phase boundary, special augmented models have been developed, notably the phase field and level set methods. 
The prototypical phase field models are described by the Allen-Cahn and Cahn-Hilliard equations. An essential ingredient of most continuum models of phase transitions is a nonconvex, multiwell energy density function. 
The DNN approach has been used to solve free boundary problems of the Stefan type \cite{wang2021} and to describe phase boundaries in the Allen-Cahn and Cahn-Hilliard equations \cite{mattey,Karniadakis2018learning}.  The method uses standard DNNs without any modification to identify phase boundaries. 

Martensitic phase transitions \cite{ericksen,ball,knowles} are often modeled using a free energy density that is a nonconvex,  multi-well  potential depending on the \textit{gradient} of the unknown deformation field.  In fact, the martensitic phase itself involves two or more symmetry-related variants, known as twins, that correspond to different energy wells.
\begin{figure}[ht]
   \begin{minipage}[]{0.28 \textwidth}
\leftline{\small\textbf{(a)}}
\centerline{\includegraphics[height=4.6cm]{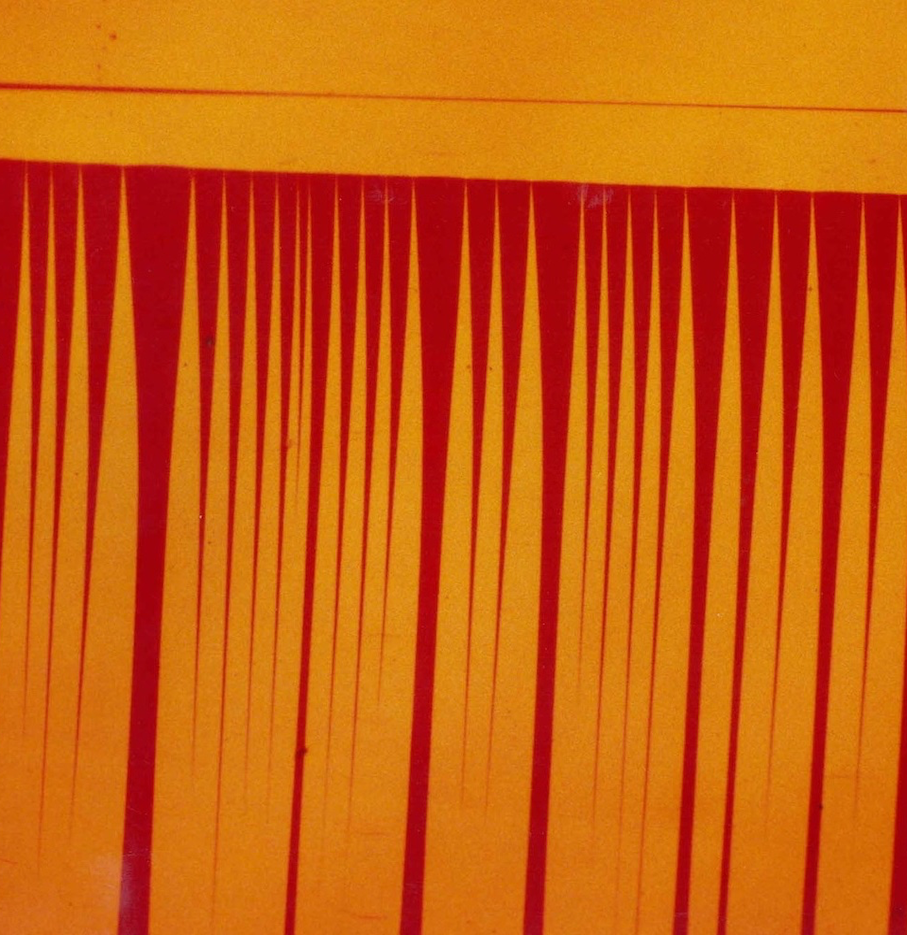}}
\end{minipage}
\hfill
 \begin{minipage}[]{0.28 \textwidth}
 \leftline{\small\textbf{(b)}}
\centerline{\includegraphics[height=4.6cm]{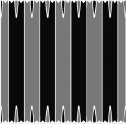}}
\end{minipage}
\hfill
   \begin{minipage}[]{0.28 \textwidth}
 \leftline{\small\textbf{(c)}}
\centerline{\includegraphics[height=4.6cm]{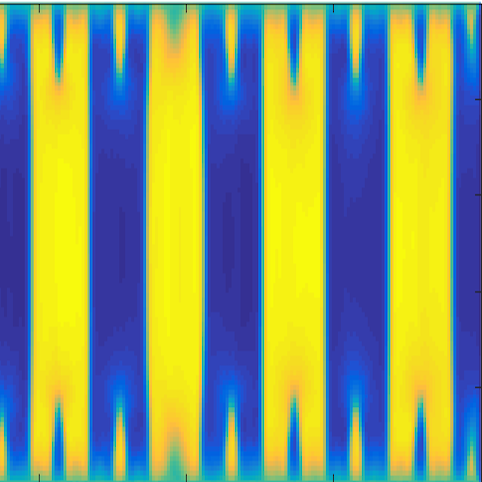}}
\end{minipage}
\caption{\textbf{Twinned Microstructures: (a)} Experimental in CuAlNi single crystal \cite{chu}(reproduced here with permission of the authors). \textbf{(b), (c):}  Comparison of results of different numerical methods for a simplified 2D model: (b) Finite Difference  solution \cite{healey} (reproduced here with permission of the authors) and (c) our DNN solution  of the 2D regularized problem \eqref{min2de}.}
\label{twins}
\end{figure}
The fact that the energy depends on the gradient of the unknown function (deformation) introduces special challenges. Local energy minimizers  involve phase boundaries  separating parts of the body in different phases. Across these interfaces, the deformation field remains continuous, but its gradient jumps. Continuity of the deformation field imposes restrictions on the jump of its gradient (Hadamard compatibility conditions \cite{knowles,ball}).  In many crystalline alloys, the austenite and martensite energy wells violate these conditions, which are nonetheless satisfied between  martensitic twin variants. As a result, there is no absolute energy minimum achievable \cite{ball}. Instead, the energy tends to its infimum in the limit, along a sequence of deformations with a laminated microstructure of parallel twin interfaces. Such  minimizing sequences involve a larger and larger number of  twin boundaries, and converge to a limit state compatible with austenite \cite{ball}. Such laminates are observed in experiments, Fig. \ref{twins}(a).
%
This microstructure is called finely twinned martensite; it can also occur because of incompatibility between twins of different orientations and/or rigid (null Dirichlet) boundary conditions, Fig. \ref{twins}.


Near an incompatible boundary,  twin layers  taper into needles, and often an individual layer is observed splitting into two or more needles in experiments, Fig. \ref{twins}(a) \cite{chu,shilo}. This is known as twin branching and has been studied analytically \cite{kohn,conti,healey,james} and numerically \cite{luskin,hourosakis,healey,dondl} as a mechanism for attainment of an energy minimum in the presence of interfacial energy. 
In order to capture this complex behaviour, it is possible to consider some lower-dimensional problems. 
Many of these  have an energy of the form
\begin{equation}\label{proble}
  \int_{\Omega}W(\nabla u(\bx))d\bx,
\end{equation}
where $\Omega\subset \mathbb{R}^n$, $u:\Omega\to \mathbb{R}^m$ and $n=2$ or $3$, while $m=n$ or $m=1$ for the vector and scalar case, respectively. In all cases, $W: \mathbb{R}^{m\times n}\to \mathbb{R}$ is a nonconvex function with multiple wells.
%
Letting the derivative of $W$ be $S: \mathbb{R}^{m\times n} \to \mathbb{R}^{m\times n}$, the Euler-Lagrange  equation of \eqref{proble} are
\begin{equation}\label{EL}\hbox{div}\, S(\nabla u)=0\quad \hbox{in $\Omega$.}\end{equation}
Because of  loss of rank one convexity \cite{ball,knowles} for values of $\nabla u$ between wells, this PDE system loses ellipticity. Because of ellipticity loss, \eqref{EL} has weak solutions with phase boundaries, across which $\nabla u$  jumps discontinuously from phase to phase \cite{knowles} and second derivatives fail to exist. This precludes using the physics-informed neural network method (PINN), which requires the strong form of \eqref{EL} to be valid \cite{Karniadakis2018learning}. 
An added problem is the extensive loss of uniqueness of weak solutions of \eqref{EL} \cite{ericksen}, many of which are energetically unstable (not even local energy minima). The alternative is  to  minimize the energy \eqref{proble} \cite{ball,kohn,dondl,james,conti}. We use the Deep Ritz method \cite{weinan2017deep}, namely energy minimization in terms of DNNs. An additional advantage is that piecewise smooth  local minimizers of \eqref{proble} are also weak solutions of \eqref{EL} which are at least metastable, in contrast to some  solutions of \eqref{EL}.

Here we demonstrate that the Deep Ritz method  spontaneously captures complex microstructures, such as the finely twinned one, including gradient jumps across multiple curved interfaces and splitting/tapering topological transitions near boundaries \cite{healey,kohn,conti,dondl,james,hourosakis,shilo}. Notably, the method is mesh-free; this allows it to accurately describe curved interfaces of arbitrary orientation, in contrast to FEM and other mesh dependent methods \cite{luskin}. No special structure of the DNN is required for this, except a specially designed activation function whose form is suggested by the structure of solutions to simple problems.

The outline of this paper is as follows: In Section \ref{sec:method}, we introduce the Deep Ritz method for energy minimization in the setting of nonlinear elasticity, and  apply it to various problems  in one and two dimensions, involving nonconvex potentials of the deformation gradient. A crucial role is played by the activation function. Our first choice is guided by the observation that in the simplest one-dimensional problem, an exact solution is provided by the piecewise linear ReLU activation function.  We then consider a regularization of the problem, by including the second derivative of the deformation in the energy \cite{cgr}. Minimizers are smooth, with transition layers replacing derivative jumps. This motivates our introduction of a smoothened version of the ReLU activation function, which we call SmReLU. In two-dimensional problems, use of SmReLU captures spatially complex microstructures with fine twinning and twin branching, Fig.~\ref{twins}(c), whereas traditional activation functions fail.
Section \ref{sec:NumericalExamle} reports on extensive numerical computations. A discussion of our results, their significance, and comparison to previous work is in Section \ref{sec:conclusion}.

\section{Methods}\label{sec:method}
\noindent
\subsection{The Minimization Problem}\label{2.1}
\noindent
The problem we study in this paper concerns the attempt to minimize a nonconvex functional that is a low-dimensional model for an austenite-martensite phase transition. The problem involves finding
\beq\label{min} \min E\{u\},  \quad E\{u\}=\int_\Omega  W(\nabla {u} (  \bm{x}))d \bm{x}\eeq
over suitable functions $u:\Om\to \mathbb{R}$,
where the function $W:\mathbb{R}^2\to \mathbb{R}$ is a two-well potential. Here we choose \cite{kohn,conti}
$$W(\nabla u)=\frac{1}{2}[u_x^2(u_x-1)^2+u_y^2],$$
with $\nabla u=(u_x,u_y)$. The wells (minima) of $W$ are at $(0,0)$ and $(1,0)$. The domain $\Om=[0,L]\times[0,1]\subset \mathbb{R}^2$. Here $u$ is subject to the Dirichlet boundary conditions
\beq\label{bc}u(x,y)=\gamma x \quad \forall \; (x,y)\in\partial\Omega.\eeq
Of particular interest is the case
$\gamma=1/2$
as the linear  function $u(x,y)=x/2$ for all $(x,y)\in\Om$  satisfying the boundary conditions has $\nabla u=(1/2,0)$ which is a saddle point of $W$. The zero-energy minima $u\equiv 0$ and  $u=x$ on $\Om$ are incompatible with the boundary conditions, whereas there are  sequences of functions approaching zero energy, namely problem \eqref{min} does not have a minimizer but only minimizing sequences, namely, one can construct sequences of  functions $u_n$ with $E\{u_n\}\to 0$ as $n\to\infty$. This occurs due to geometric incompatibility of the energy-minimal states
$u=\text{const} $ and  $u=x+\text{const}$ on $\Om$ with the boundary conditions \eqref{bc}.

On the other hand, the mixed boundary conditions
\beq\label{bcf}u(0,y)=0, \quad u(1,y)=\gamma, \quad 0<y<1,\eeq
at the vertical sides $x=0$, $L$ with the horizontal sides $y=0$, $1$ free, are compatible with the ansatz
$$u(x,y)=u(x), \quad \; 0<y<1,$$
(using $u$ for both functions by notation abuse) which reduces \eqref{min} to the following one-dimensional problem.
Let $\Omega=[0,1]$, define:
$$W(z)=z^2(1-z)^2,\quad z\in \mathbb{R},$$
so that $W$ is a double well potential with minima at 0,1.  Suppose $u:[0,1]\to \mathbb{R}$ satisfies  $u(0)=0$ and $u(1)=\g$ for  a given constant $\g\in \mathbb{R}$. Then,  minimize
\beq\label{min1d}\min\int_0^1  W(u'(x) )d x,\eeq
with $u'$ the derivative of $u$.
Letting
\beq\label{frelu} f(x)={ \frac{x+|x|}{2}}, \quad x\in \mathbb{R},\eeq
a solution of the one-dimensional problem \eqref{min1d} (with zero energy) is
\beq\label{fsol}u(x)=f\left(x-1+\g\right).\eeq
This is piecewise smooth with a derivative discontinuity at $x=1-\g$. \begin{remark} \label{r1}Any partition of $[0,1]$ into intervals where the slope of $u$ alternates between  0 and 1 (while $u$ remains continuous) is a minimizer (with zero energy). This shows the massive loss of uniqueness of minimizers of the one dimension (1D) problem \eqref{min1d}, but also the two dimension (2D) problem with boundary conditions \eqref{bcf}. \end{remark}
\begin{remark} \label{r2}In order to make problem \eqref{min1d} well posed, it is common \cite{cgr} to add a higher gradient regularization, also known as capillarity, to the energy which becomes
\beq\label{min1de}\min E_\eps\{u\}, \quad E_\eps\{u\}= \int_0^1  \left[  W(u'(x) )d x +{\frac{\eps^2}{2} }[ u'' (x) ]^2 \right]dx,\eeq
where $\eps>0$ is the higher-gradient coefficient, a small parameter. It is known \cite{cgr} that \eqref{min1de} has an essentially unique solution, where the gradient discontinuity of \eqref{fsol} is replaced by a single,  smooth transition layer between values 0 and 1 of the derivative $u'$. For small $\eps$ the thickness of this layer is of order $\eps$ and the gradient term contributes an interfacial or surface energy to the interface, also of order $\eps$.\end{remark}




The analogous 2D regularized energy is
\beq\label{min2de} E_\eps\{u\}=\int_\Omega \left[ W(\nabla {u} (x,y))+ \frac{\eps^2}{2}\Bigl[u_{xx}(x,y)\Bigr]^2\right]dxdy. \eeq

\subsection{Definition of deep neural networks (DNNs)}
\noindent
We briefly recall the definition and properties of a DNN. There are two ingredients in defining a DNN. The first one is a (vector) linear function of the form $T:\mathbb{R}^n\rightarrow \mathbb{R}^m$, defined as $T(x)=Ax+b$, where $A=(a_{ij})\in \mathbb{R}^{m\times n}$, $x\in \mathbb{R}^{n}$ and $b\in \mathbb{R}^m$. The second one is a nonlinear activation function $\sigma:\mathbb{R}\rightarrow \mathbb{R}$.
A frequently used activation function in the artificial neural network literature, the sigmoid function, is  defined as $\sigma(x)=(1+e^{-x})^{-1}$.
Here, for reasons that will become clear below, we will use another activation function, known as the rectified linear unit (ReLU),  defined by $\sigma(x)=\max(0,x)$ \cite{lecun2015deep}.
Observe that $\sigma(x)=f(x)$ defined in \eqref{frelu}, which provides a solution  \eqref{fsol} of the one-dimensional minimization problem \eqref{min1d}. By applying the activation function to each component $x_j$ ($j=1,\ldots n$) of the vector $x\in \mathbb{R}^n$, one can define a vector activation function $\sigma: \mathbb{R}^n\rightarrow \mathbb{R}^n$ with components $\sigma(x_j)$, $j=1,\ldots n$.

Equipped with those definitions, we are able to construct a continuous function $F(x)$ as an alternating composition of
linear transformations and (vector) activation functions as follows:
\begin{equation}\label{eqn:eg3layernet}
F(x)=T^{k}\circ\sigma\circ T^{k-1}\circ\sigma
\cdot\cdot\cdot\circ T^{1}\circ\sigma\circ T^{0}(x).
\end{equation}
Here $T^{i}(x)=W_ix+b_i$ is an affine transformation, with the weights $W_i$ as of yet unspecified matrices and the biases $b_i$  unspecified vectors. Also $\sigma(\cdot)$ is the component-wise defined activation function. Dimensions of $W_i$ and $b_i$ are chosen to make \eqref{eqn:eg3layernet} meaningful. The parametrization \eqref{eqn:eg3layernet}   is called a $(k+1)$-layer DNN, which has $k$ hidden layers. Denoting all the undetermined coefficients (namely $W_i$ and $b_i$) in \eqref{eqn:eg3layernet} as $\theta\in\Theta$, where $\theta$ is a high-dimensional vector and $\Theta$ is the space of $\theta$. The DNN representation of a continuous function can be viewed as
\begin{align}\label{eqn:solution_DNN}
F=F(x;\theta).
\end{align}
Let $\mathbb{F}=\{ F(\cdot,\theta)|\theta\in\Theta\}$ denote the set of all functions expressible by DNNs parametrized by $\theta\in\Theta$ as in \eqref{eqn:eg3layernet}. Then $\mathbb{F}$ provides an efficient way to represent unknown continuous functions \cite{JinchaoXu:2018}, compared to a linear solution space used in classic numerical methods, e.g., a trial space spaced by linear nodal basis functions in the FEM.

\begin{remark} In the computation, we propose a Smoothened Rectified Linear Unit (SmReLU) activation function to approximate $u$ for the regularized energy problems \eqref{min1de} and \eqref{min2de}. We refer readers to Section \ref{sec:NumericalExamle} for details. \end{remark}

\subsection{The DNN representation of the nonconvex energy minimization}\label{sec:deepRitz}
\noindent
Given the energy functional $E\{u\}$ from \eqref{min}, we can derive a DNN-based numerical method to solve the minimization problem in order to compute the solution $u$, in the spirit of \cite{weinan2017deep}.
Here we seek the solution of the energy minimization problem
\begin{equation}
u=\argmin_{v\in \mathbb{H}^1(\Omega)}E\{v\}, \label{energeminproblem}
\end{equation}
subject to the specified boundary condition $u(x)=g(x)$ on $\partial \Omega$.

From the perspective of scientific computing, the energy minimization problem \eqref{energeminproblem} can be solved using numerical methods, such as FDMs or FEMs. From the perspective of machine learning however, the numerical solution of $u(x)$ is interpreted as a function with $x\in \mathbb{R}^2$ as its input and $u \in \mathbb{R}^1$ as its output, that can be approximated by a DNN $F(x)$ defined in \eqref{eqn:eg3layernet}.
As a result, problem \eqref{energeminproblem} is replaced by the  problem
\begin{equation}\label{eqn:lag_representation}
\tilde{u}=\argmin_{F\in \mathbb{F}_b}E\{F\},
\end{equation}
Here $\tilde{u}$ denotes the  DNN solution  of the problem of  minimizing  the energy  over the DNNs \eqref{eqn:eg3layernet} in  $\mathbb{F}$. Let
\begin{equation}\label{J}
J(\theta)=E\{F(\cdot,\theta)\}= \int_{\Omega}  W(\nabla F(x,\theta))dx, \quad \theta\in \Theta,
\end{equation}
where $F\in \mathbb{F}$ is as in \eqref{eqn:eg3layernet} and  \eqref{eqn:solution_DNN}. Then, the minimization problem \eqref{eqn:lag_representation} reduces to the (finite-dimensional) optimization problem
\begin{equation}\label{eqn:lag_representation_para}
 \tilde\theta=\argmin_{\theta\in\Theta}J(\theta), \quad \tilde u(x)= F(x,\tilde\theta).
\end{equation}
This yields $\tilde u$, the DNN representation of the solution of \eqref{eqn:lag_representation}.

Notice that the original energy minimization problem \eqref{energeminproblem} is non-convex. The
variational problem \eqref{eqn:lag_representation_para} is non-convex as well. Clearly, the issue of local minima and saddle points is nontrivial, which brings essential challenges to many existing optimization methods. Since the parameter space $\Theta$ is typically very large, one often uses the Stochastic Gradient Descent (SGD) method \cite{bottou2010large} to solve \eqref{eqn:lag_representation_para}.

How to impose boundary conditions  in the DNN method is an important issue \cite{Lagaris1998}.
For the Dirichlet boundary condition, we use the equivalent of the Nitsche method in FEM for DNN, known as the Deep Nitsche Method \cite{liao2019deep}. We add a boundary  penalty term in the energy function  to address this issue. Specifically, one adds a soft constraint (a boundary integral term) to the
energy $J(\cdot)$ defined in \eqref{J} and obtains
\begin{equation}\label{eqn:lag_representation_bdd}
\tilde{u}_\tau=\argmin_{\theta\in \Theta}\Big[J(\theta)+\tau\int_{\partial \Omega}\big(F(x,\theta)-g(x)\big)^2 dx\Big],
\end{equation}
where $\tau>0$ is the penalty parameter to scale and balance the losses from the energy part and the boundary part. The idea is that the term $\int_{\partial \Omega}\big(F(x,\theta)-g(x)\big)^2 dx$ will approach zero when we increase the parameter $\tau$ in the calculation, whereas the Dirichlet boundary condition $u=g$ on $\partial \Om$ will be satisfied in a weak $L^2$ sense, and one hopes that $\tilde{u}_\tau\to \tilde{u}$ in $\mathbb{F}$ as $\tau \to\infty$.

It has been recently pointed out that the penalty parameter $\tau$ should not be chosen arbitrarily but adaptively in a rational way stemming from the specific problem \cite{georgoulis}.  Here we also point out that the $\tau$ can be given physical meaning as the stiffness of the austenite phase that exists beyond the boundary $y=1$, whereas the martensite phase occupies $\Om$.  The exact satisfaction of boundary conditions such as \eqref{bc} would correspond to infinitely rigid austenite.

\subsection{Implementation  of the Deep Ritz  DNN method}\label{sec:ImplementationDNN}
\noindent
Since the number $\hbox{dim}\,\Theta$ of degrees of freedom in the optimization problem \eqref{eqn:lag_representation_para} is quite large,
we apply the SGD method on the parameter space $\Theta$ to solve it. Notice that $\theta$ is a high-dimensional vector and let $\theta_k$ be any component of $\theta$. We approximate the corresponding component $\partial J(\theta)/\partial\theta_k$ of the gradient of $J(\theta)$ as follows:
\begin{align}
	\frac{\partial J(\theta)}{\partial \theta_k}&=\int_{\Omega} \frac{\partial\big(W(\nabla F(x,\theta)) \big)}{\partial\theta_k} dx +\tau\int_{\partial \Omega} \frac{\partial\big(F(x,\theta)-g(x)\big)^2}{\partial\theta_k} ds \nonumber\\
	&\approx\frac{vol(\Omega)}{N}\sum_{i=i}^{N}\frac{\partial\big(W(\nabla F(x_i,\theta))\big)}{\partial\theta_k}
	+ \frac{\tau S(\partial \Omega)}{N_b} \frac{\partial\big((F(y_j,\theta)-g(y_j))^2\big)}{\partial\theta_k}\label{num_lag},
\end{align}
where the collocation points  $x_i\overset{i.i.d.}{\sim} Unif(\Omega)$ are uniform random points that are sampled from the physical domain $\Omega$,
$y_j\overset{i.i.d.}{\sim} Unif(\partial \Omega)$ are uniform random points that are sampled from the boundary $\partial \Omega$, $vol(\Omega)$ is the volume of the domain, and $S(\partial \Omega)$ is the perimeter of the domain.
In the context of the deep learning method, $N$ and $N_b$ are called batch numbers, which means the number of training samples utilized in one iteration.

After we compute the approximation of the gradient with respect to $\theta_k$, we can update each component of $\theta$ as
\begin{equation}\label{eqn:sgd_update}
	\theta_k^{n+1} =  \theta_k^{n} - \eta \frac{\partial J(F(\cdot,\theta))}{\partial \theta_k}|_{\theta_k=\theta_k^{n}},
\end{equation}
where $\eta$ is the learning rate. In our numerical computation, we use the Adam optimizer \cite{kingma2014adam}, a modified version of the SGD method. Our results indicate that the Adam optimizer is very effective in solving this non-convex optimization problem.
\begin{remark}
From the derivation of the DNN formulation, one can see that the proposed method automatically deals with the
phase boundary (or discontinuity in the gradient of the solution) without knowing the locations of the discontinuity \textit{a priori}.
\end{remark}

 \begin{figure}[ht]
\centerline{\includegraphics[width=13cm]{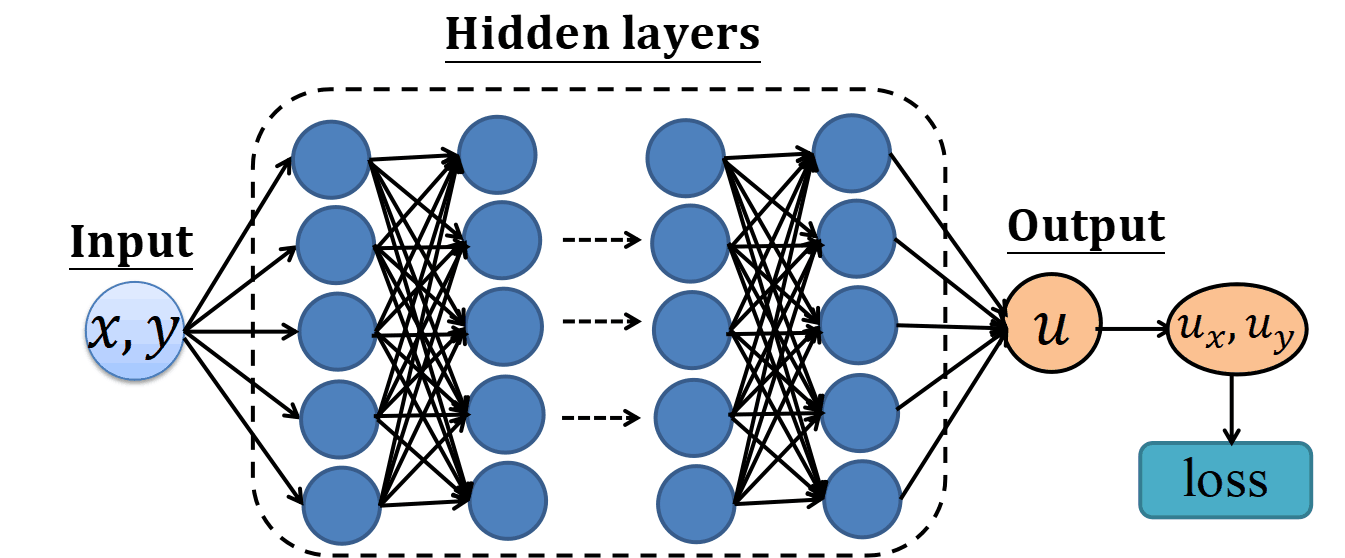}}
\caption{\textbf{Schematic of a DNN. }}\label{NN_structure}
\end{figure}

Fig.~\ref{NN_structure} shows a possible network layout for approximating $u$ with input $(x,y)$. For the output $u$, we can use automatic differentiation  techniques to compute derivatives of $u$ that appear in the energy in the minimization problem. 

\section{Numerical Results}\label{sec:NumericalExamle}
\noindent
We  use a fully connected  DNN (Fig.\ref{NN_structure}) to perform our computations. The weights are initialized with truncated normal distributions with variances set as two over the sum of input and output unit numbers. The biases are initialized as zero. We use the Adam optimizer with a learning rate of $10^{-3}$.
In the main text, the training takes $100,000$ iterations for one-dimensional problems, while it takes $300,000$ iterations for two-dimensional problems.

\subsection{One-dimensional problem}
We demonstrate how to use DNNs to minimize the energy in a one-dimensional problem. We consider the following energy minimization  problem:
\begin{equation}\label{problem}
   \min \int_0^1W( u'(x))dx,
\end{equation}
where
\beq\label{wbc}W(z)=z^2(1-z)^2  \;\; \forall z\in \mathbb{R},  \quad u(0)=0, \quad u(1)=\gamma.\eeq
The integrand $W$ has two wells at 0 and 1 and we choose typically choose $\gamma=1/2$.
The functions $u$ are approximated by DNNs (see Methods)
$$u_{NN}(x)=F(x;\theta),$$
 where $x$ is the input of the DNN and $\theta$ is the vector of weights $w_i$ and biases $b_i$ as in \eqref{eqn:eg3layernet}, which are the trainable parameters in the DNN.

On one hand, the parameters $\theta$ of the DNN $u_{NN}(x)$ should  minimize the energy in \eqref{problem} (Deep Ritz Method \cite{weinan2017deep}). We  construct the corresponding loss function as follows:
\begin{equation}\label{loss_energy}
loss_e(\theta)=\frac{1}{N}\sum_{i=1}^N W\left(\frac{\partial F}{\partial x} (x_i,\theta)\right),
\end{equation}
where $\{x_i\}_{i=1}^N$ are the collocation points in $\Omega$. The sum here approximates the integral in \eqref{problem}. The derivative  $u'_{NN}=\partial F/\partial x$ is computed with automatic differentiation.

On the other hand, $u_{NN}(x)$ should satisfy the boundary conditions, which corresponds to minimizing the loss function
\begin{equation}\label{loss_boundary}
loss_b(\theta)= (F(0,\theta)-0)^2+(F(1,\theta))-\gamma)^2.
\end{equation}
This corresponds to the Deep Nitsche method  \cite{liao2019deep}.
Combining the two loss functions \eqref{loss_energy} and \eqref{loss_boundary}, the total loss function to be minimized is
\begin{equation}\label{loss_all_1d}
loss(\theta)=loss_e(\theta)+\tau loss_b(\theta),
\end{equation}
where $\tau$ is the penalty parameter to scale and balance the losses from the energy part and the boundary part.
During training, the shared parameters $\theta$ are adjusted by back-propagating the error obtained by minimizing a loss function \eqref{loss_all_1d} that is the weighted sum of the above two constraints.

\subsubsection{Activation function}
\begin{figure}
\centerline{\includegraphics[width=6cm,height=5cm]{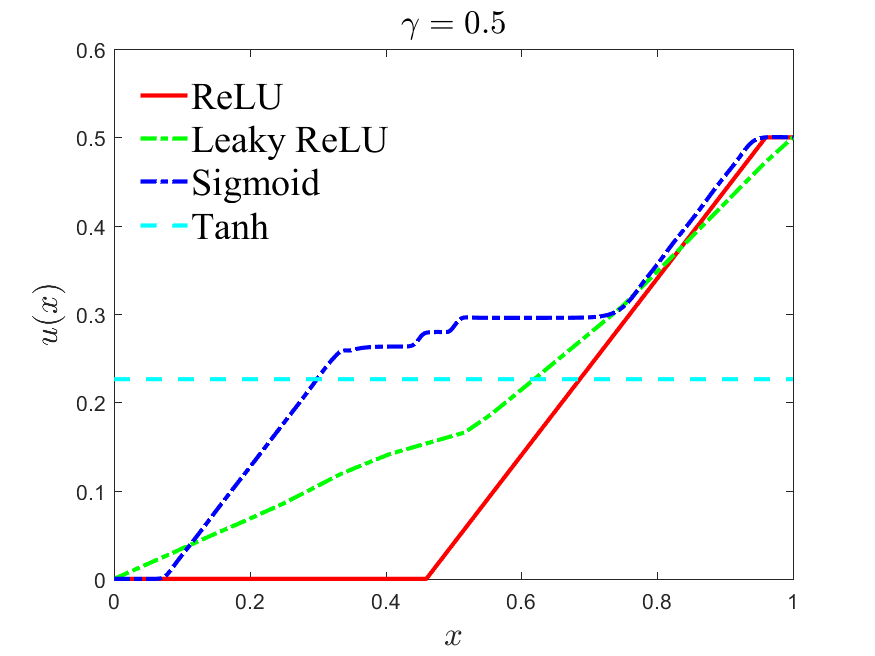}}
\caption{\textbf{Results of solving  problem \eqref{problem} with different activation function.} Shown is the plot of $u(x)$.}
\label{ac_result}
\end{figure}

In this problem, an important role is played by the activation function. As is observed in Methods, Section \ref{2.1}, an exact solution of \eqref{problem} is provided by \eqref{fsol}. We observe that in \eqref{fsol}, the function $f(x)$ defined by \eqref{frelu} is identical with the well-known Rectified Linear Unit activation function  \beq\label{ReLU}\hbox{ReLU:  }\;\sigma(x)=\max\{0,x\}={ \frac{x+|x|}{2}},\quad x\in \mathbb{R}.\eeq
Thus by choosing one weight equal to 1 and one bias equal to $\g-1$,  the DNN can provide the exact solution \eqref{fsol} to this simple one-dimensional problem.  This strongly suggests that we choose ReLU as the activation function until further notice. To test the suitability of our choice, we explore how the activation function affects the learning results. The results are shown in Fig.~\ref{ac_result}.  It is clear that only ReLU captures an exact minimizer of the problem (any piecewise smooth function with slope 0 or 1 almost everywhere that satisfies the boundary conditions), whereas other choices including Tanh, Sigmoid, and even Leaky ReLU perform poorly.  It is remarkable here that the Deep Ritz method with ReLU captures exact global minimizers in this simple problem, due to the piecewise smooth character of this activation function. The other activation functions struggle and are indeed unable to capture these weak solutions with derivative jumps. Later on we will propose a modification of ReLU to treat more complex problems.

\subsubsection{Effect of learning rate for 1D problem}
\begin{figure}[ht]
   \begin{minipage}[]{0.28 \textwidth}
\leftline{\small\textbf{(a)}}
\centerline{\includegraphics[height=4.2cm]{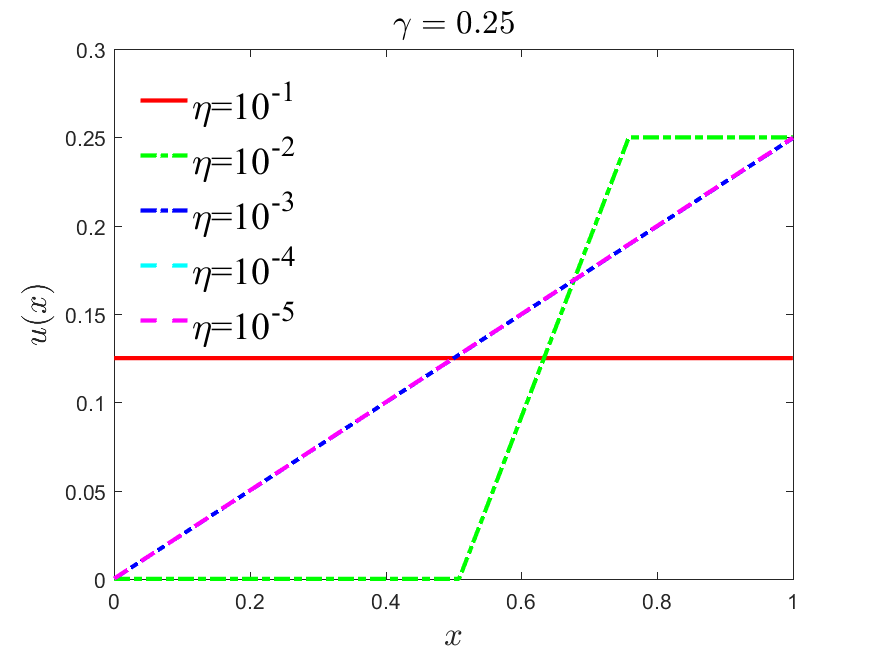}}
\end{minipage}
\hfill
 \begin{minipage}[]{0.28 \textwidth}
 \leftline{\small\textbf{(b)}}
\centerline{\includegraphics[height=4.2cm]{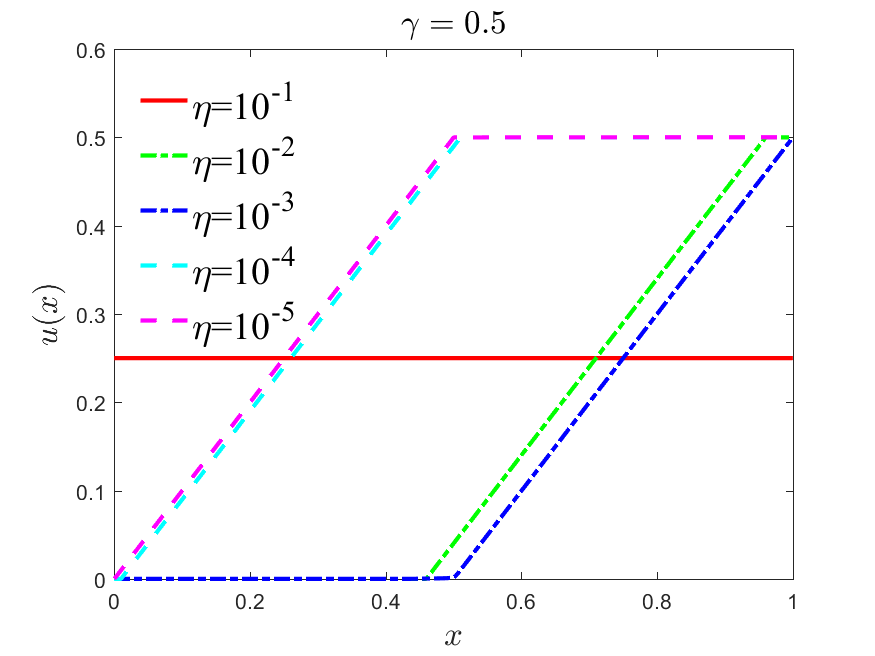}}
\end{minipage}
\hfill
   \begin{minipage}[]{0.28 \textwidth}
 \leftline{\small\textbf{(c)}}
\centerline{\includegraphics[height=4.2cm]{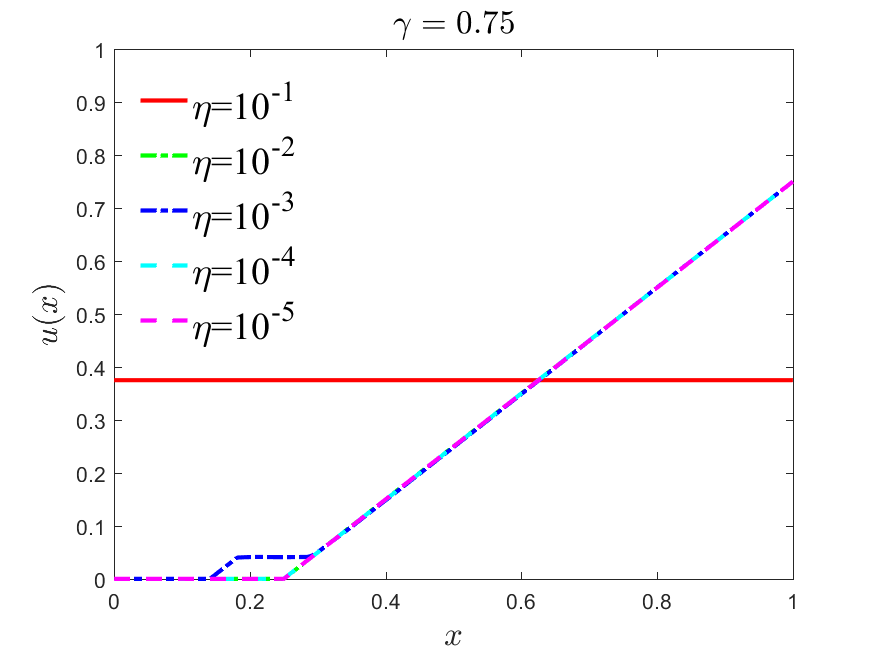}}
\end{minipage}
\caption{\textbf{ 1d results with ReLU activation function and different learning rates} with $iteration=100,000$, NN: $3\times 128$. (a) $\gamma=0.25$; (b) $\gamma=0.5$; (c) $\gamma=0.75$. }
\label{1d_lr}
\end{figure}

We  investigate how the learning rate affects the results for 3 values of the parameter
$\gamma=0.25,0.5,0.75$. We use the ReLU activation function. The results are shown in Fig.~\ref{1d_lr}. We can see if we use very large learning rate  $10^{-1}$ to train the loss function, the solution will go to a constant $u(x)=\frac{\gamma}{2}$, which is  a local minimum of the elastic energy, but does not satisfy the boundary conditions. If we use very small learning rate, the solution will get stuck in another local minimum $u(x)=\gamma x$ for the case $\gamma=0.25$, which however is not a global minimum, although it satisfies the boundary conditions. For intermediate values of the learning rate, a global minimum is achieved. We therefore should exercise some care in choosing the learning rate carefully.

\subsubsection{Effect of DNN size for 1D problem}
\begin{figure}[ht]
   \begin{minipage}[]{0.28 \textwidth}
 \leftline{\small\textbf{(a)}}
\centerline{\includegraphics[height=4.2cm]{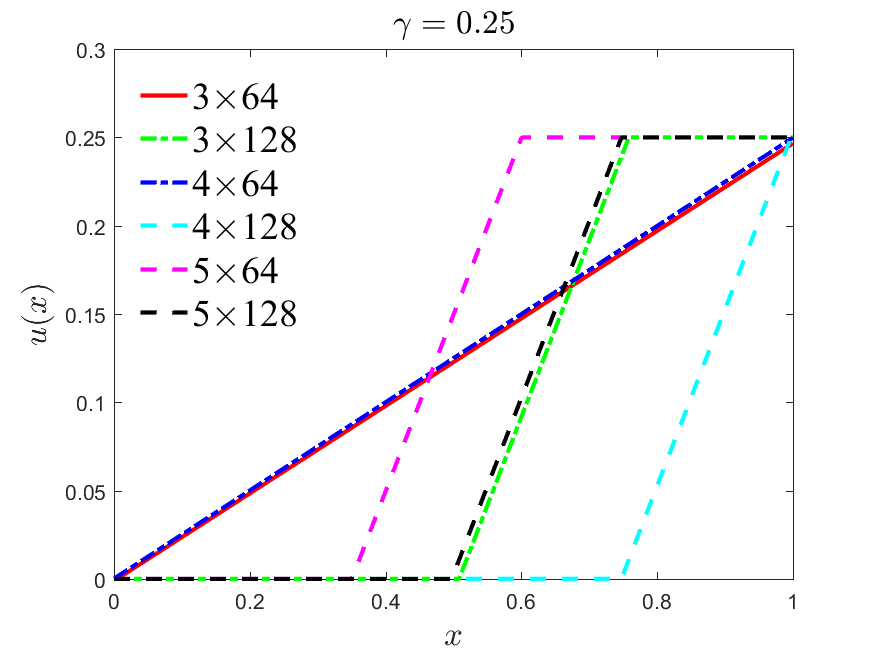}}
\end{minipage}
\hfill
 \begin{minipage}[]{0.28 \textwidth}
 \leftline{\small\textbf{(b)}}
\centerline{\includegraphics[height=4.2cm]{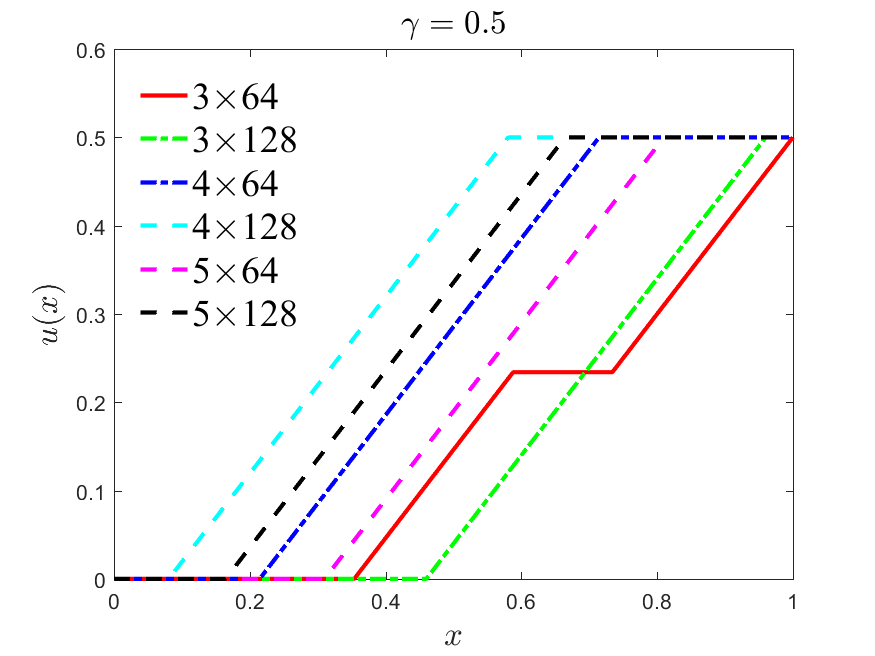}}
\end{minipage}
\hfill
   \begin{minipage}[]{0.28 \textwidth}
 \leftline{\small\textbf{(c)}}
\centerline{\includegraphics[height=4.2cm]{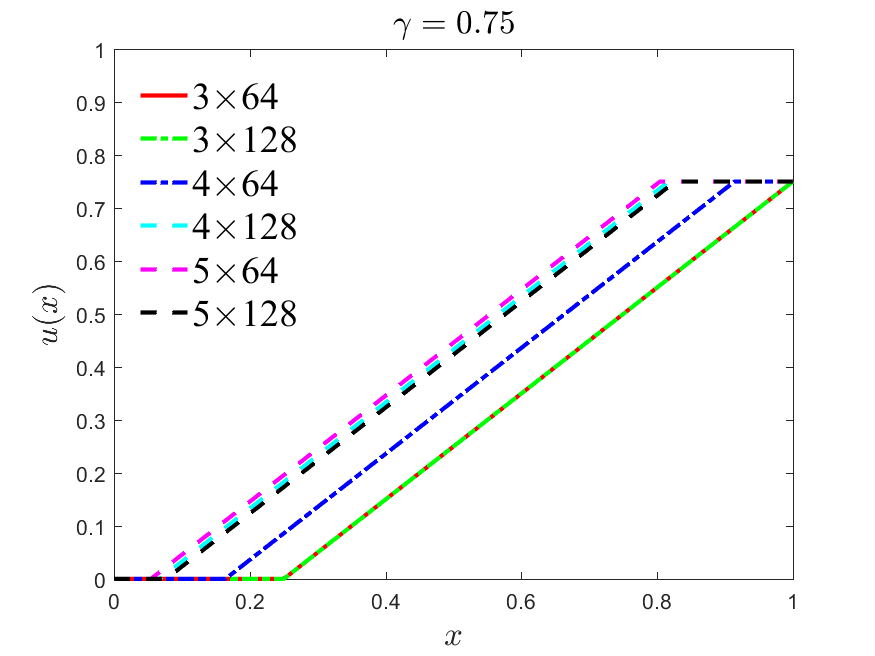}}
\end{minipage}
\caption{\textbf{ 1d results with different size DNN.} Here $iteration=100,000$ and learning rate $\eta=10^{-2}$. (a) $\gamma=0.25$; (b) $\gamma=0.5$; (c) $\gamma=0.75$. }
\label{1d_NN_structure}
\end{figure}

Then we explore how the depth and width affect the results when $\gamma=0.25,0.5,0.75$. We use the ReLU activation function. The results are shown in Fig.~\ref{1d_NN_structure}. For small $\gamma$, we need to use a larger DNN to approximate the solution $u$.

\subsubsection{Regularized 1D problem}
In order to cure the massive loss of uniqueness of minimizers for problem \eqref{problem} it is common to add higher gradients to the energy; see Remarks \ref{r1} and \ref{r2}, also \cite{cgr}. Thus we consider the  minimization problem
\beq\label{problem-rp}\min E_\eps\{u\}, \quad E_\eps\{u\}= \int_0^1  \left[  W(u'(x) )d x +{\frac{\eps^2}{2} }[ u'' (x) ]^2 \right]dx,\eeq
where $\varepsilon>0$ is a small  parameter, with boundary conditions $u(0)=0$ and $u(1)=\gamma$.

It is well known that the higher gradients in the energy \eqref{problem-rp} smoothen derivative discontinuities, and replace them with a smooth transition whose width is of order $\eps$, e.g. \cite{cgr}.  Thus we expect that the DNN with a ReLU activation function will not be able to  capture this accurately due to its lack of smoothness.  This leads us to construct a new activation function by smoothening ReLU. In \eqref{ReLU}, replace the absolute value  $|x|=\sqrt{x^2}$ by $\sqrt{x^2+\rho^2}$,  where $\rho$ is a small parameter. We call the resulting function a Smoothened Rectified Linear Unit or
\beq\label{SmReLU}\hbox{SmReLU:  }\;\sigma(x)=\frac{x+\sqrt{x^2+\rho^2}}{2},\quad x\in \mathbb{R}.\eeq
SmReLU is a smooth function for $\rho>0$; it reduces to ReLU for $\rho=0$. We typically use $\rho=0.1$ in computations.

Next, we consider how the initial condition, the value of the higher gradient coefficient $\varepsilon$,  and the activation function (ReLU vs SmReLU) affect our computations.

\begin{figure}[ht]
   \begin{minipage}[]{0.2 \textwidth}
 \leftline{\small\textbf{(a1)}}
\centerline{\includegraphics[height=3.4cm]{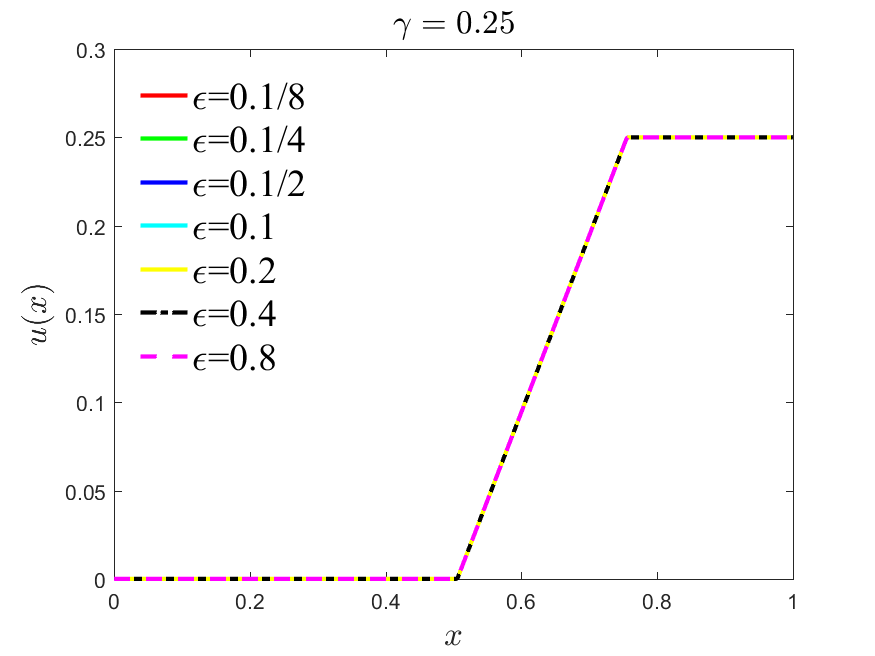}}
\end{minipage}
\hfill
 \begin{minipage}[]{0.2 \textwidth}
 \leftline{\small\textbf{(a2)}}
\centerline{\includegraphics[height=3.4cm]{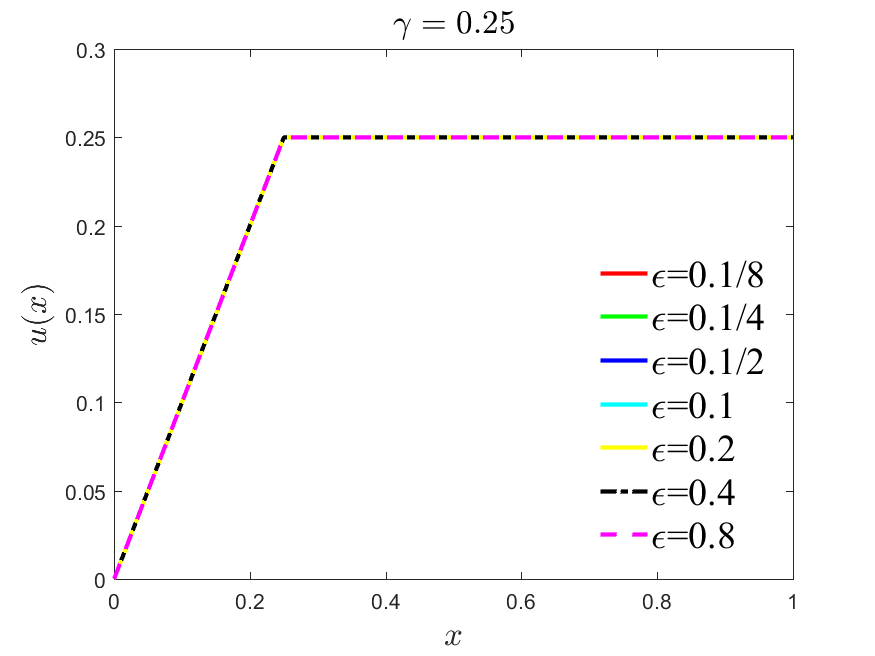}}
\end{minipage}
\hfill
   \begin{minipage}[]{0.2 \textwidth}
 \leftline{\small\textbf{(a3)}}
\centerline{\includegraphics[height=3.4cm]{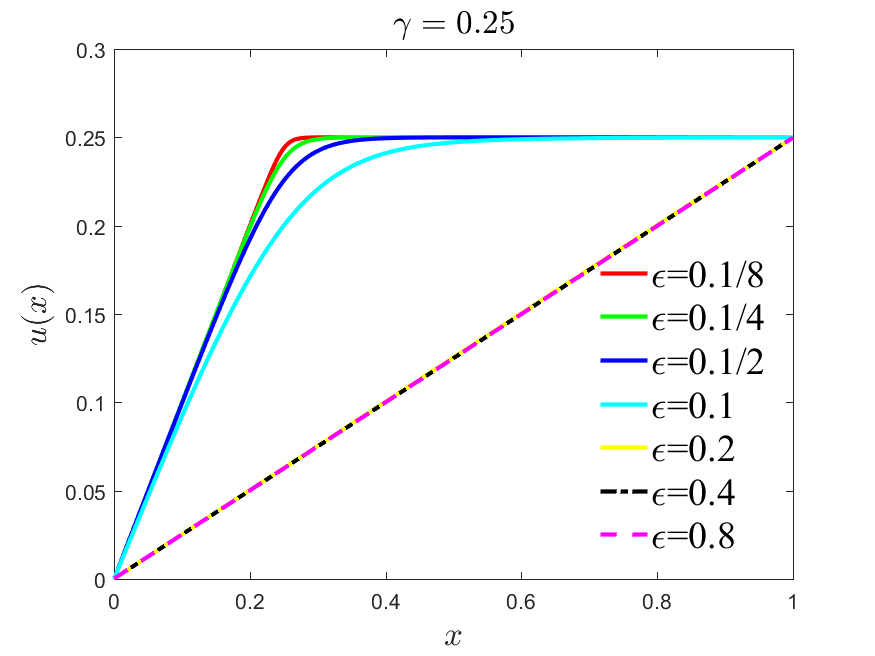}}
\end{minipage}
\hfill
 \begin{minipage}[]{0.2 \textwidth}
 \leftline{\small\textbf{(a4)}}
\centerline{\includegraphics[height=3.4cm]{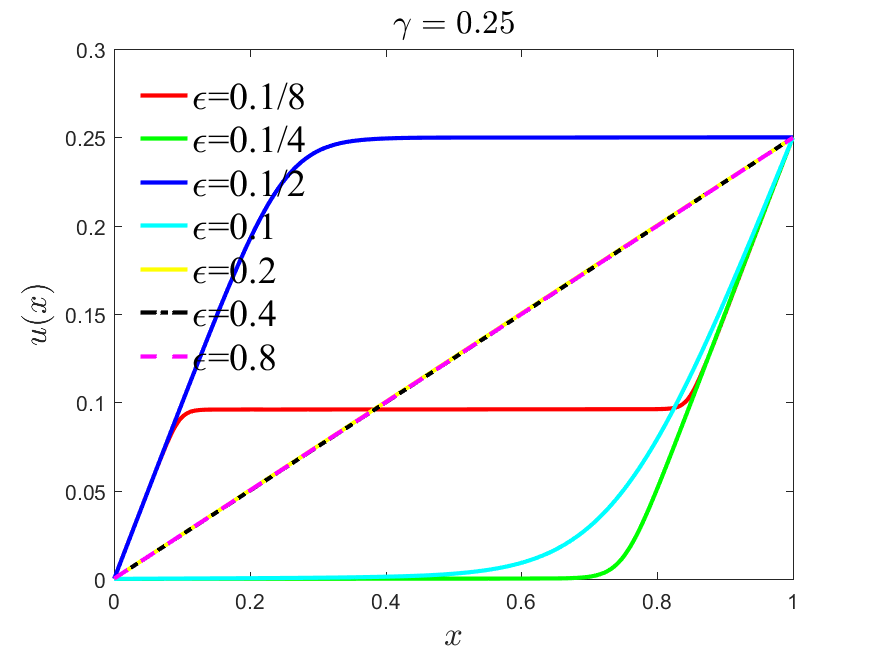}}
\end{minipage}
   \begin{minipage}[]{0.2 \textwidth}
 \leftline{\small\textbf{(b1)}}
\centerline{\includegraphics[height=3.4cm]{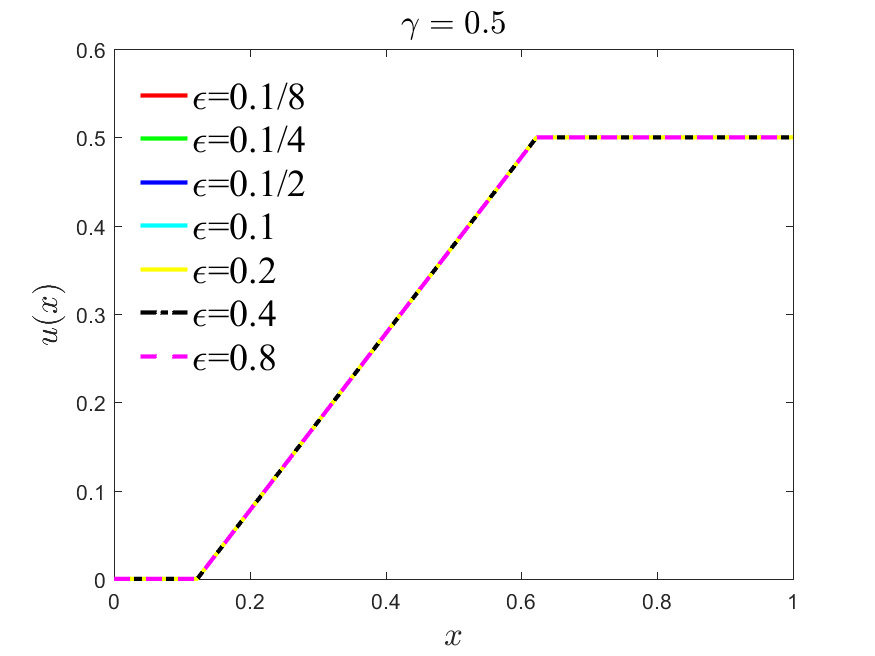}}
\end{minipage}
\hfill
 \begin{minipage}[]{0.2 \textwidth}
 \leftline{\small\textbf{(b2)}}
\centerline{\includegraphics[height=3.4cm]{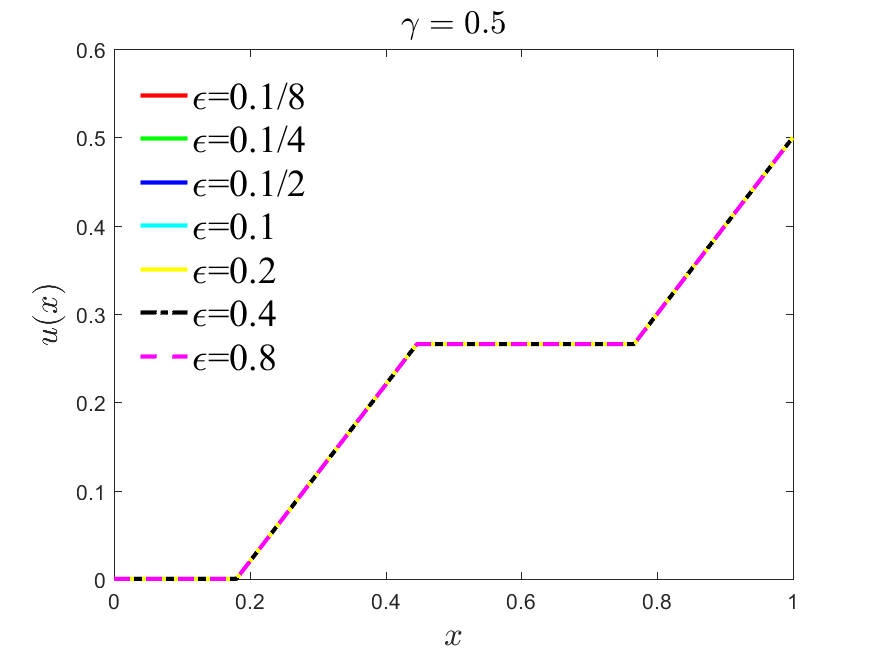}}
\end{minipage}
\hfill
   \begin{minipage}[]{0.2 \textwidth}
 \leftline{\small\textbf{(b3)}}
\centerline{\includegraphics[height=3.4cm]{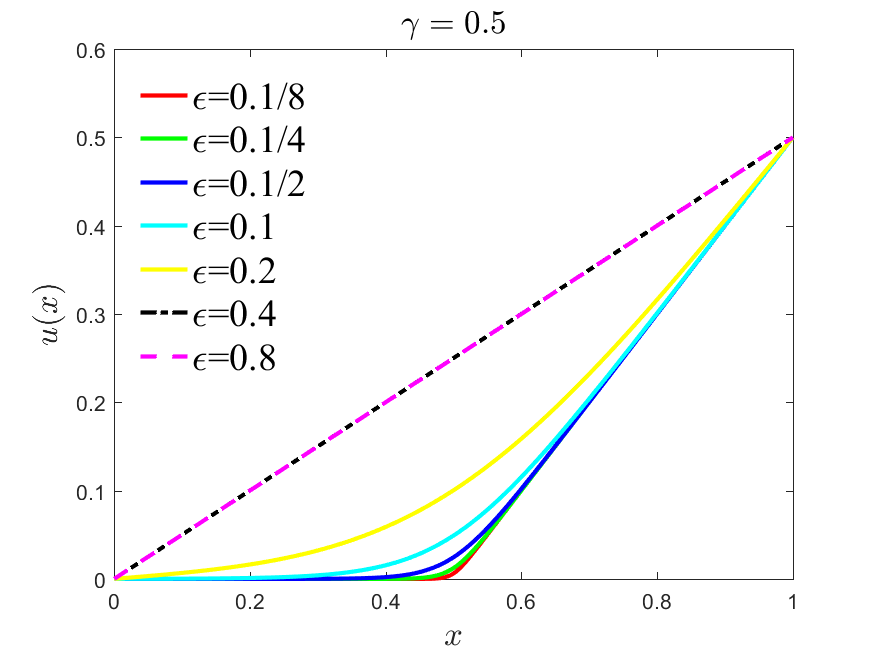}}
\end{minipage}
\hfill
 \begin{minipage}[]{0.2 \textwidth}
 \leftline{\small\textbf{(b4)}}
\centerline{\includegraphics[height=3.4cm]{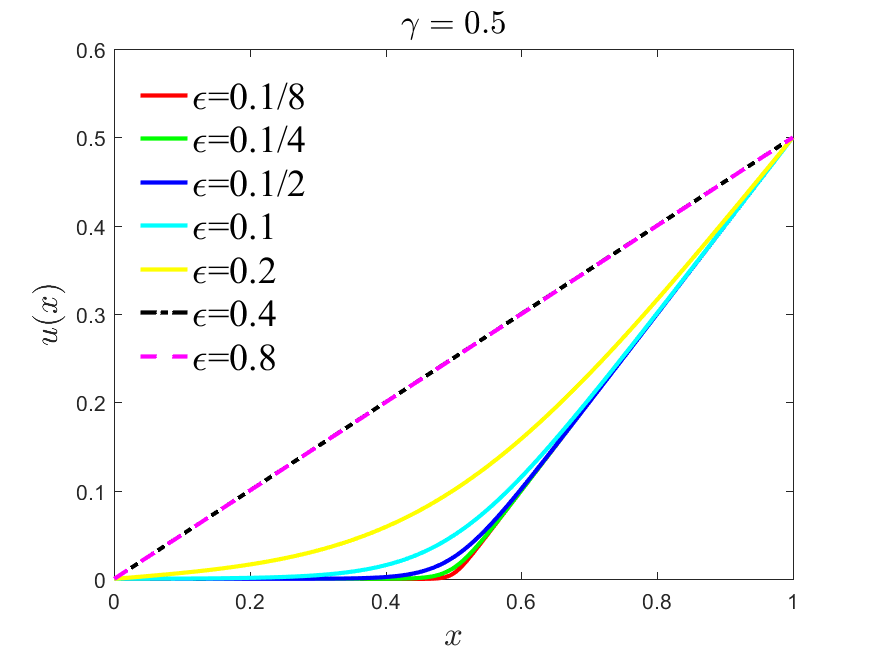}}
\end{minipage}
   \begin{minipage}[]{0.2 \textwidth}
 \leftline{\small\textbf{(c1)}}
\centerline{\includegraphics[height=3.4cm]{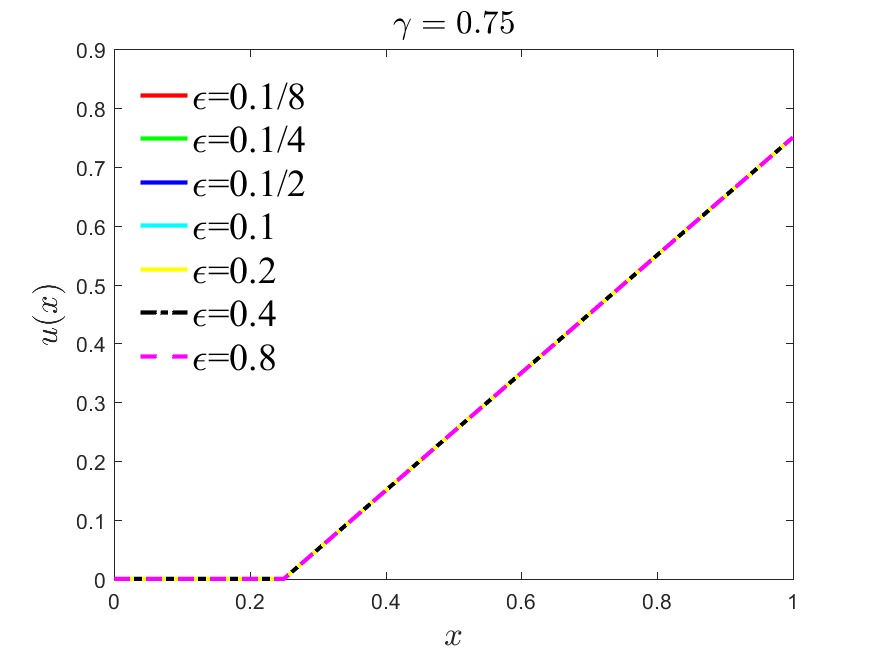}}
\end{minipage}
\hfill
 \begin{minipage}[]{0.2 \textwidth}
 \leftline{\small\textbf{(c2)}}
\centerline{\includegraphics[height=3.4cm]{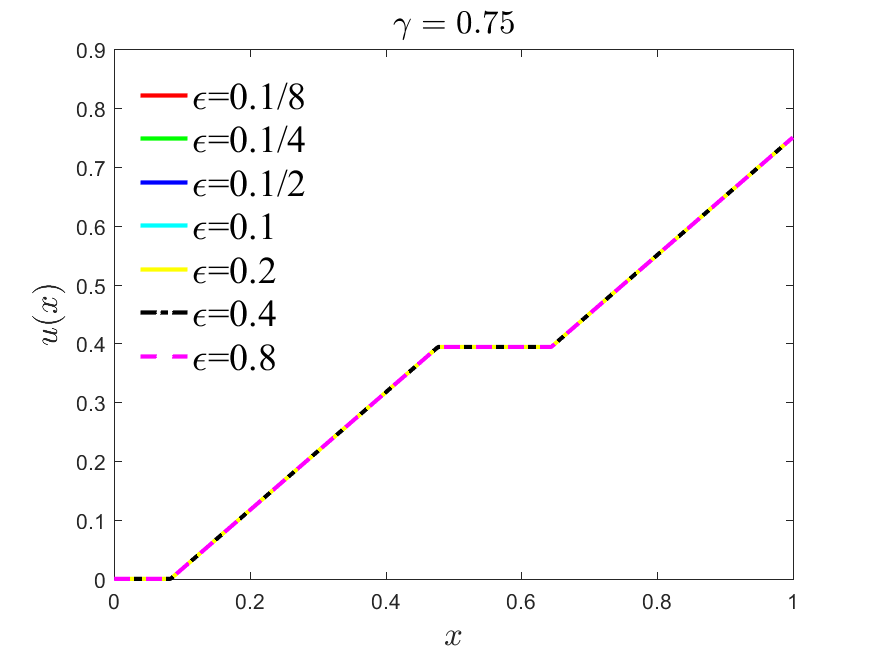}}
\end{minipage}
\hfill
   \begin{minipage}[]{0.2 \textwidth}
 \leftline{\small\textbf{(c3)}}
\centerline{\includegraphics[height=3.4cm]{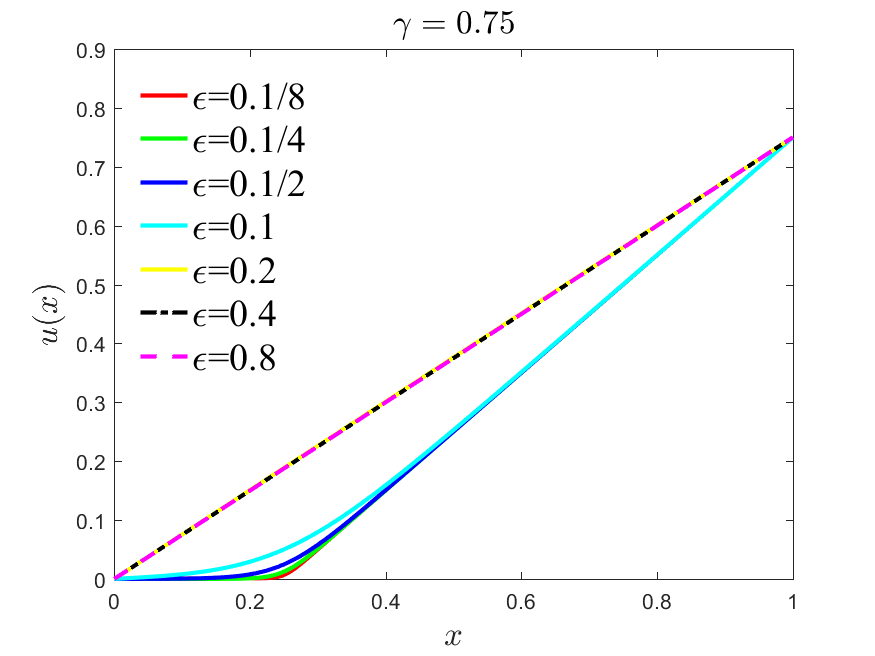}}
\end{minipage}
\hfill
 \begin{minipage}[]{0.2 \textwidth}
 \leftline{\small\textbf{(c4)}}
\centerline{\includegraphics[height=3.4cm]{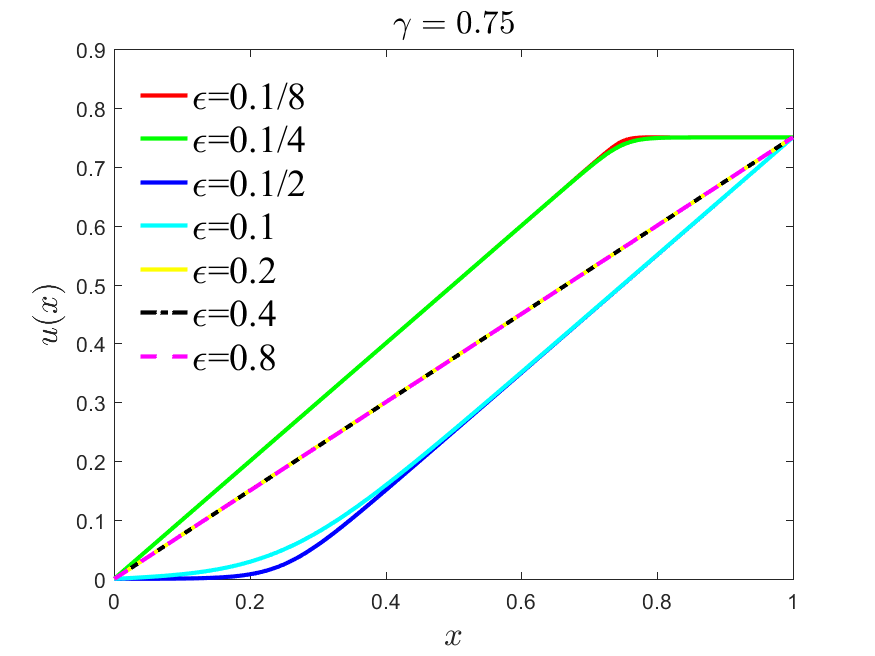}}
\end{minipage}
\caption{\textbf{ 1d regularized problem  results.} Here $iteration=100,000$, $\eta=10^{-2}$ and NN: $3\times128$. (a) $\gamma=0.25$; (b) $\gamma=0.5$; (c) $\gamma=0.75$.
First column: Random initial condition and ReLU activation function; Second column: $\gamma x+0.1\sin(4x)$  initial condition and ReLU activation function; Third column: Random initial condition and SmReLU activation function; Last column: $\gamma x+0.1\sin(4x)$ initial condition and SmReLU activation function.}
\label{1d_reg}
\end{figure}

The results are shown in Fig.~\ref{1d_reg}. We use a $3\times128$ DNN with ReLU and SmReLU activation function to approximate the solution $u$. The learning rate is $\eta=10^{-2}$ and $\tau=500$.

If we use the ReLU activation function, the optimal $u$ is not differentiable at some points, and moreover, more than one derivative jumps (kinks) occur when the initial condition has oscillations; see Fig.~\ref{1d_reg}. On the other hand,  problem \eqref{problem-rp}  has a unique smooth minimizer (modulo reflection about $x=1/2$) \cite{cgr} with a single smoothened kink (transition layer between slopes 0 and 1). This is well captured by the SmReLU DNN, which gives solutions with one smooth kink. The only exception which has a local minimum with two kinks occurs in Fig.~\ref{1d_reg} (a4).  For larger $\varepsilon$, the DNN learns a more smooth solution with a wider transition layer, in agreement with theory \cite{cgr}. If  $\varepsilon$ is large enough, the regularization term dominates the loss function. So to minimize the second derivative energy term,  $u$ is very close to a linear function for this case. We conclude that the SmReLU activation function \eqref{SmReLU} we proposed works well for this problem and captures the behavior expected from theory.

\clearpage{}
\subsection{Two-dimensional problem}

Our main goal is to consider the following two dimensional nonconvex problem:
\begin{equation}\label{problem_2d}
   \min \int_{\Omega}W(\nabla u(x,y))dxdy,
\end{equation}
 and its regularized counterpart \eqref{problem_2d_reg}. Here
\beq \label{W} W(\nabla u)=\frac{1}{2}[u_x^2(1-u_x)^2+u_y^2],\eeq and $\Omega=[0,L]\times[0,1]$ with $L=1$ or $2$.
We construct a fully connected DNN for $u$ with input $(x,y)$. The loss function is defined as follows:
\begin{equation}\label{loss_all_2d}
loss=loss_e+\tau loss_b,
\end{equation}
where
\begin{align}\label{loss_min}
loss_{e} &=\frac{1}{N}\sum_{i=1}^N W(\nabla u_{NN}(x_i,y_i)),\nonumber\\
loss_b&=\frac{1}{N_b}\sum_{j=1}^{N_b}(u_{NN}(x_j^{b},y_j^{b})-u_{true}(x_j^{b},y_j^{b}))^2,\nonumber\
\end{align}
$\{(x_i,y_i)\}_{i=1}^N$ are the collocation points in the domain $\Omega$, $\{(x_j^{b},y_j^{b})\}_{i=1}^{N_b}$ are the collocation points at the parts of the  boundary where Dirichlet boundary conditions are specified,  while $\tau$ is the weight to scale and balance the losses from the energy ($loss_e$) and the boundary conditions ($loss_b$).

\subsubsection{Mixed Boundary Conditions}\label{mixed}
\begin{figure}
\begin{minipage}[]{0.3 \textwidth}
 \leftline{\small\textbf{(a)}}
\centerline{\includegraphics[height=4cm]{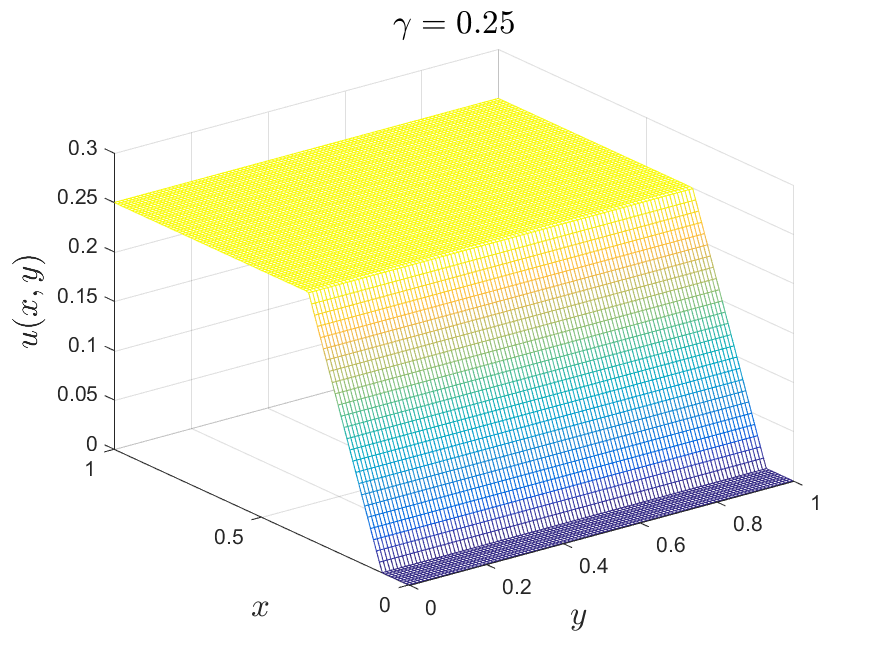}}
\end{minipage}
\hfill
\begin{minipage}[]{0.3 \textwidth}
 \leftline{\small\textbf{(b)}}
\centerline{\includegraphics[height=4cm]{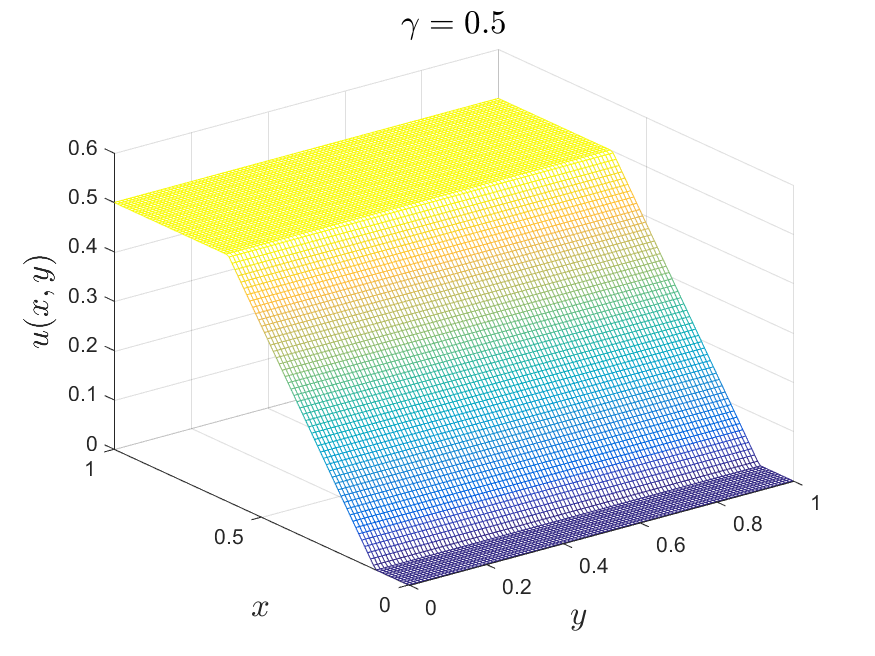}}
\end{minipage}
\hfill
\begin{minipage}[]{0.3 \textwidth}
 \leftline{\small\textbf{(c)}}
\centerline{\includegraphics[height=4cm]{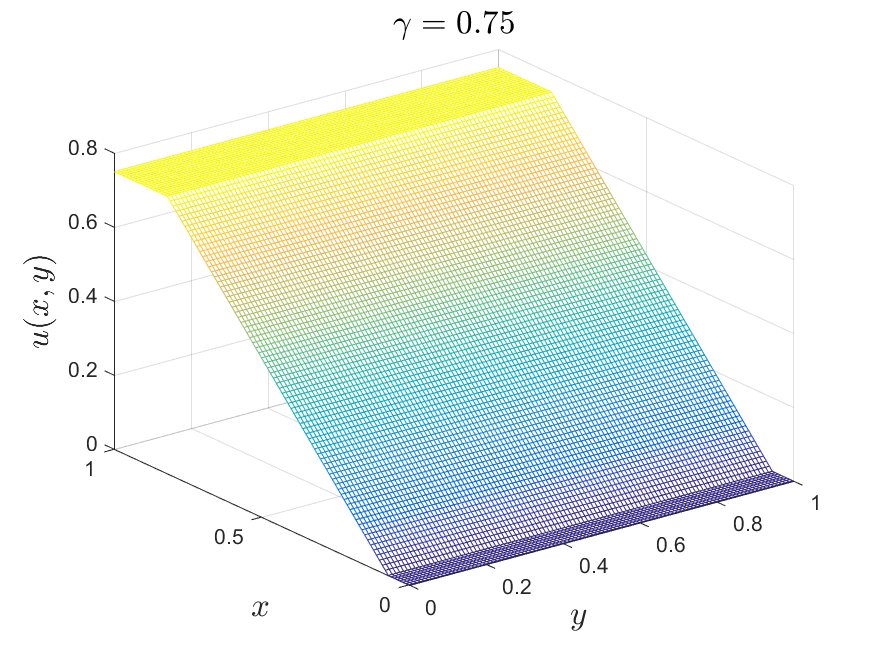}}
\end{minipage}
\begin{minipage}[]{0.3 \textwidth}
\centerline{\includegraphics[height=4cm]{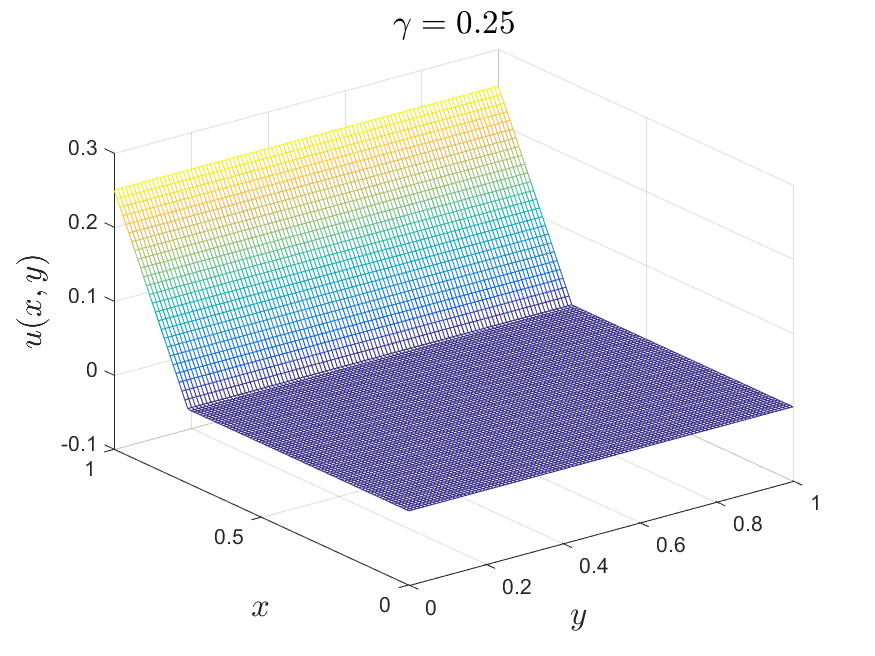}}
\end{minipage}
\hfill
\begin{minipage}[]{0.3 \textwidth}
\centerline{\includegraphics[height=4cm]{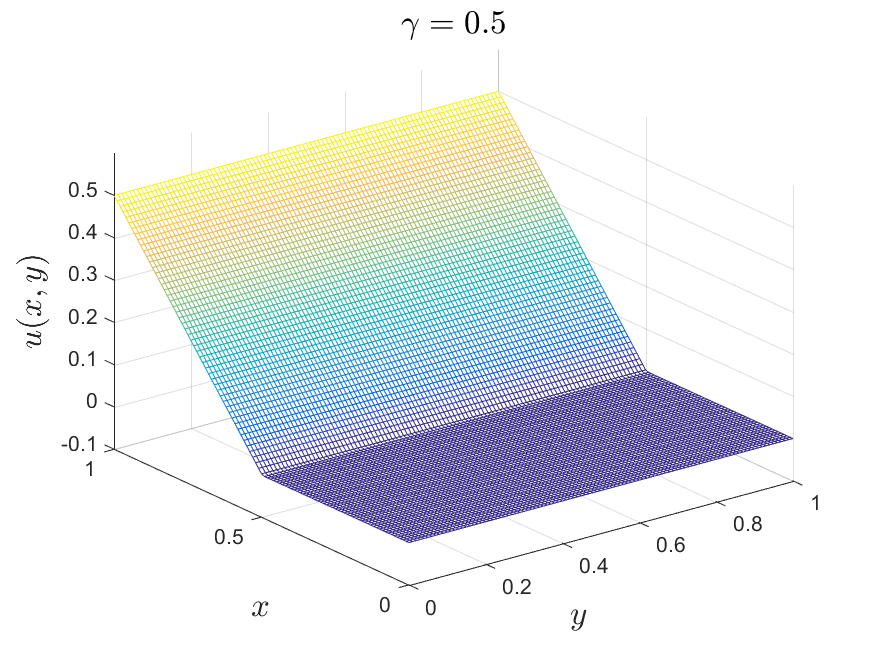}}
\end{minipage}
\hfill
\begin{minipage}[]{0.3 \textwidth}
\centerline{\includegraphics[height=4cm]{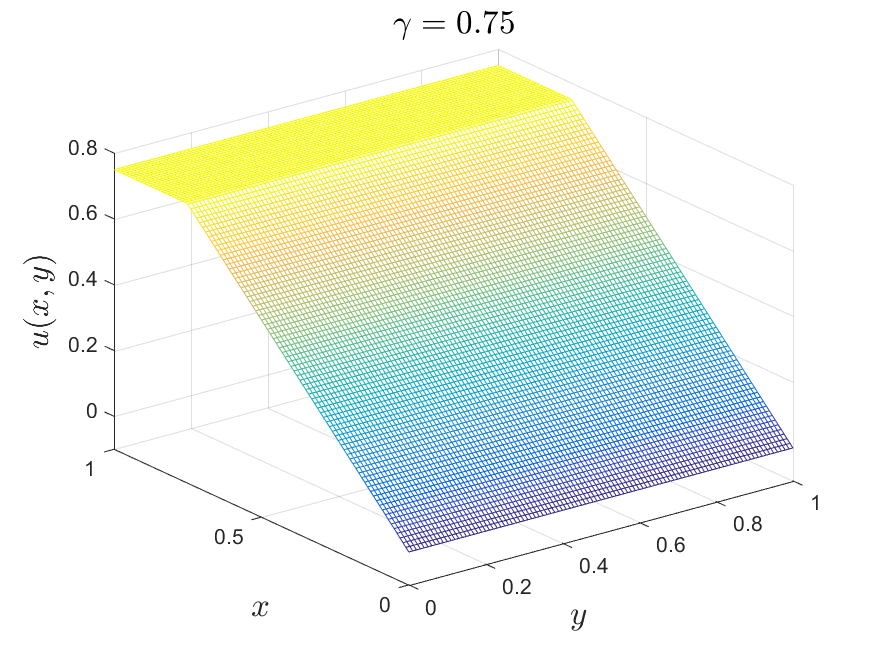}}
\end{minipage}
\caption{\textbf{ 2d mixed boundary conditions.} (a) $\gamma=0.25$; (b) $\gamma=0.5$; (c) $\gamma=0.75$. Top: ReLU; bottom: SmReLU activation function.}
\label{2d_MB}
\end{figure}

Consider the mixed boundary conditions:
\begin{equation}\label{DB}
u(0,y)=0,~ u(1,y)=\gamma, ~~~0<\gamma<1,
\end{equation}
and the other two sides are free. Functions $u$ that are independent of $y$ minimize the energy provided that they are solutions of the one-dimensional problem \eqref{problem}. We use both ReLU and SmReLU activation function, and  $3\times 128$ DNNs to approximate $u$ when $\gamma=0.25,~0.5,~0.75$.
The learned solutions are shown in Fig.~\ref{2d_MB}.
In the top row of Fig.~\ref{2d_MB}, we show results from the ReLU DNN.  The bottom row of Fig.~\ref{2d_MB} shows the SmReLU DNN minimizers. In the latter case solutions only have one gradient jump, whereas ReLU solutions typically have two.

Our numerical results match the exact $y$-independent solutions of the one-dimensional problem.

\subsubsection{Dirichlet Boundary Conditions}

In the following, we will explore Dirichlet boundary condition
\begin{equation}\label{2d_DB}
u(x,y)=\gamma x, \quad \forall (x,y)\in \partial \Omega.
\end{equation}
A typical value of interest is $\g=1/2$. In order to get a zero energy deformation  $u_x$ should take values  0 or 1 with $u_y=0$, which is what happens in Fig.~\ref{2d_MB}, the mixed boundary condition case considered in subsection \ref{mixed}.  On the other hand, here the boundary conditions \eqref{2d_DB} dictate that $u_x$ be close to 1/2 near the boundary so there is an incompatibility between the energy wells and the boundary conditions. In fact it is possible to show that problem \eqref{problem_2d} subject to \eqref{2d_DB} does not have a minimizer \cite{ball,kohn}. Instead, the energy can be effectively minimized in the limit by a sequence of deformations that involve a laminated microstructure, consisting of vertical parallel interfaces between domains with gradients in alternating  wells $(u_x,u_y)=(0,0)$ and $(1,0)$. Such a minimizing sequence involves a larger and larger number of twin boundaries (jumps in $\nabla u$ along lines $x=$const.), in effect converging to a mixture of phases, where in the limit $u_x$ converges weakly to the average value $1/2$, so that it becomes compatible with the boundary conditions $u=x/2$ on $\partial\Om$.

\begin{figure}[ht]
\centerline{\includegraphics[height=3.5cm]{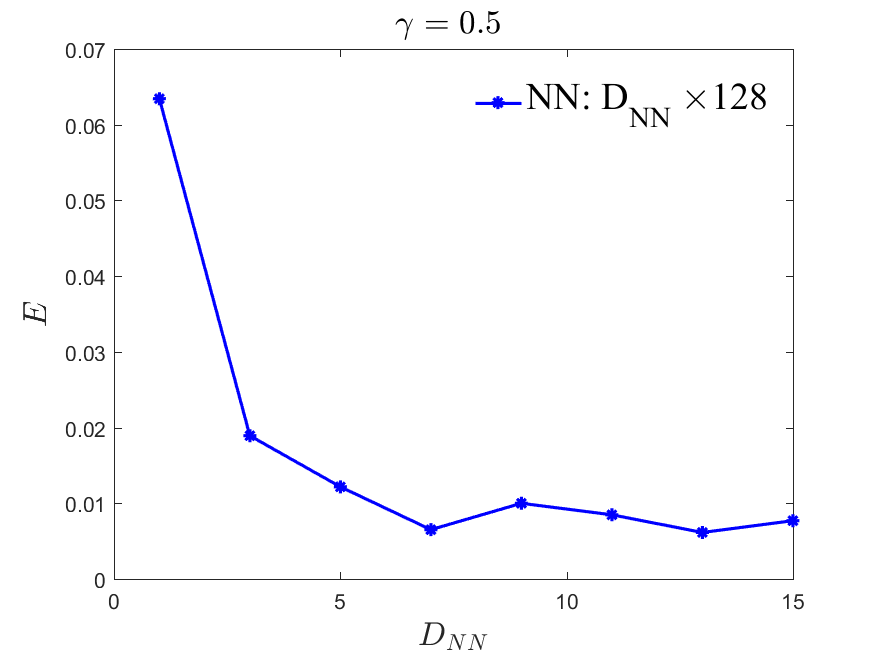}}
   \begin{minipage}[]{0.2 \textwidth}
 \leftline{\small\textbf{(a)}}
\centerline{\includegraphics[height=3.5cm]{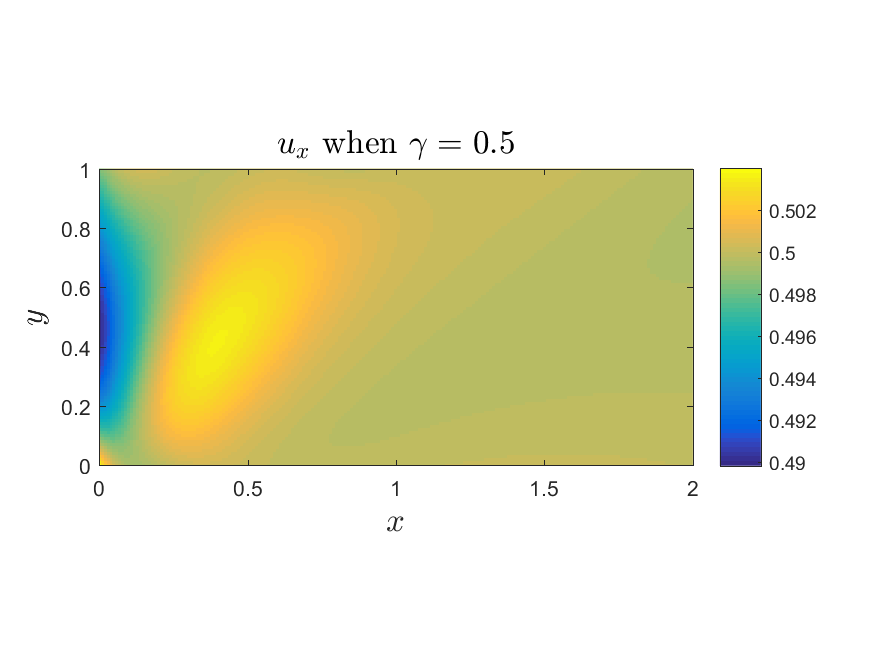}}
\end{minipage}
\hfill
 \begin{minipage}[]{0.2 \textwidth}
 \leftline{\small\textbf{(b)}}
\centerline{\includegraphics[height=3.5cm]{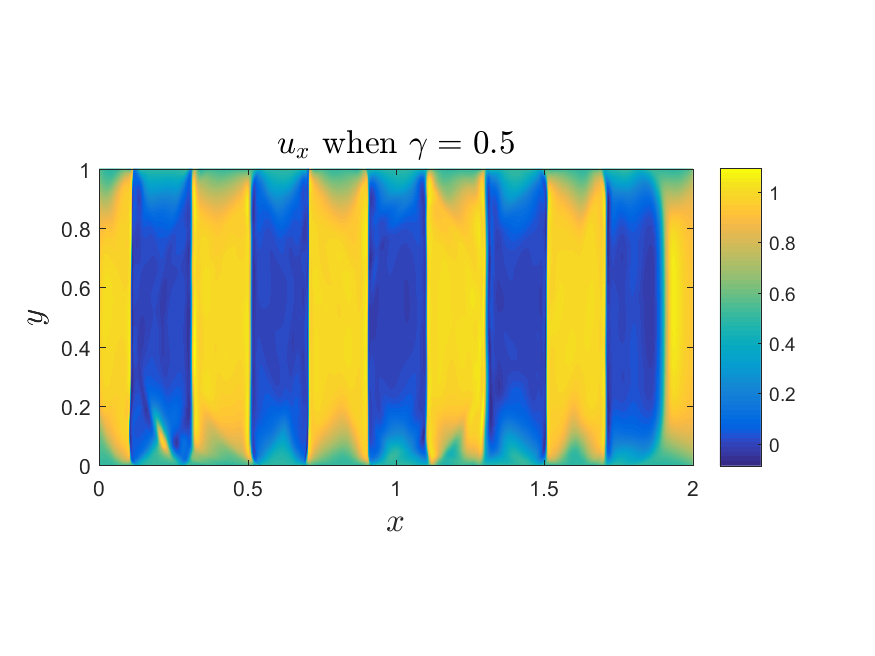}}
\end{minipage}
\hfill
\begin{minipage}[]{0.2 \textwidth}
 \leftline{\small\textbf{(c)}}
\centerline{\includegraphics[height=3.5cm]{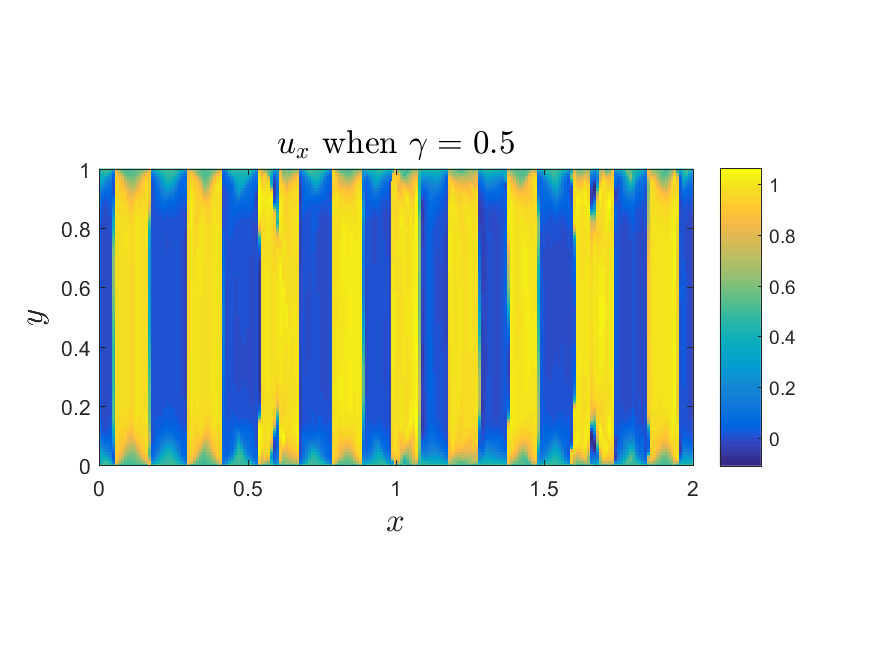}}
\end{minipage}
\hfill
\begin{minipage}[]{0.2 \textwidth}
 \leftline{\small\textbf{(d)}}
\centerline{\includegraphics[height=3.5cm]{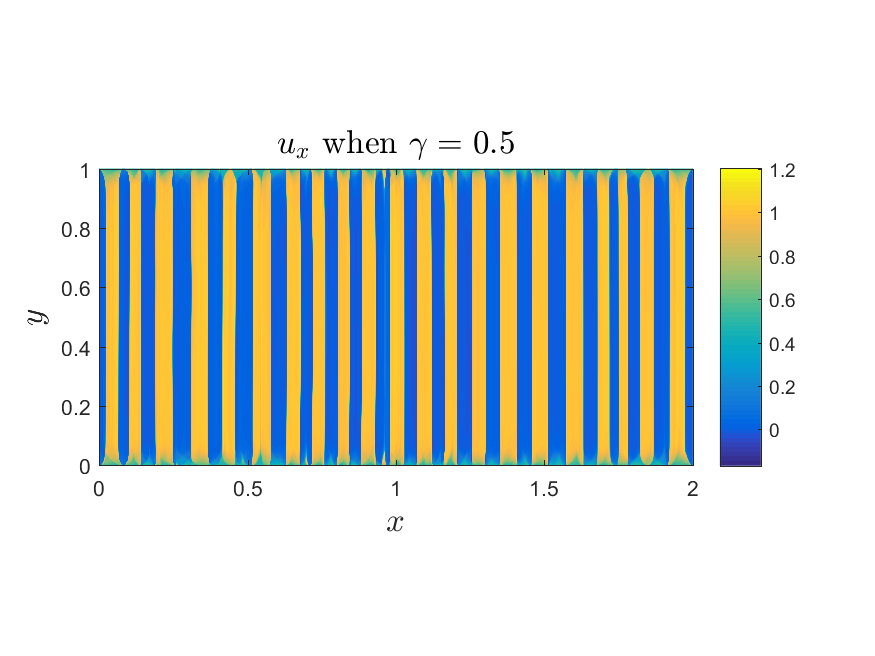}}
\end{minipage}
 \begin{minipage}[]{0.2 \textwidth}
 \leftline{\small\textbf{(e)}}
\centerline{\includegraphics[height=3.5cm]{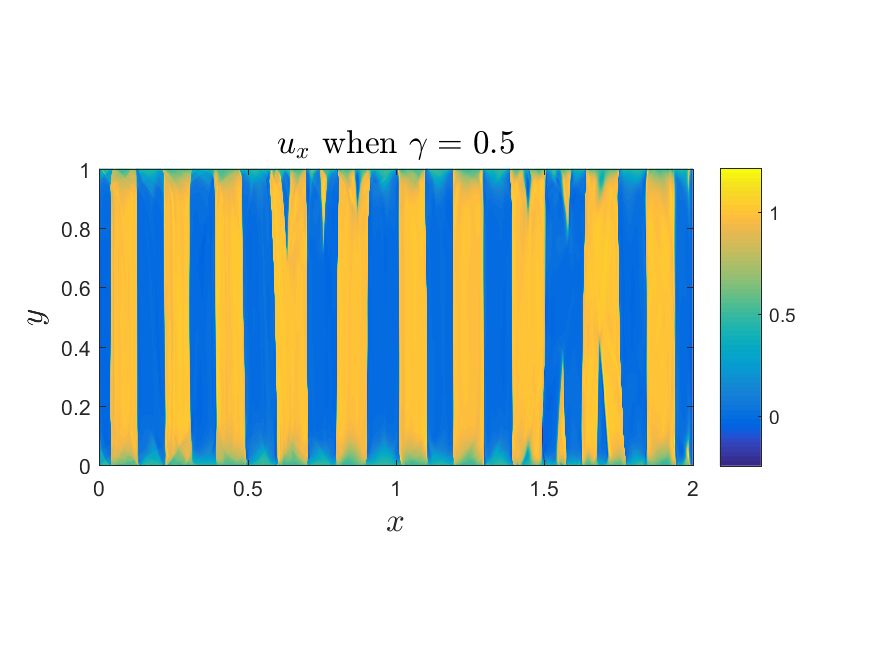}}
\end{minipage}
\hfill
\begin{minipage}[]{0.2 \textwidth}
 \leftline{\small\textbf{(f)}}
\centerline{\includegraphics[height=3.5cm]{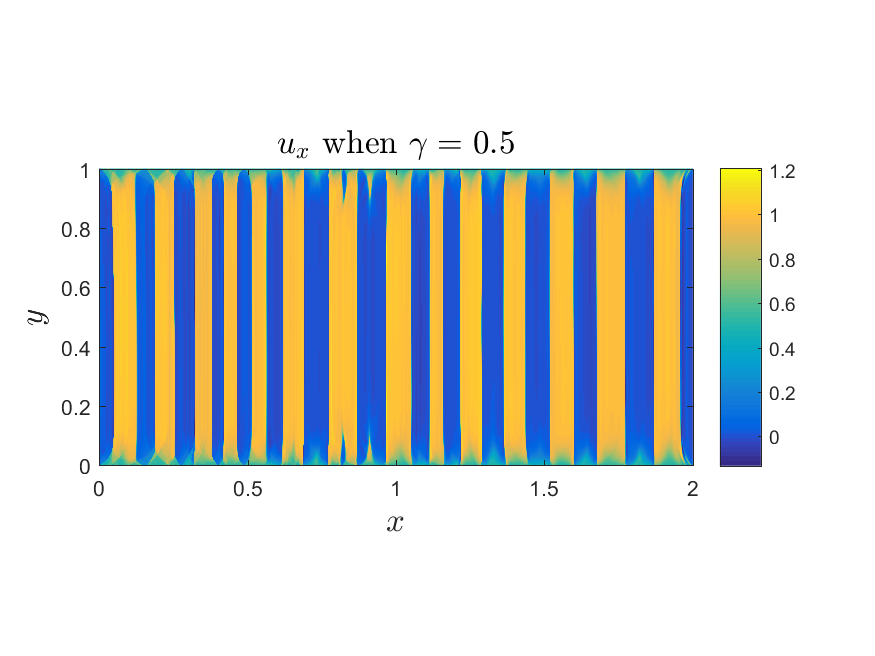}}
\end{minipage}
\hfill
\begin{minipage}[]{0.2 \textwidth}
 \leftline{\small\textbf{(g)}}
\centerline{\includegraphics[height=3.5cm]{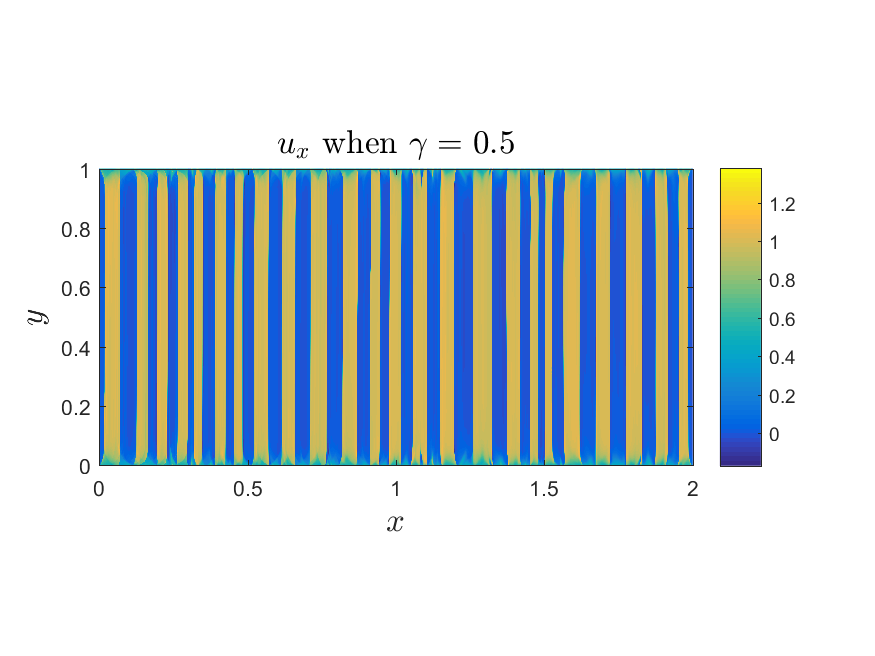}}
\end{minipage}
\hfill
\begin{minipage}[]{0.2 \textwidth}
 \leftline{\small\textbf{(h)}}
\centerline{\includegraphics[height=3.5cm]{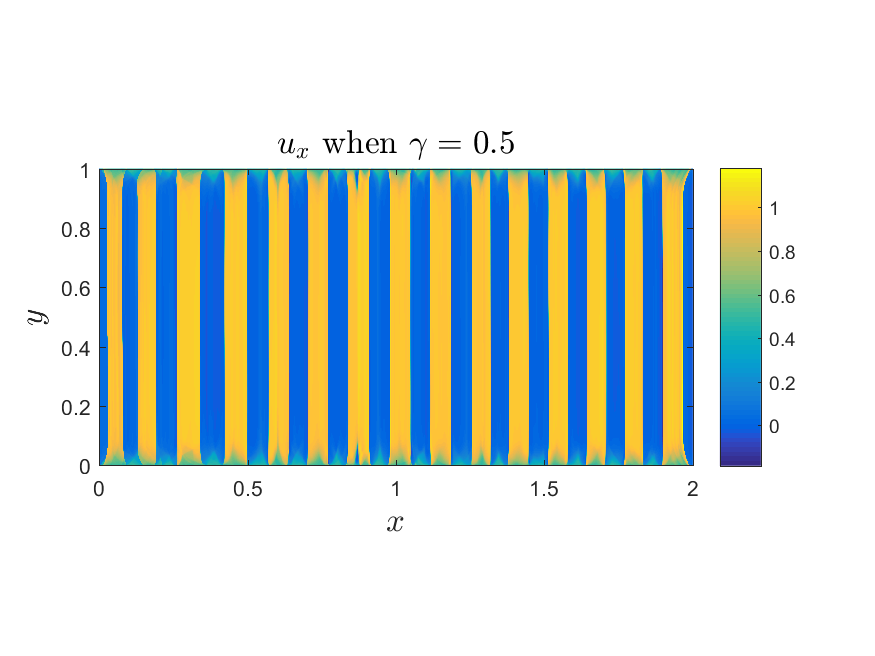}}
\end{minipage}
\caption{\textbf{ 2d Dirichlet boundary condition with SmReLU activation function.} Here $\gamma=0.5$, $iteration=300,000$, $\eta=10^{-3}$.(a) NN: $1\times 128$; (b) NN: $3\times 128$; (c) NN: $5\times 128$; (d) NN: $7\times 128$; (e) NN: $9\times 128$; (f) NN: $11\times 128$;  (g) NN: $13\times 128$;  (h) NN: $15\times 128$.}
\label{2d_depth}
\end{figure}

\begin{figure}[ht]
    \centerline{\includegraphics[height=3.5cm]{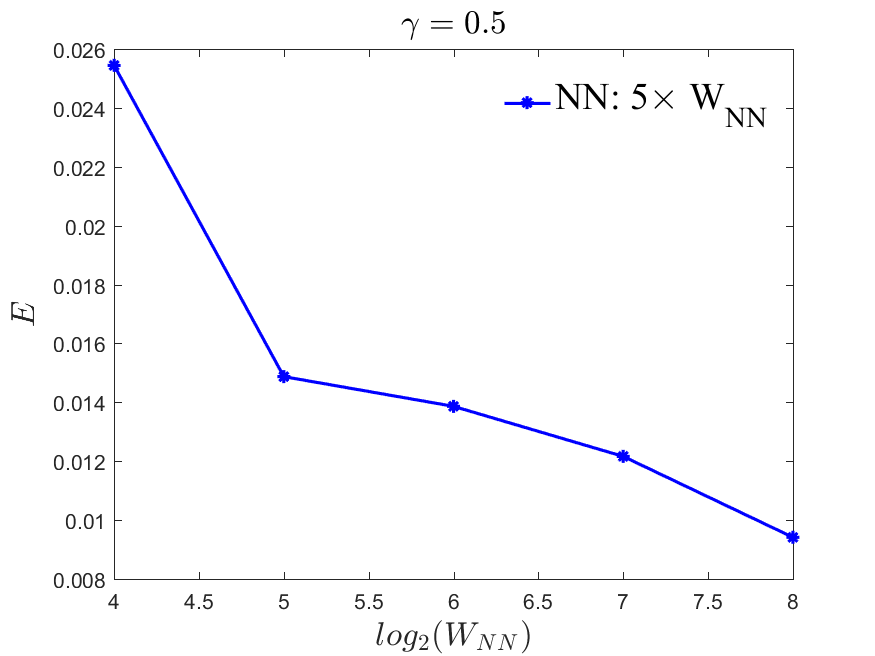}}
   \begin{minipage}[]{0.2 \textwidth}
 \leftline{\small\textbf{(a)}}
\centerline{\includegraphics[height=3.5cm]{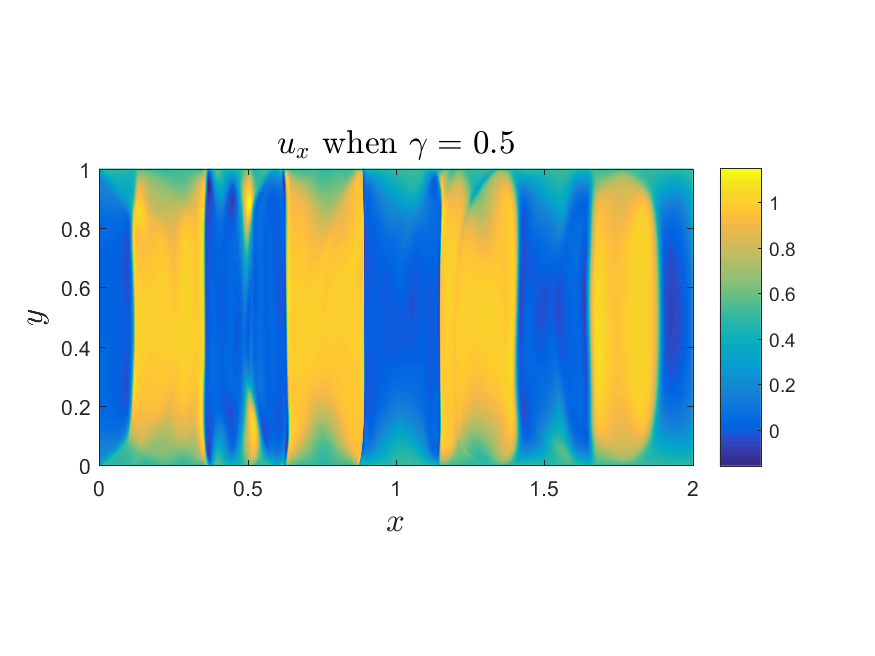}}
\end{minipage}
\hfill
 \begin{minipage}[]{0.2 \textwidth}
 \leftline{\small\textbf{(b)}}
\centerline{\includegraphics[height=3.5cm]{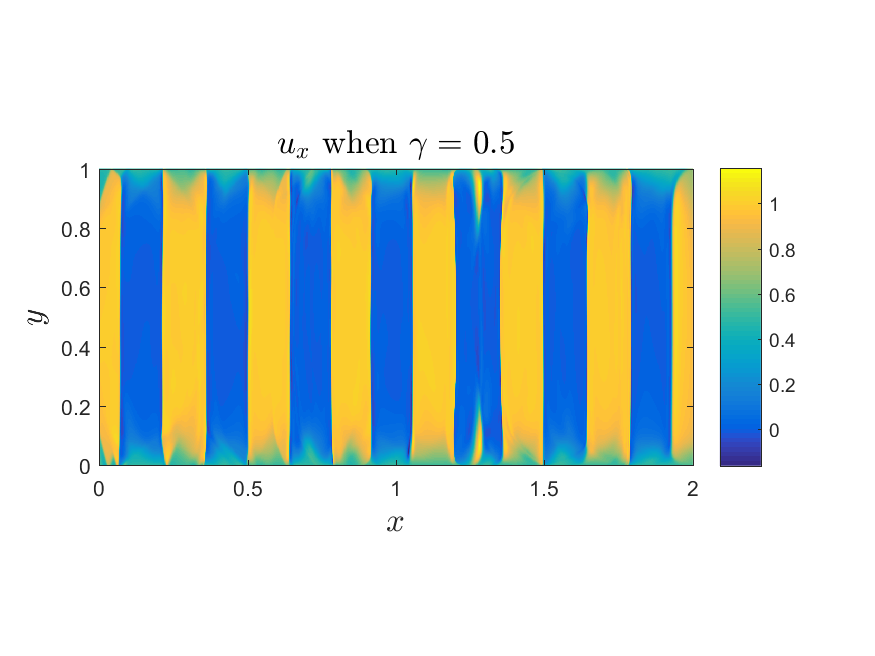}}
\end{minipage}
\hfill
\begin{minipage}[]{0.2 \textwidth}
 \leftline{\small\textbf{(c)}}
\centerline{\includegraphics[height=3.5cm]{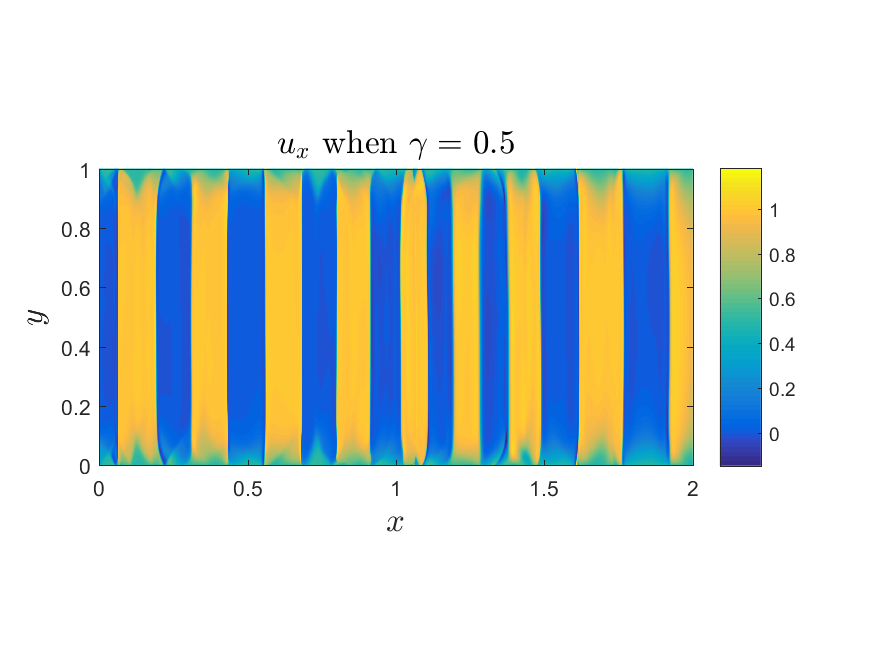}}
\end{minipage}
\hfill
\begin{minipage}[]{0.2 \textwidth}
 \leftline{\small\textbf{(d)}}
\centerline{\includegraphics[height=3.5cm]{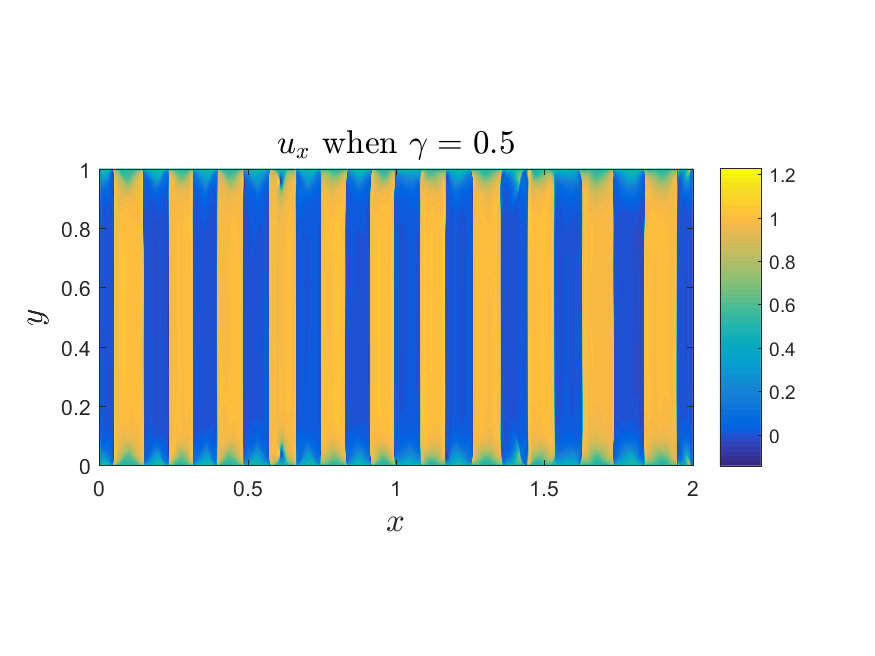}}
\end{minipage}
\caption{\textbf{ 2d Dirichlet boundary condition with SmReLU activation function.} Here $\gamma=0.5$, $iteration=300,000$, $\eta=10^{-3}$. Top: the energy with different (a) NN: $5\times 16$; (b) NN: $5\times 32$; (c) NN: $5\times 64$; (d) NN: $5\times 256$.}
\label{2d_width}
\end{figure}

We  show results using the SmReLU activation function and $\gamma=0.5$ in Fig.~\ref{2d_depth} and \ref{2d_width}. Laminated microstructures prevail with alternating bands having $u_x$ jumping between values near 0 (blue) and 1 (yellow) to minimize the energy density $W$, except near horizontal boundaries where these values are incompatible with boundary conditions \eqref{2d_DB} that require $u_x=1/2$. This causes each band to split into two or more near the boundary as is also observed in experiments  Fig.~\ref{twins}(a) \cite{chu}.  In Fig.~\ref{2d_depth}, we study the effect of DNN depth (number of layers) in capturing local minima with more and more bands. With one layer only, the solution is stuck at an unstable state (saddle point of $W$) with $u_x\approx 0.5$, Fig.~\ref{2d_depth}(a). The number of bands increases as we increase the number of  layers (depth) of the DNN, albeit with some oscillations. The energy also decreases and then appears to tend to a constant. In Fig.~\ref{2d_width}, we fix the depth as $5$ and increase the width of the DNNs. The energy will decrease for wider DNNs. If we use larger DNNs (deeper or wider), we can get more refined microstructures in our results, but we recall that there is no limit to the number and fineness of bands in this problem all the way to infinity \cite{ball,kohn}, as there is no global minimum, but apparently a sequence of local minima with an increasing number of bands, like the ones observed here,  that approach the infimum of energy. In a wider network, there will be more parameters (weights and biases) to be trained, rendering optimization more costly.

\clearpage{}
\subsubsection{Regularized Scalar Problem with Dirichlet Boundary Conditions}
In order to study a problem that actually has a global minimum we use the common approach of adding higher gradients and studying the regularized problem
\begin{equation}\label{problem_2d_reg}
   \min \int_{\Omega}\bigg[ W(\nabla u(x,y))+\frac{\varepsilon^2}{2}\big[\frac{\partial^2u(x,y)}{\partial x^2}\big]^2 \bigg]dxdy,
\end{equation}
where $W$ is given by \eqref{W}, $\Omega=[0,2]\times[0,1]$ and the boundary condition is the linear Dirichlet one \eqref{2d_DB}.  The higher gradients render local minima smooth and gradient discontinuities become smooth transition layers with width of order $\eps$ \cite{cgr,healey, kohn,dondl,conti}. Also there is a global minimizer, but a careful bifurcation analysis \cite{healey} has revealed multitudes of stable local minima with multiple bands. A comparison of one of our solutions with one of the equilibria found in \cite{healey} is shown in Fig.~\ref{twins}(b), (c).

\begin{figure}[ht]
 \centerline{\includegraphics[height=3.5cm]{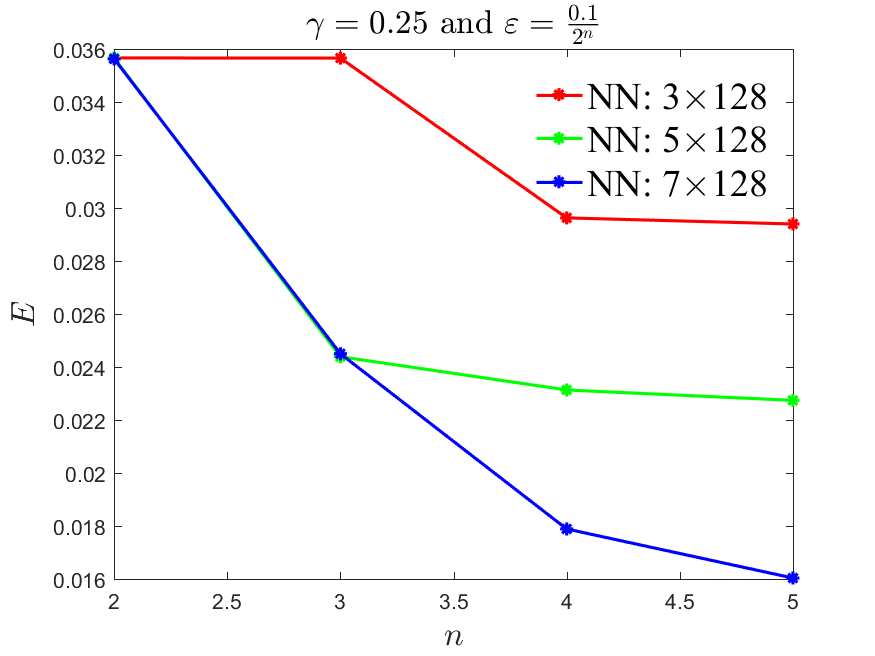}}
   \begin{minipage}[]{0.2 \textwidth}
 \leftline{\small\textbf{(a1)}}
\centerline{\includegraphics[height=3.5cm]{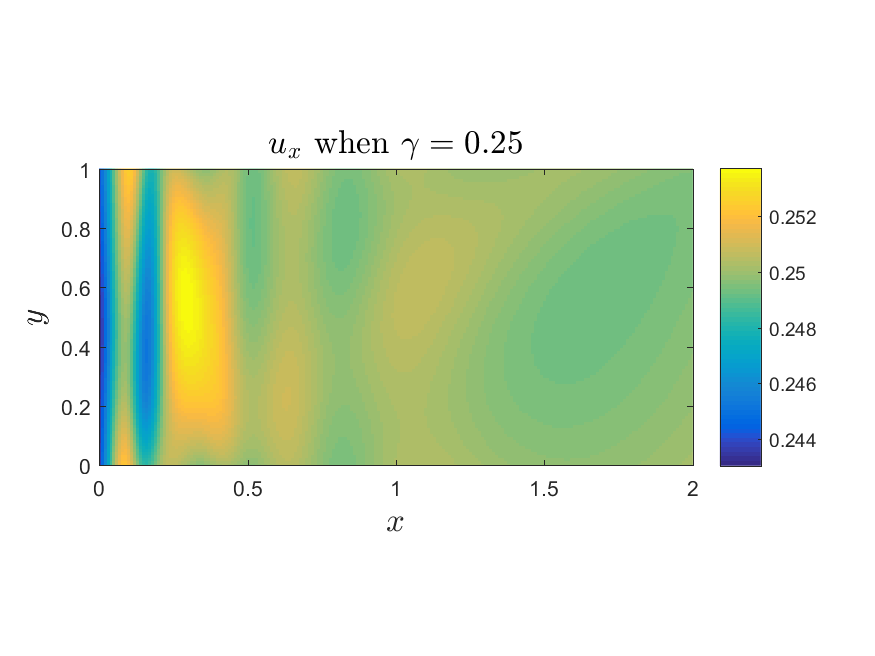}}
\end{minipage}
\hfill
 \begin{minipage}[]{0.2 \textwidth}
 \leftline{\small\textbf{(a2)}}
\centerline{\includegraphics[height=3.5cm]{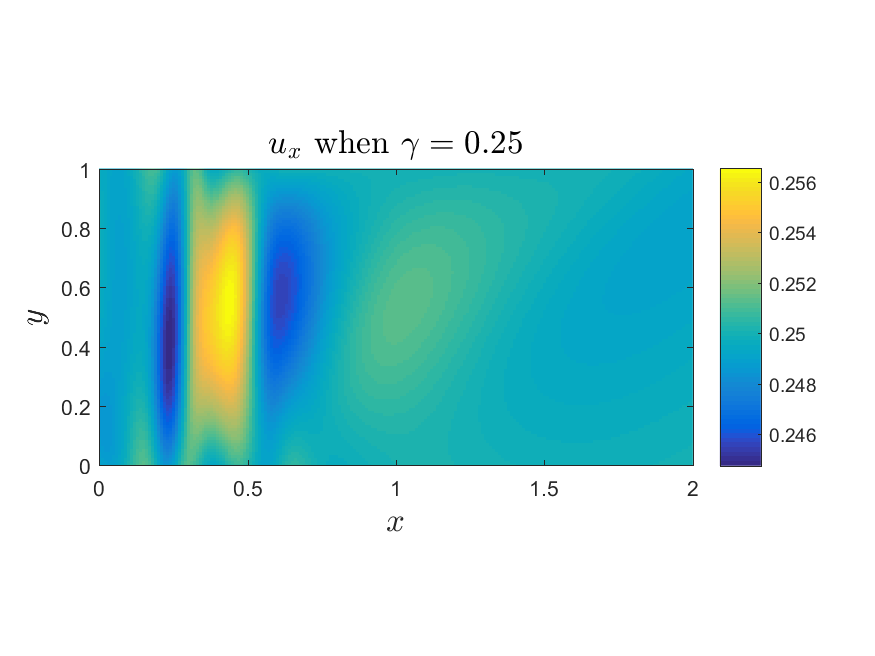}}
\end{minipage}
\hfill
\begin{minipage}[]{0.2 \textwidth}
 \leftline{\small\textbf{(a3)}}
\centerline{\includegraphics[height=3.5cm]{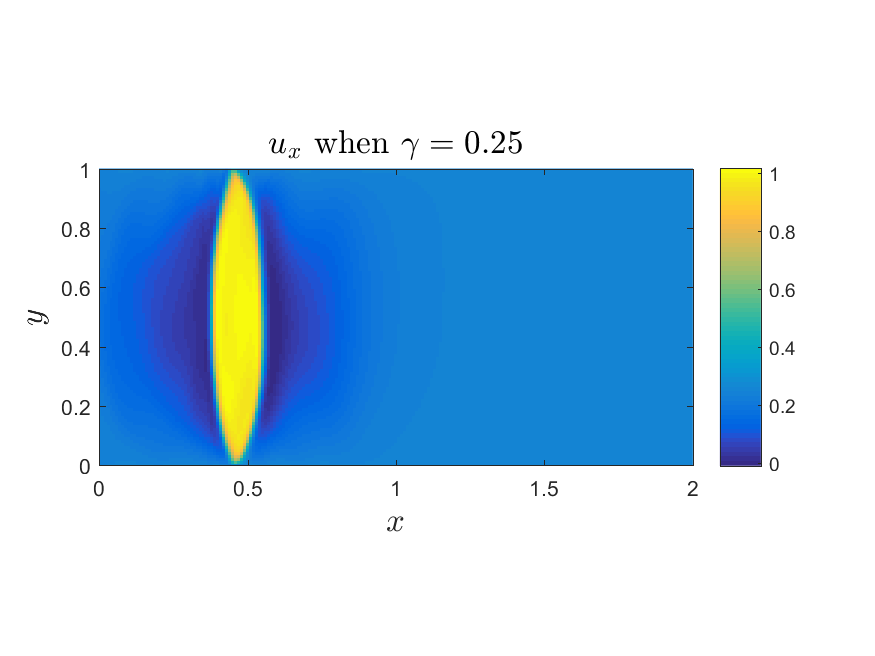}}
\end{minipage}
\hfill
\begin{minipage}[]{0.2 \textwidth}
 \leftline{\small\textbf{(a4)}}
\centerline{\includegraphics[height=3.5cm]{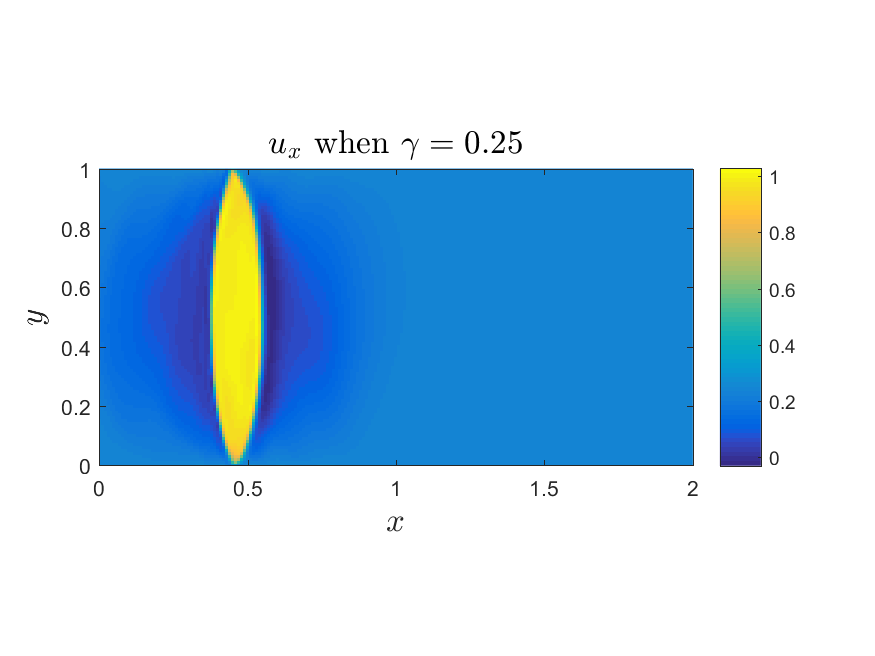}}
\end{minipage}
   \begin{minipage}[]{0.2 \textwidth}
 \leftline{\small\textbf{(b1)}}
\centerline{\includegraphics[height=3.5cm]{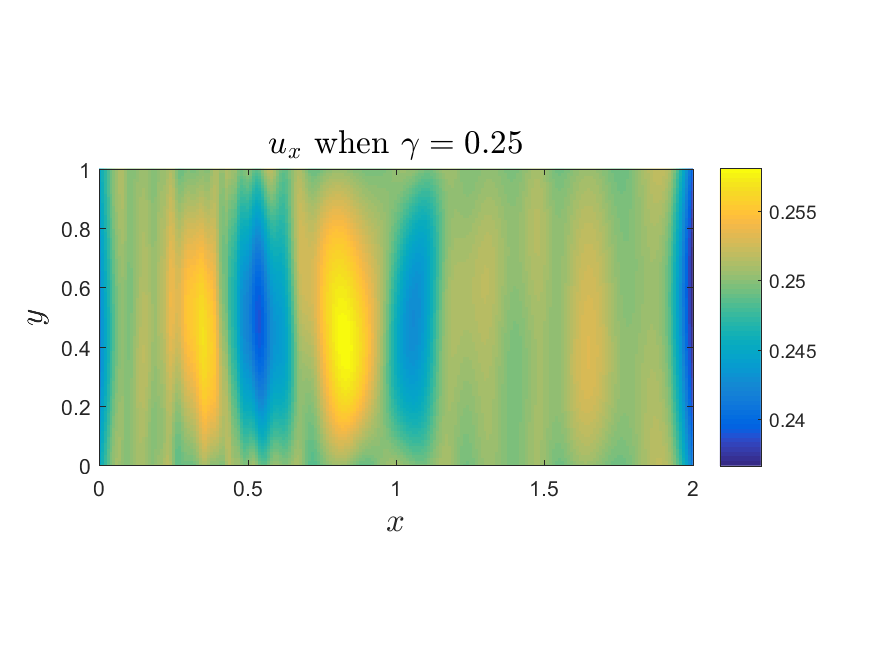}}
\end{minipage}
\hfill
 \begin{minipage}[]{0.2 \textwidth}
 \leftline{\small\textbf{(b2)}}
\centerline{\includegraphics[height=3.5cm]{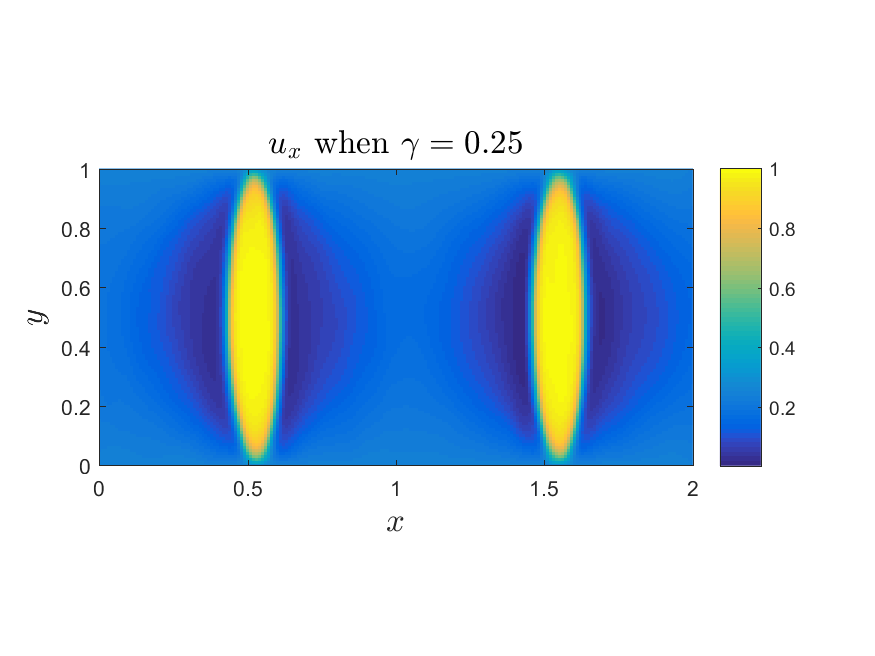}}
\end{minipage}
\hfill
\begin{minipage}[]{0.2 \textwidth}
 \leftline{\small\textbf{(b3)}}
\centerline{\includegraphics[height=3.5cm]{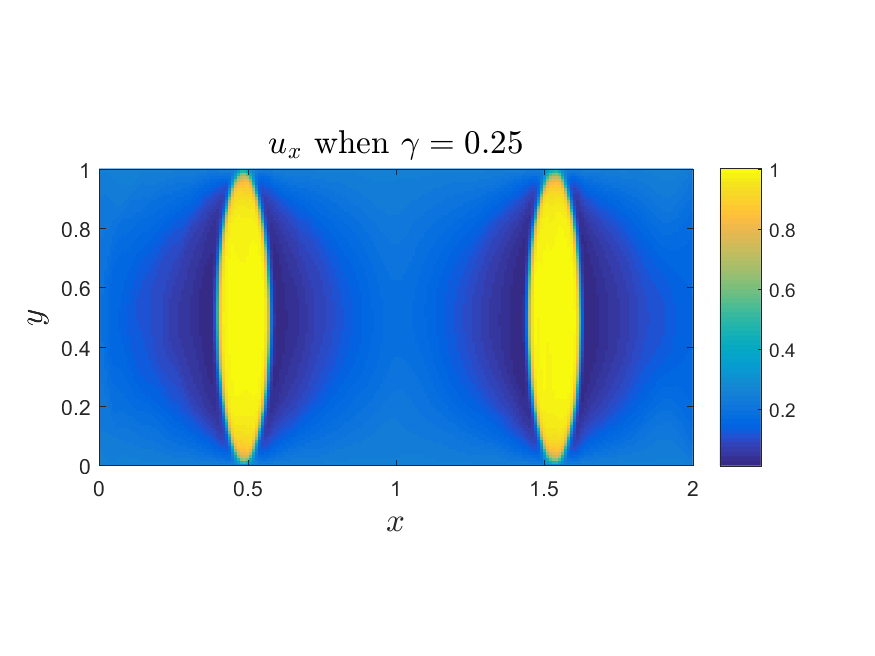}}
\end{minipage}
\hfill
\begin{minipage}[]{0.2 \textwidth}
 \leftline{\small\textbf{(b4)}}
\centerline{\includegraphics[height=3.5cm]{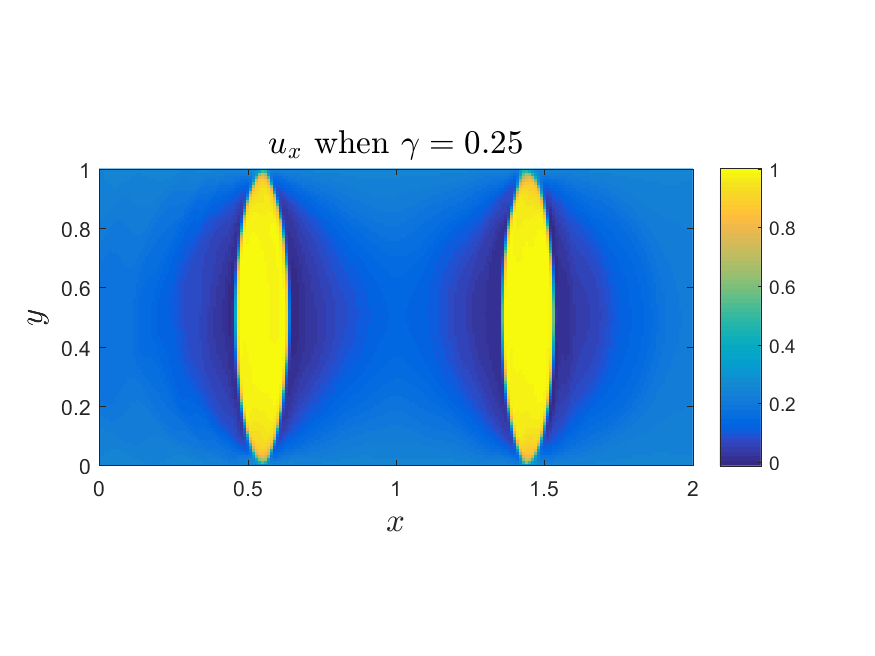}}
\end{minipage}
   \begin{minipage}[]{0.2 \textwidth}
 \leftline{\small\textbf{(c1)}}
\centerline{\includegraphics[height=3.5cm]{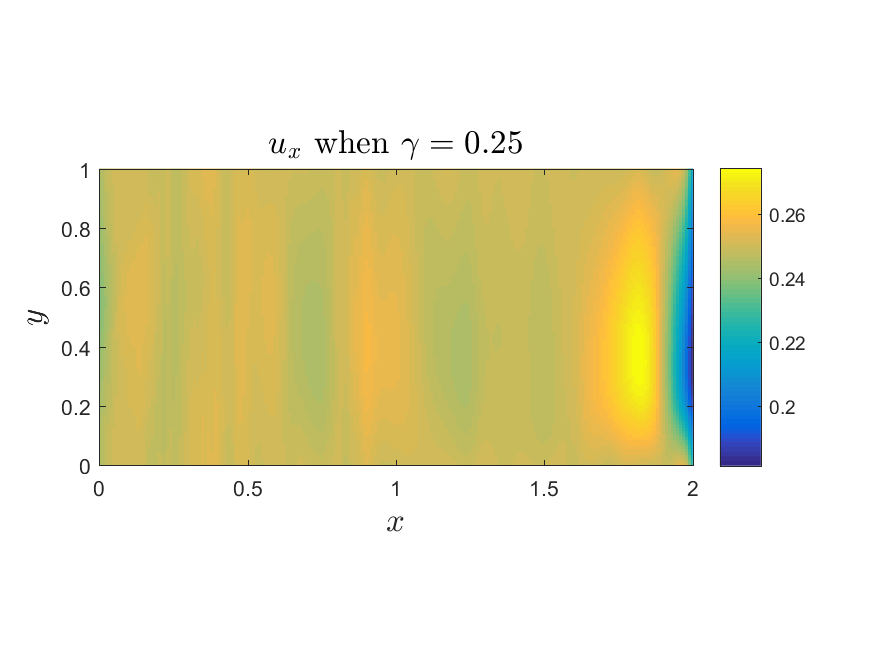}}
\end{minipage}
\hfill
 \begin{minipage}[]{0.2 \textwidth}
 \leftline{\small\textbf{(c2)}}
\centerline{\includegraphics[height=3.5cm]{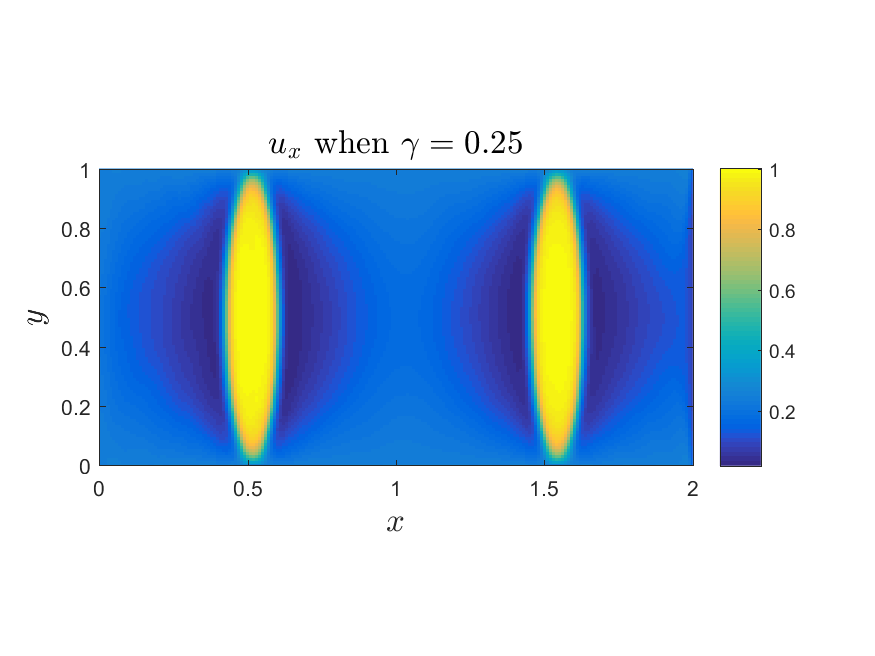}}
\end{minipage}
\hfill
\begin{minipage}[]{0.2 \textwidth}
 \leftline{\small\textbf{(c3)}}
\centerline{\includegraphics[height=3.5cm]{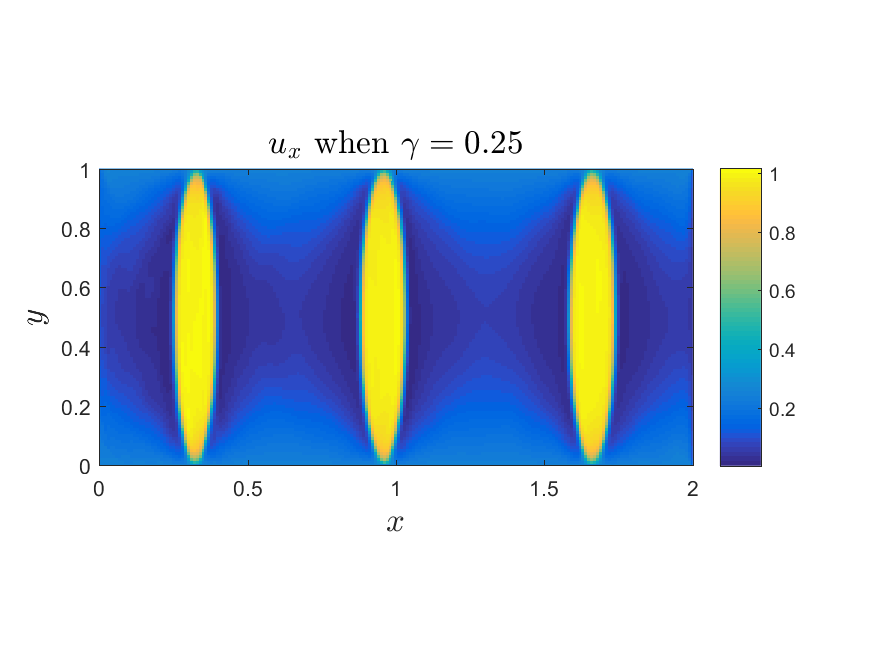}}
\end{minipage}
\hfill
\begin{minipage}[]{0.2 \textwidth}
 \leftline{\small\textbf{(c4)}}
\centerline{\includegraphics[height=3.5cm]{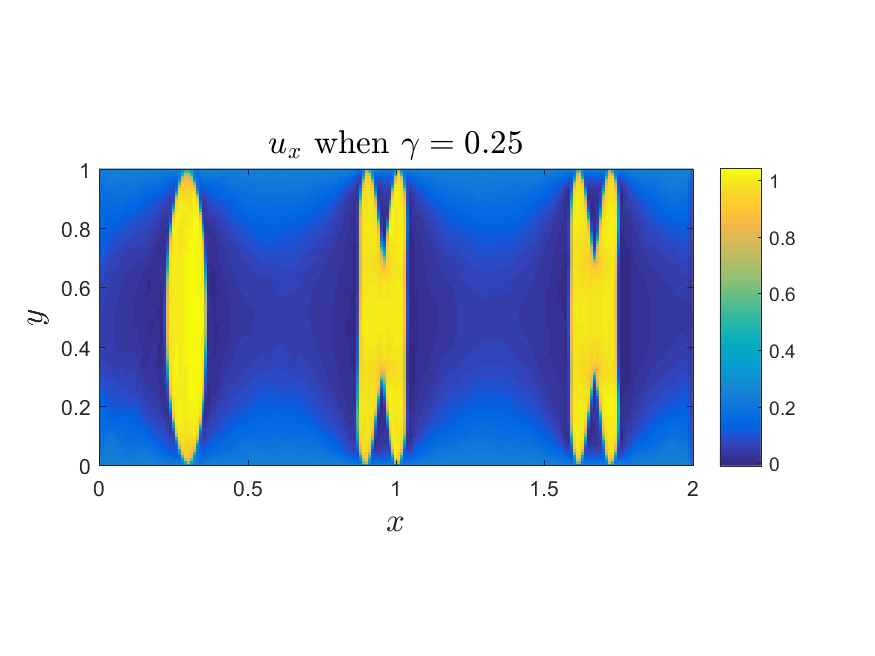}}
\end{minipage}
\caption{\textbf{ 2d regularized problem with SmReLU activation function.} Here $\gamma=0.25$, $iteration=300,000$ and learning rate $\eta=10^{-3}$. Top: the energy with different (a) NN: $3\times 128$; (b) NN: $5\times 128$; (c) NN: $7\times 128$. First column: $\varepsilon=\frac{0.1}{4}$; Second column: $\varepsilon=\frac{0.1}{8}$; Third column: $\varepsilon=\frac{0.1}{16}$; Last column: $\varepsilon=\frac{0.1}{32}$. }
\label{2d_reg_eps_gamma025}
\end{figure}

\begin{figure}[ht]
  \centerline{\includegraphics[height=3.5cm]{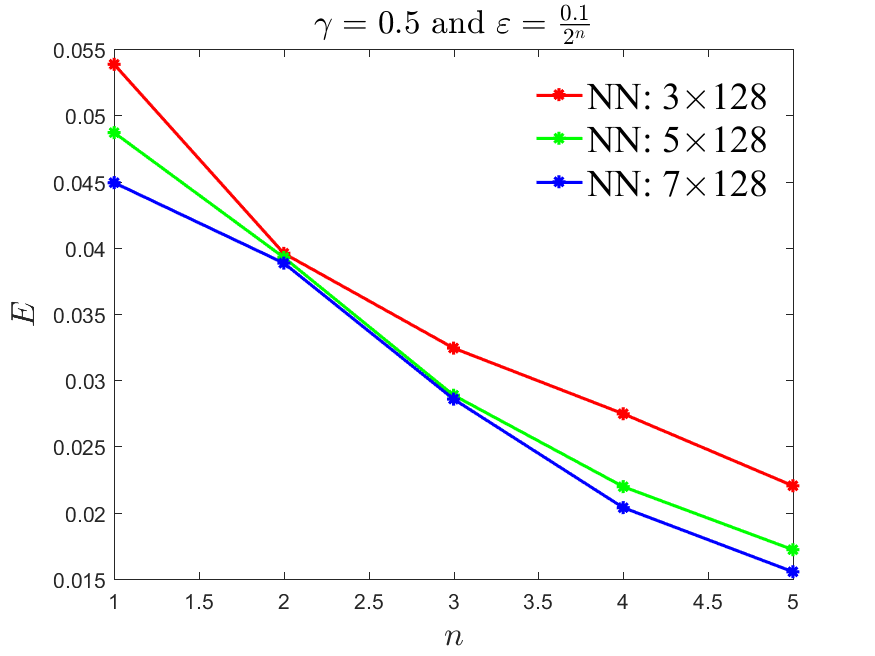}}
   \begin{minipage}[]{0.2 \textwidth}
 \leftline{\small\textbf{(a1)}}
\centerline{\includegraphics[height=3.5cm]{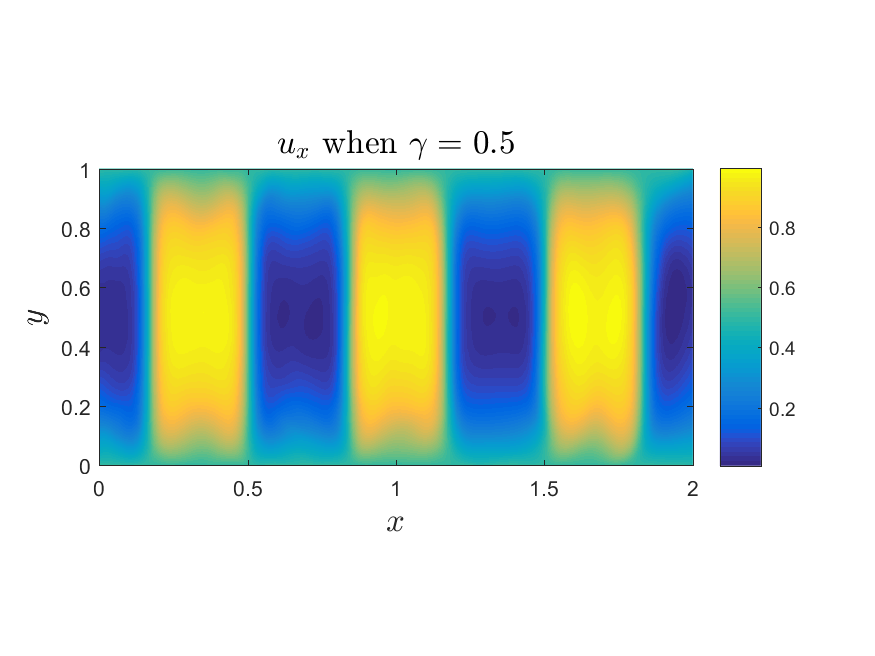}}
\end{minipage}
\hfill
 \begin{minipage}[]{0.2 \textwidth}
 \leftline{\small\textbf{(a2)}}
\centerline{\includegraphics[height=3.5cm]{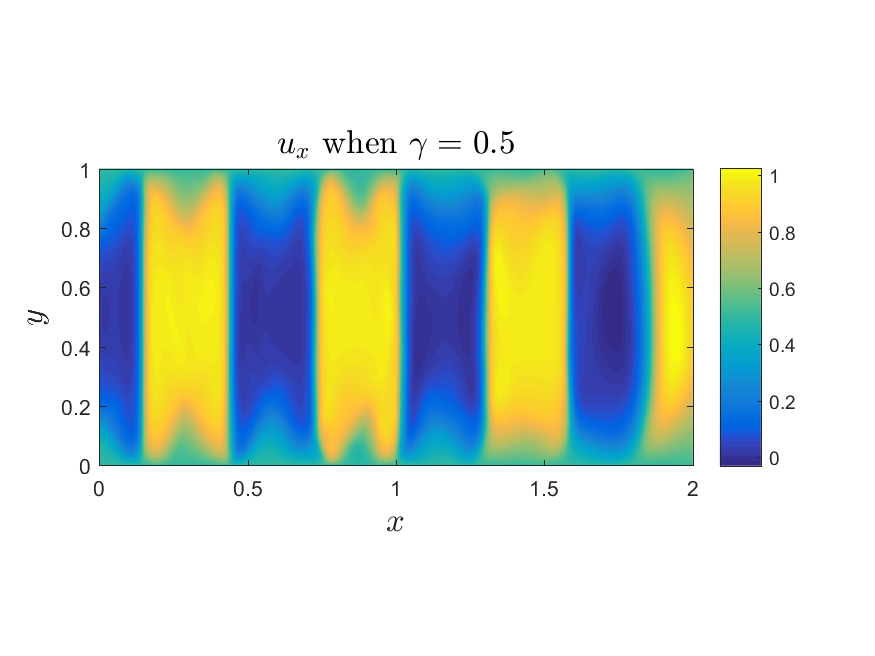}}
\end{minipage}
\hfill
\begin{minipage}[]{0.2 \textwidth}
 \leftline{\small\textbf{(a3)}}
\centerline{\includegraphics[height=3.5cm]{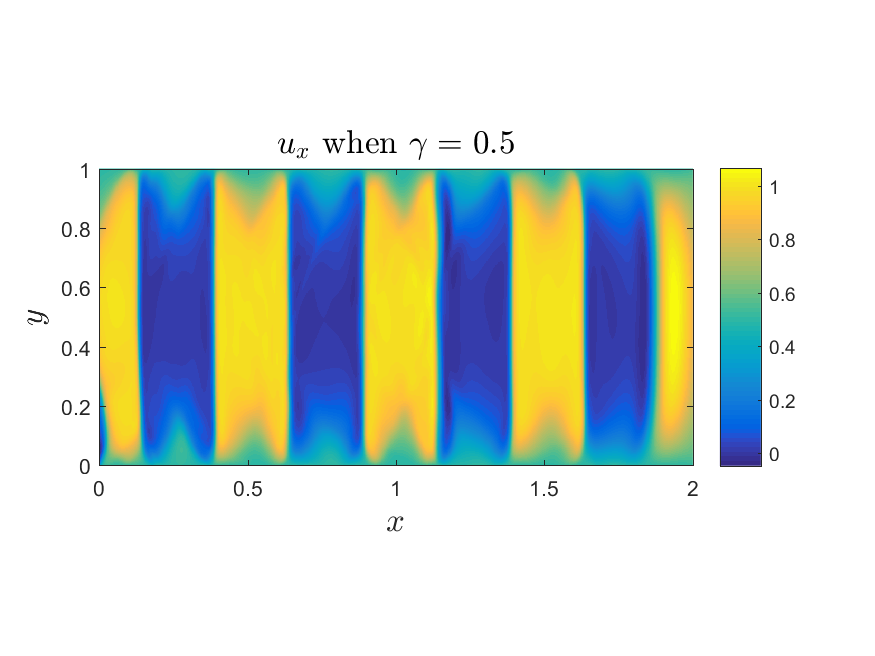}}
\end{minipage}
\hfill
\begin{minipage}[]{0.2 \textwidth}
 \leftline{\small\textbf{(a4)}}
\centerline{\includegraphics[height=3.5cm]{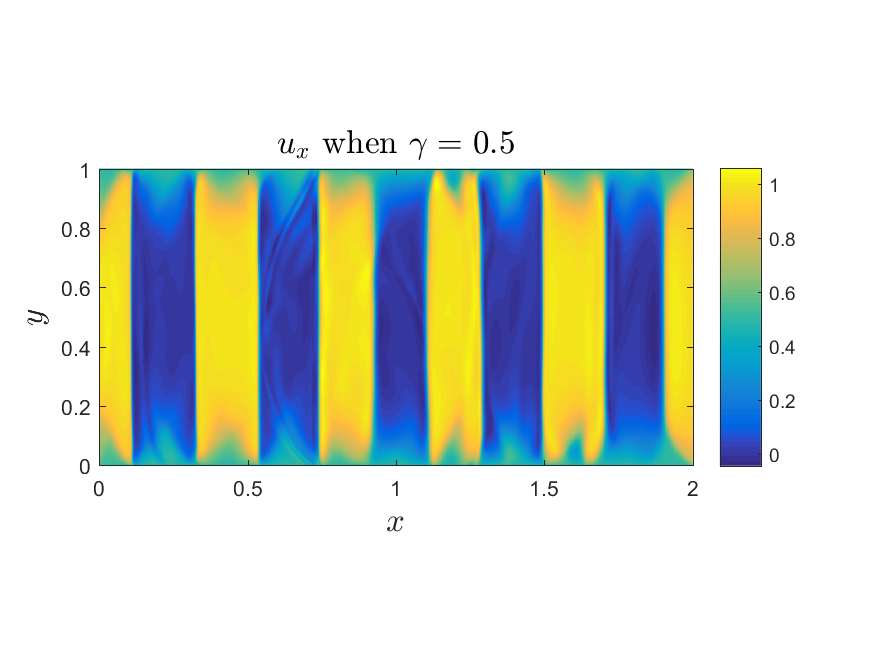}}
\end{minipage}
   \begin{minipage}[]{0.2 \textwidth}
 \leftline{\small\textbf{(b1)}}
\centerline{\includegraphics[height=3.5cm]{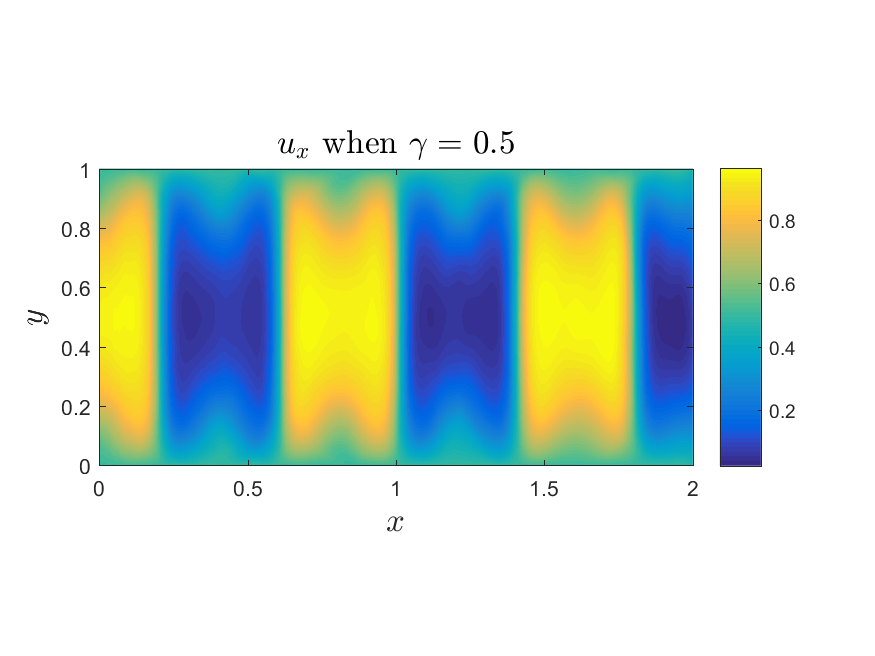}}
\end{minipage}
\hfill
 \begin{minipage}[]{0.2 \textwidth}
 \leftline{\small\textbf{(b2)}}
\centerline{\includegraphics[height=3.5cm]{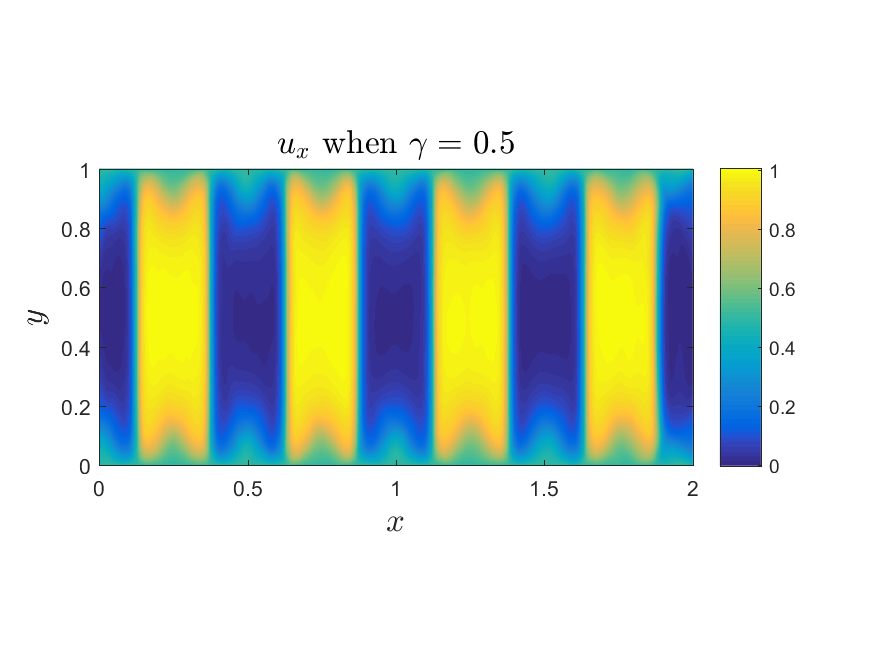}}
\end{minipage}
\hfill
\begin{minipage}[]{0.2 \textwidth}
 \leftline{\small\textbf{(b3)}}
\centerline{\includegraphics[height=3.5cm]{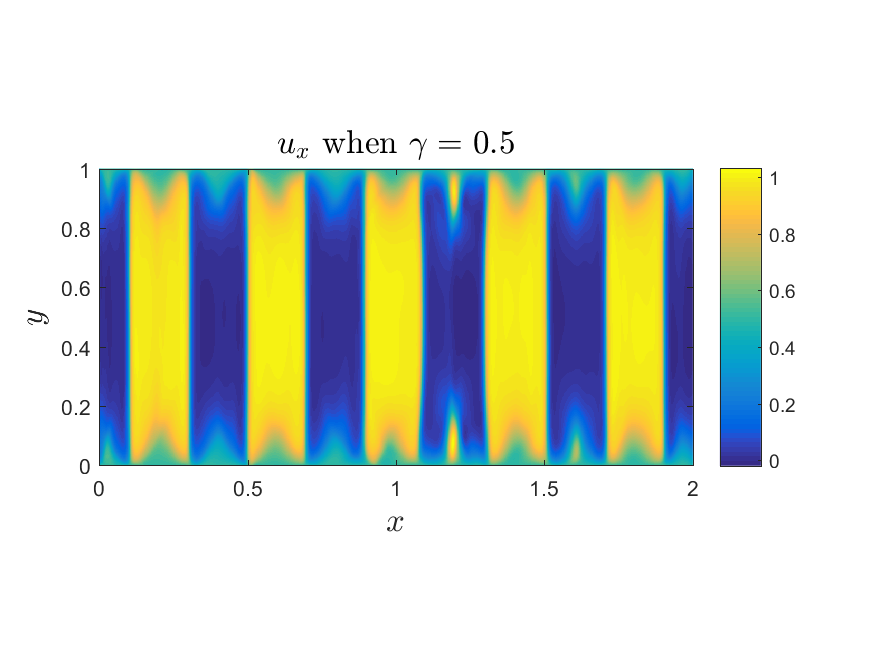}}
\end{minipage}
\hfill
\begin{minipage}[]{0.2 \textwidth}
 \leftline{\small\textbf{(b4)}}
\centerline{\includegraphics[height=3.5cm]{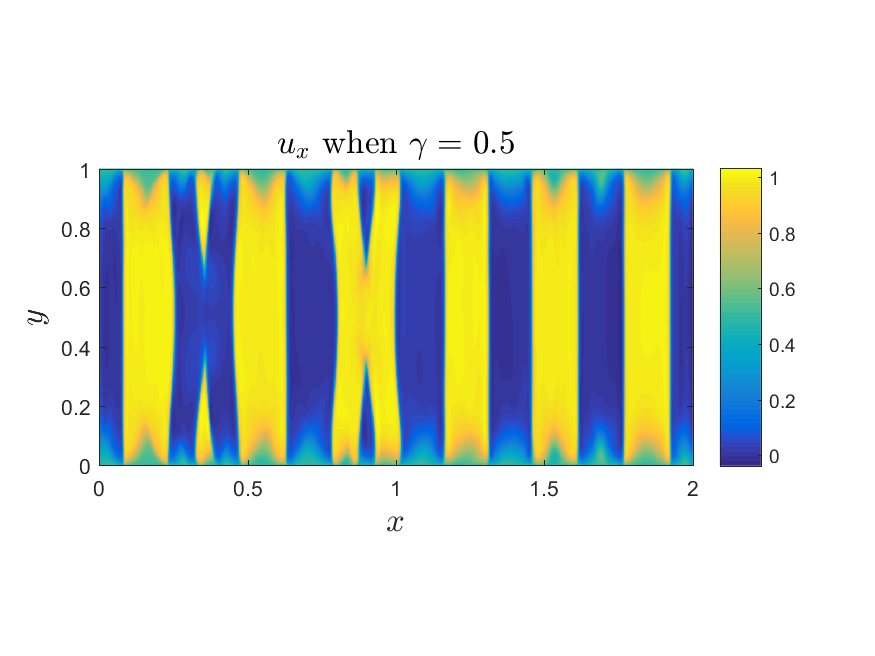}}
\end{minipage}
   \begin{minipage}[]{0.2 \textwidth}
 \leftline{\small\textbf{(c1)}}
\centerline{\includegraphics[height=3.5cm]{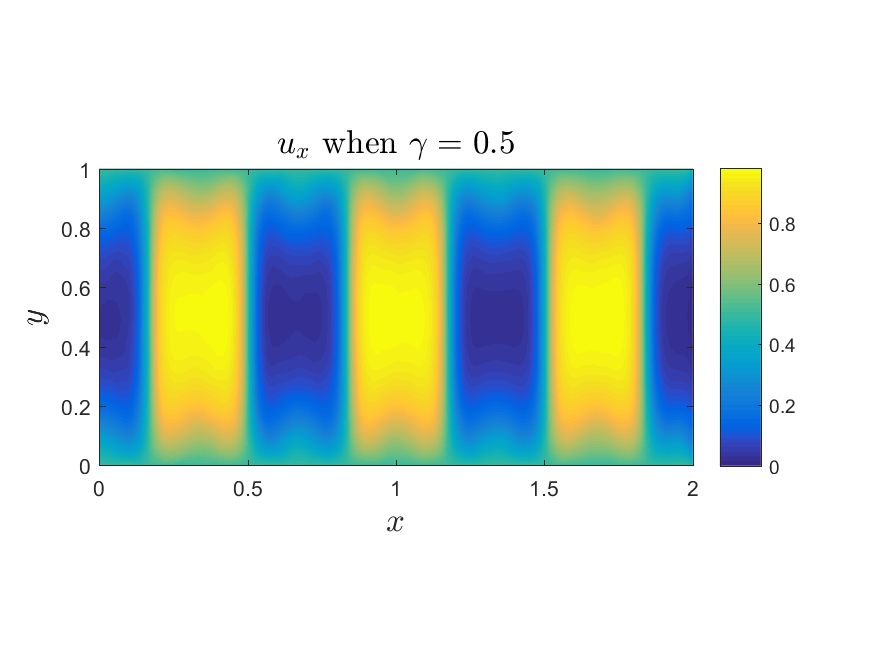}}
\end{minipage}
\hfill
 \begin{minipage}[]{0.2 \textwidth}
 \leftline{\small\textbf{(c2)}}
\centerline{\includegraphics[height=3.5cm]{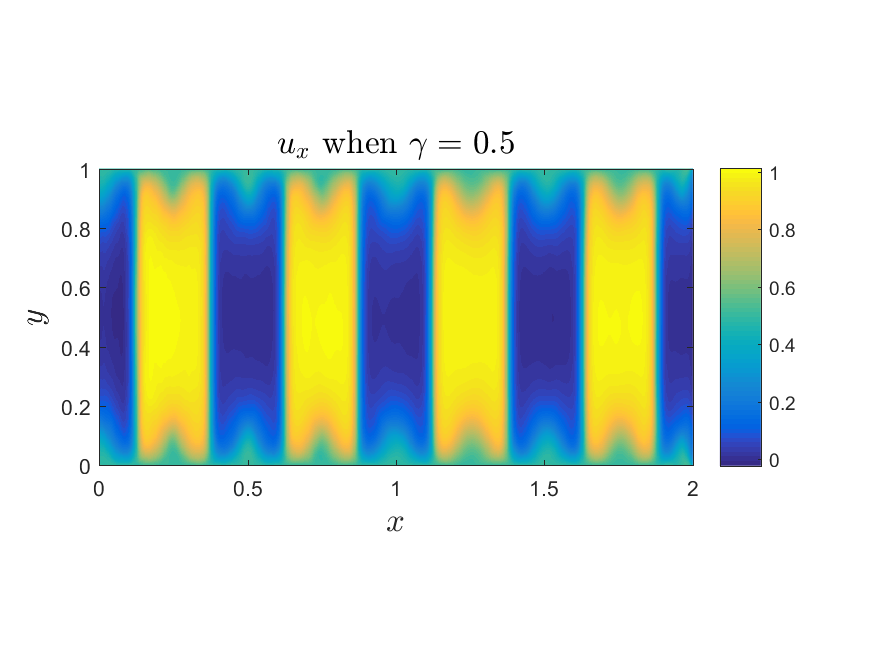}}
\end{minipage}
\hfill
\begin{minipage}[]{0.2 \textwidth}
 \leftline{\small\textbf{(c3)}}
\centerline{\includegraphics[height=3.5cm]{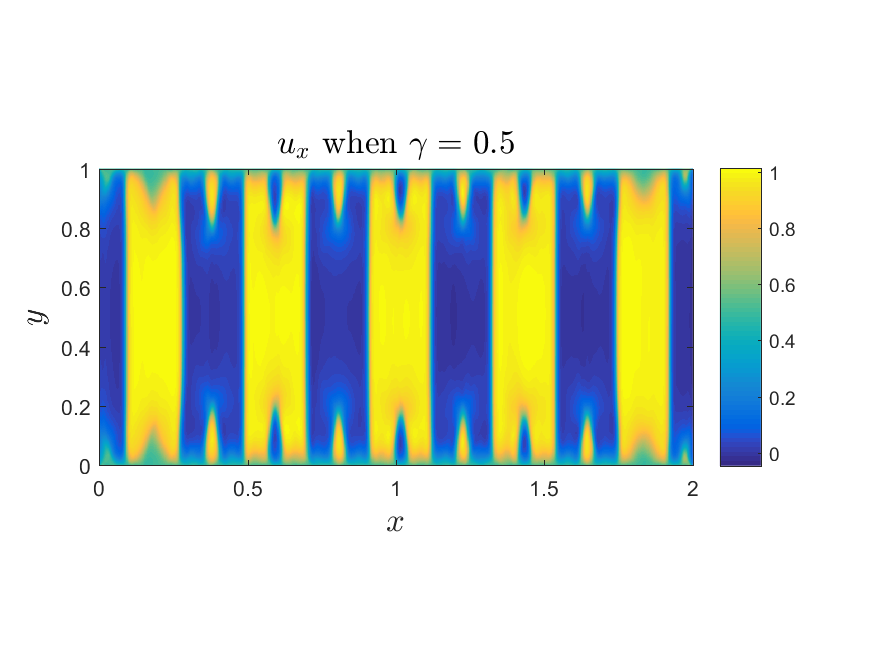}}
\end{minipage}
\hfill
\begin{minipage}[]{0.2 \textwidth}
 \leftline{\small\textbf{(c4)}}
\centerline{\includegraphics[height=3.5cm]{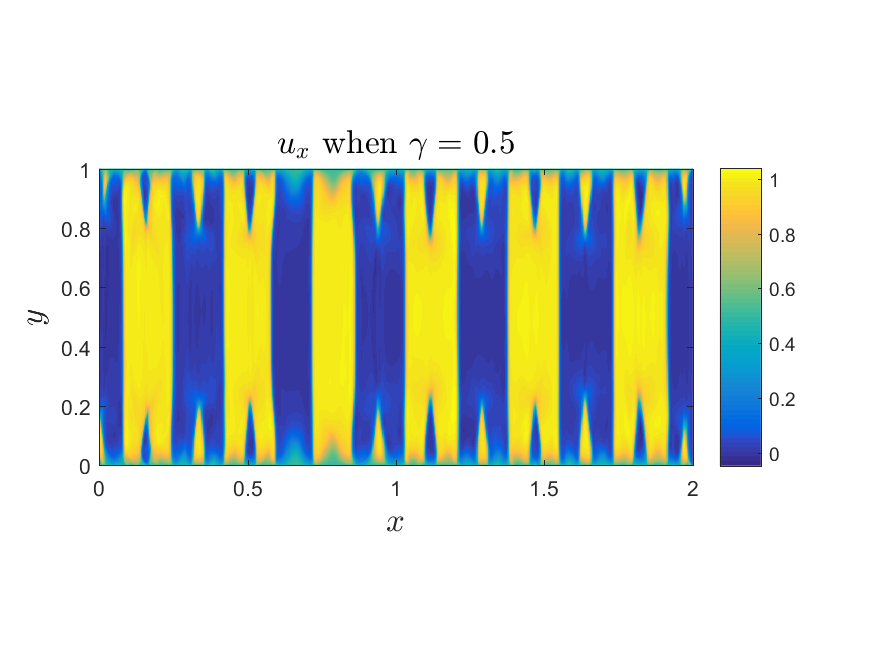}}
\end{minipage}
\caption{\textbf{ 2d regularized problem with SmReLU activation function.} Here $\gamma=0.5$, $iteration=300,000$ and learning rate $\eta=10^{-3}$. Top: the energy with different (a) NN: $3\times 128$; (b) NN: $5\times 128$; (c) NN: $7\times 128$.  First column: $\varepsilon=\frac{0.1}{4}$; Second column: $\varepsilon=\frac{0.1}{8}$; Third column: $\varepsilon=\frac{0.1}{16}$; Last column: $\varepsilon=\frac{0.1}{32}$.}
\label{2d_reg_eps_gamma05}
\end{figure}

\begin{figure}[ht]
  \centerline{\includegraphics[height=3.5cm]{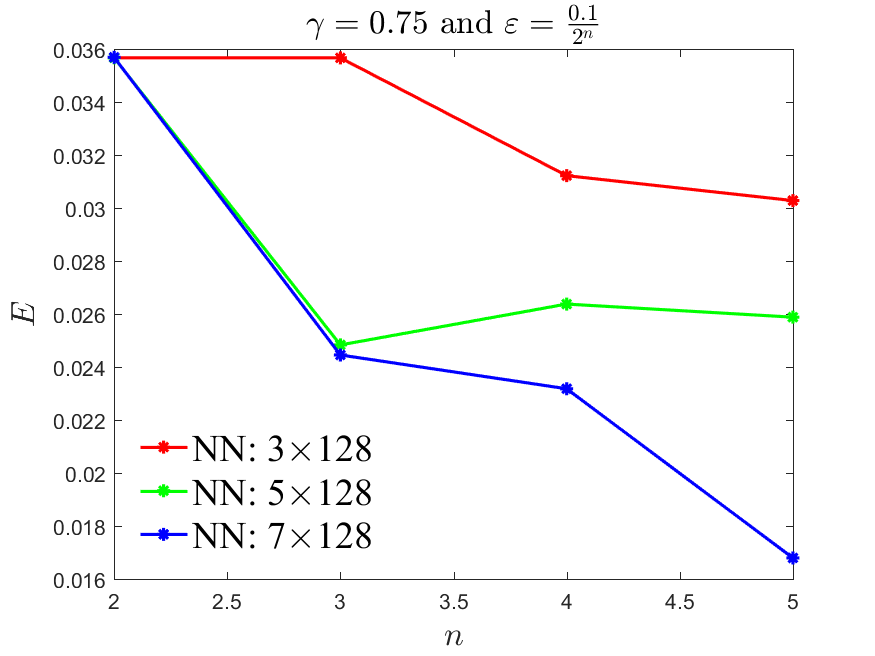}}
   \begin{minipage}[]{0.2 \textwidth}
 \leftline{\small\textbf{(a1)}}
\centerline{\includegraphics[height=3.5cm]{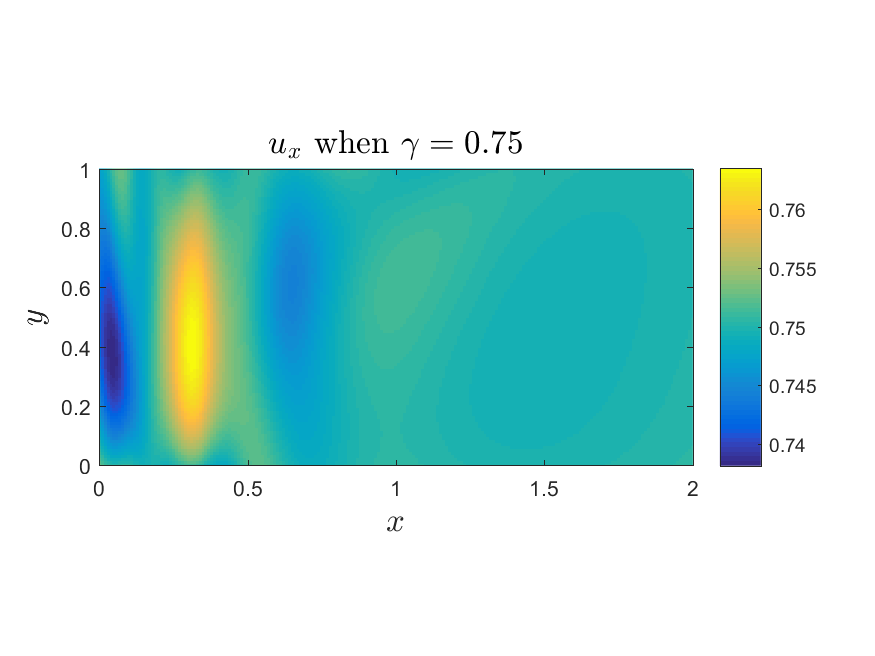}}
\end{minipage}
\hfill
 \begin{minipage}[]{0.2 \textwidth}
 \leftline{\small\textbf{(a2)}}
\centerline{\includegraphics[height=3.5cm]{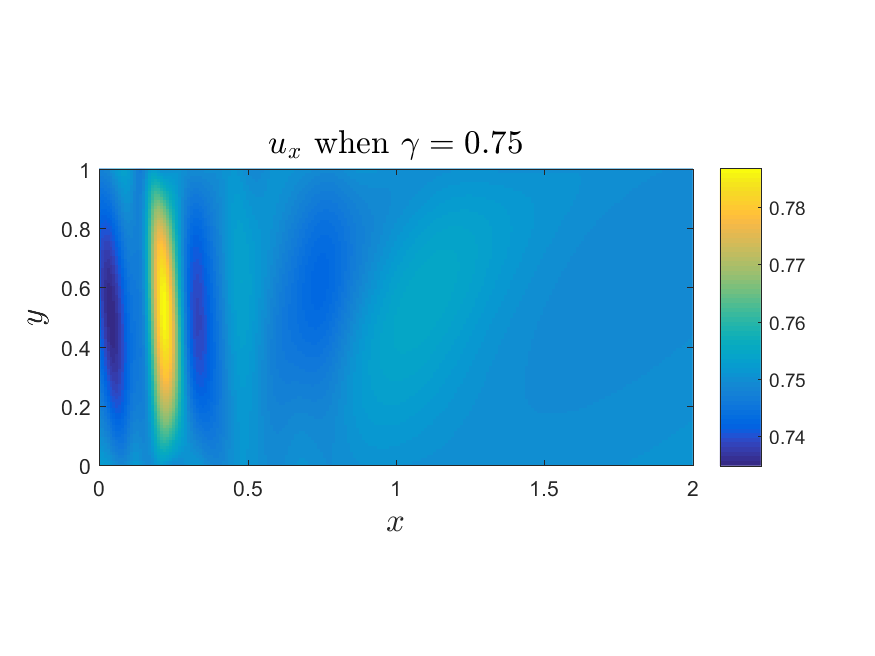}}
\end{minipage}
\hfill
\begin{minipage}[]{0.2 \textwidth}
 \leftline{\small\textbf{(a3)}}
\centerline{\includegraphics[height=3.5cm]{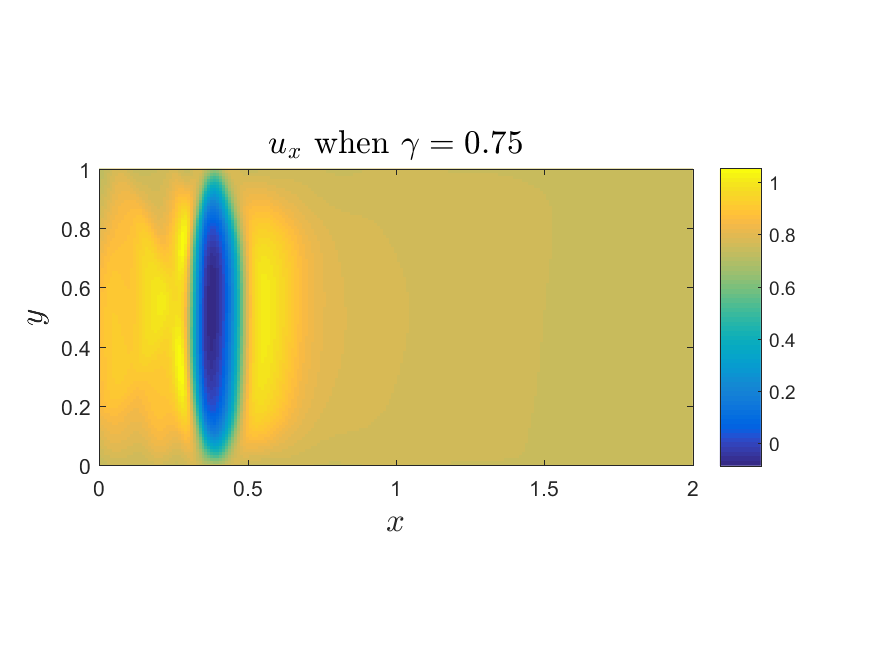}}
\end{minipage}
\hfill
\begin{minipage}[]{0.2 \textwidth}
 \leftline{\small\textbf{(a4)}}
\centerline{\includegraphics[height=3.5cm]{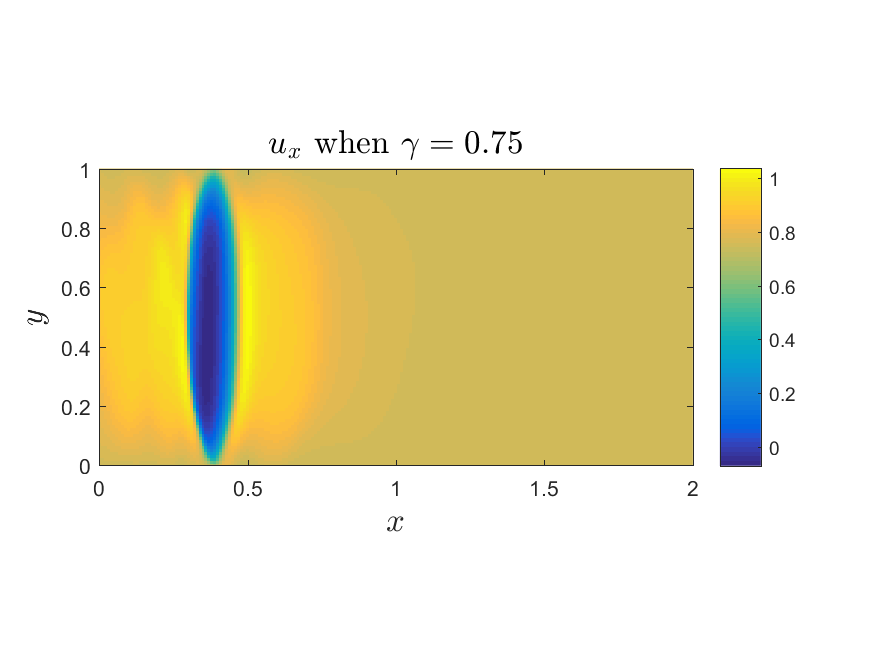}}
\end{minipage}
   \begin{minipage}[]{0.2 \textwidth}
 \leftline{\small\textbf{(b1)}}
\centerline{\includegraphics[height=3.5cm]{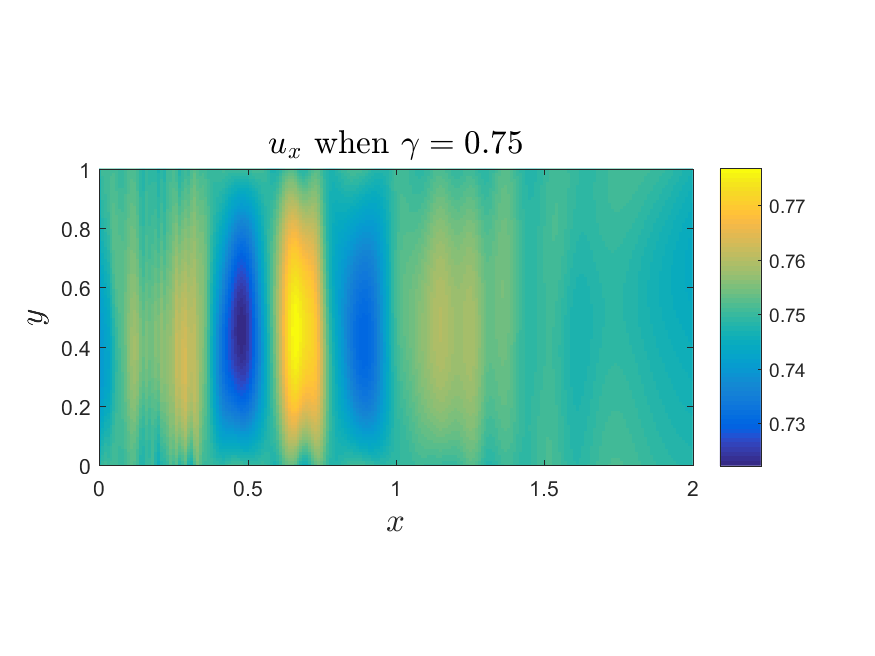}}
\end{minipage}
\hfill
 \begin{minipage}[]{0.2 \textwidth}
 \leftline{\small\textbf{(b2)}}
\centerline{\includegraphics[height=3.5cm]{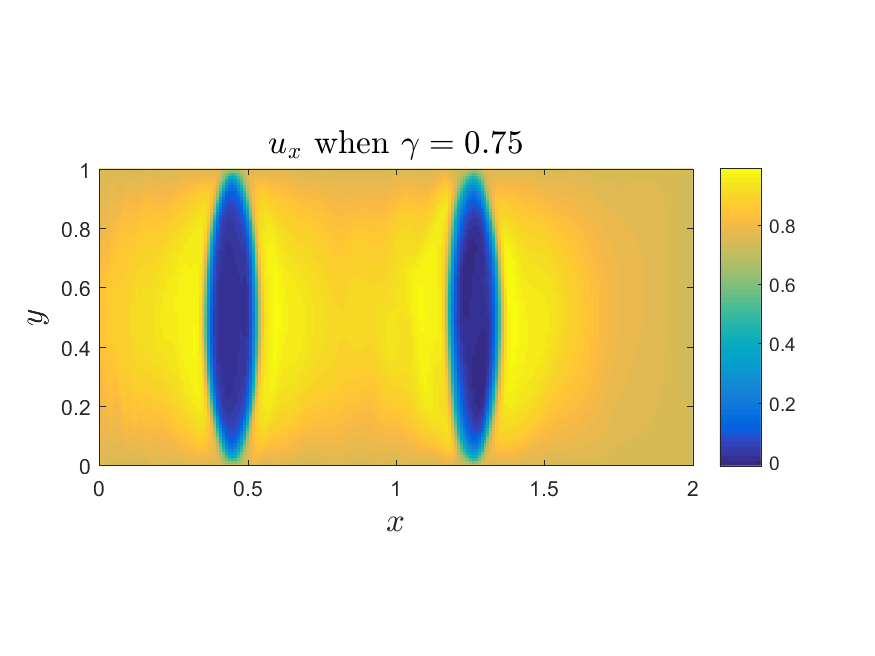}}
\end{minipage}
\hfill
\begin{minipage}[]{0.2 \textwidth}
 \leftline{\small\textbf{(b3)}}
\centerline{\includegraphics[height=3.5cm]{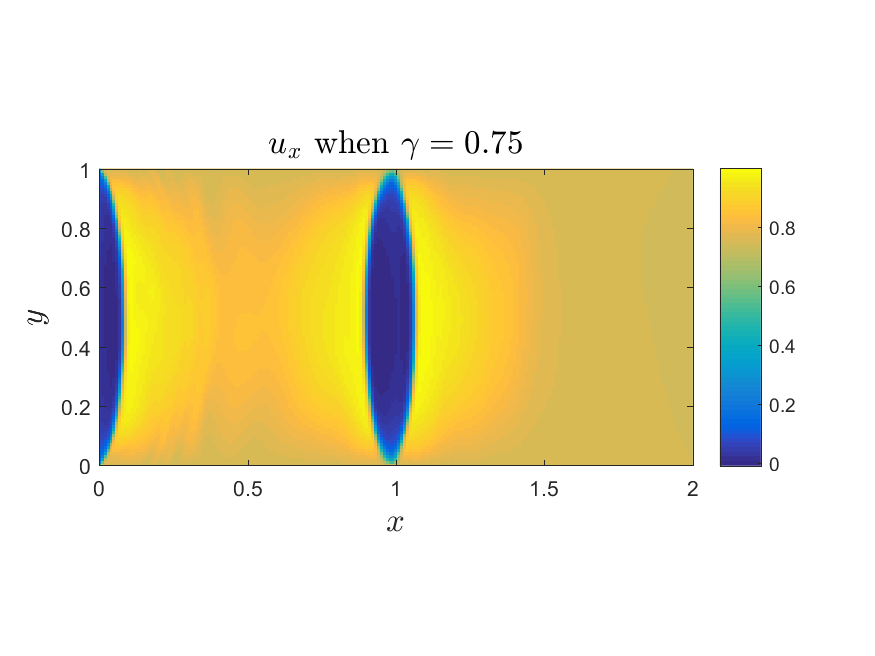}}
\end{minipage}
\hfill
\begin{minipage}[]{0.2 \textwidth}
 \leftline{\small\textbf{(b4)}}
\centerline{\includegraphics[height=3.5cm]{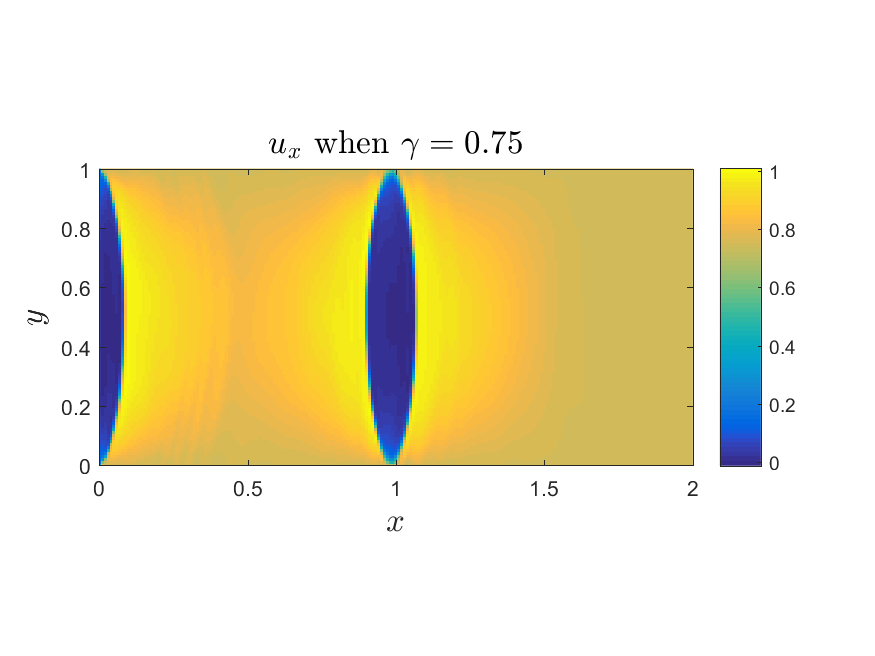}}
\end{minipage}
   \begin{minipage}[]{0.2 \textwidth}
 \leftline{\small\textbf{(c1)}}
\centerline{\includegraphics[height=3.5cm]{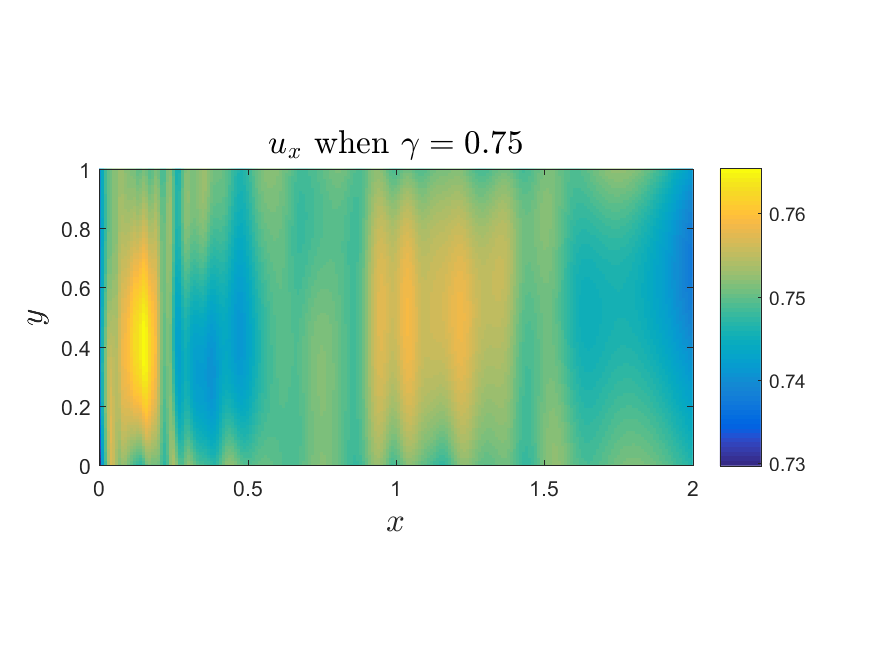}}
\end{minipage}
\hfill
 \begin{minipage}[]{0.2 \textwidth}
 \leftline{\small\textbf{(c2)}}
\centerline{\includegraphics[height=3.5cm]{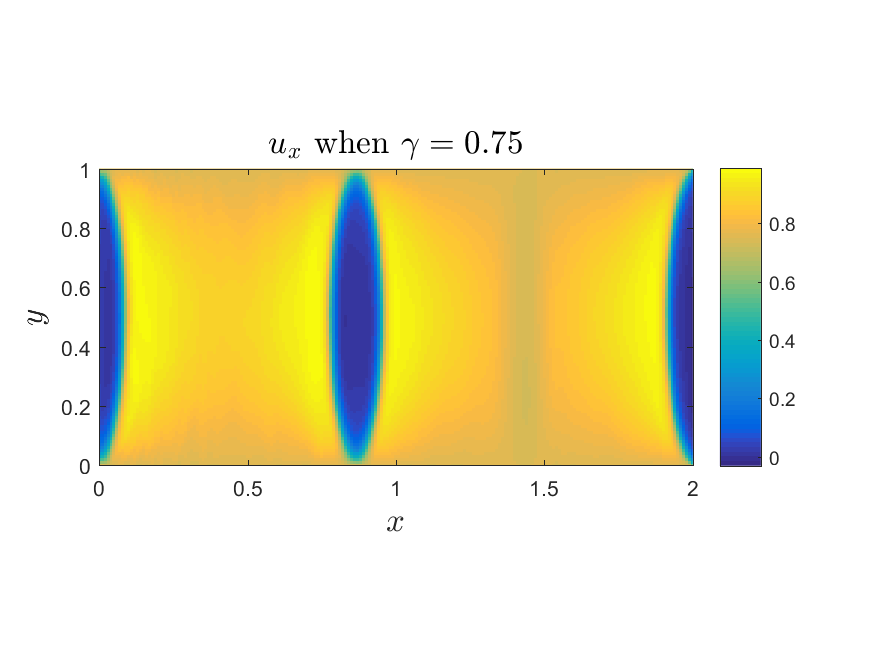}}
\end{minipage}
\hfill
\begin{minipage}[]{0.2 \textwidth}
 \leftline{\small\textbf{(c3)}}
\centerline{\includegraphics[height=3.5cm]{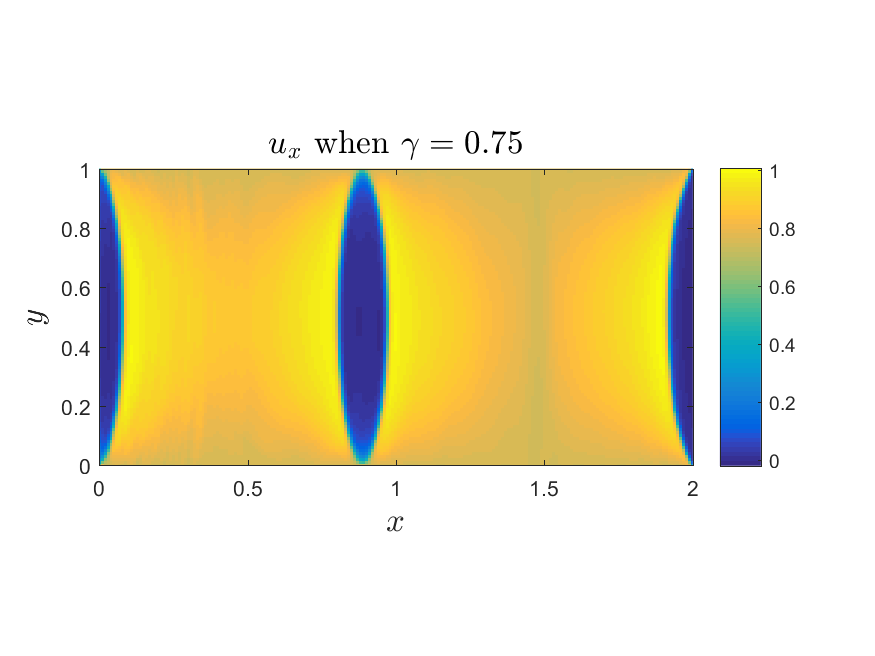}}
\end{minipage}
\hfill
\begin{minipage}[]{0.2 \textwidth}
 \leftline{\small\textbf{(c4)}}
\centerline{\includegraphics[height=3.5cm]{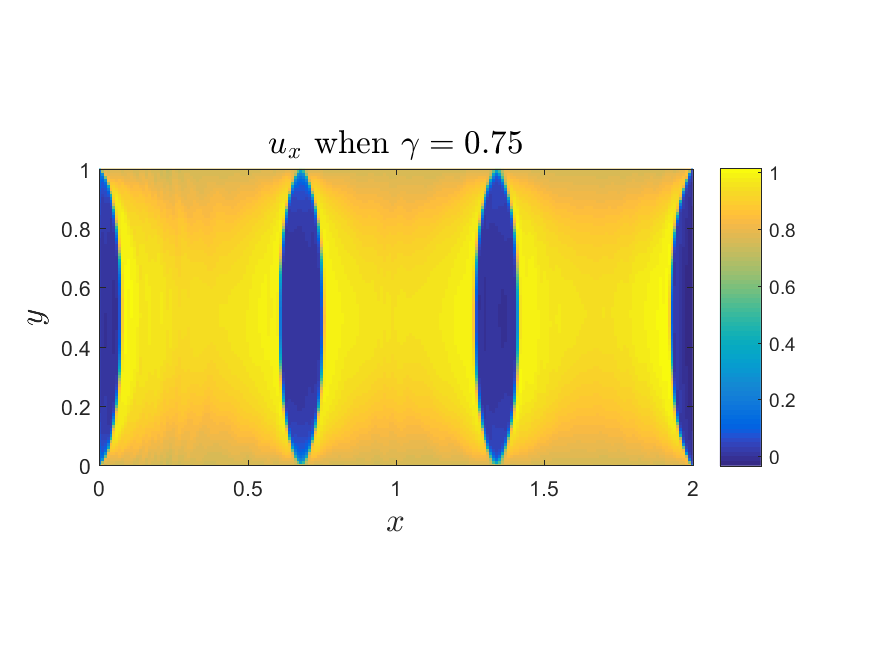}}
\end{minipage}
\caption{\textbf{ 2d regularized problem with SmReLU activation function.} Here $\gamma=0.75$, $iteration=200,000$ and learning rate $\eta=10^{-3}$. Top: the energy with different (a) NN: $3\times 128$; (b) NN: $5\times 128$; (c) NN: $7\times 128$.  First column: $\varepsilon=\frac{0.1}{4}$; Second column: $\varepsilon=\frac{0.1}{8}$; Third column: $\varepsilon=\frac{0.1}{16}$; Last column: $\varepsilon=\frac{0.1}{32}$.}
\label{2d_reg_eps_gamma075}
\end{figure}

We explore how the regularization parameter $\eps$ affects the energy. The results are shown in Fig.~\ref{2d_reg_eps_gamma025}-\ref{2d_reg_eps_gamma075}, where $\gamma=0.25,0.5,0.75$ respectively.
The energy decreases when the regularization parameter $\eps$ is smaller, while solutions for smaller $\eps$  have a finer microstructure with more bands, e.g., Fig.~\ref{2d_reg_eps_gamma05}, as is also observed in \cite{healey}.
Also if we use a larger DNN, the energy will be lower, since the greater number of parameters allows capturing a local minimum with lower energy and eventually hopefully a global minimum. In the following, we will explore how the structure of DNNs affect the results.

\subsubsection{Effect of the structure of the DNNs}
\begin{figure}[ht]
   \centerline{\includegraphics[height=3.5cm]{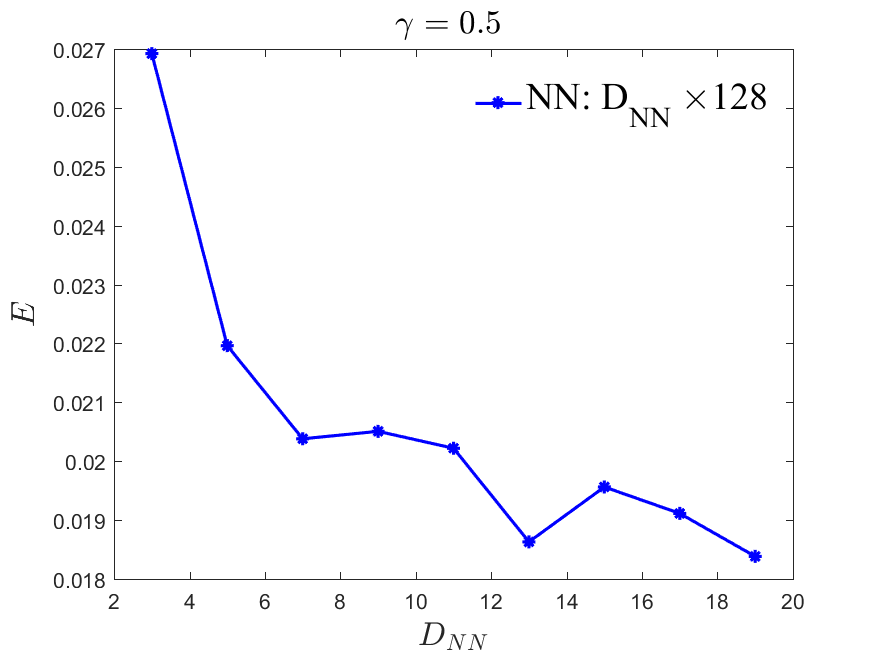}}
   \begin{minipage}[]{0.2 \textwidth}
 \leftline{\small\textbf{(a)}}
\centerline{\includegraphics[height=3.5cm]{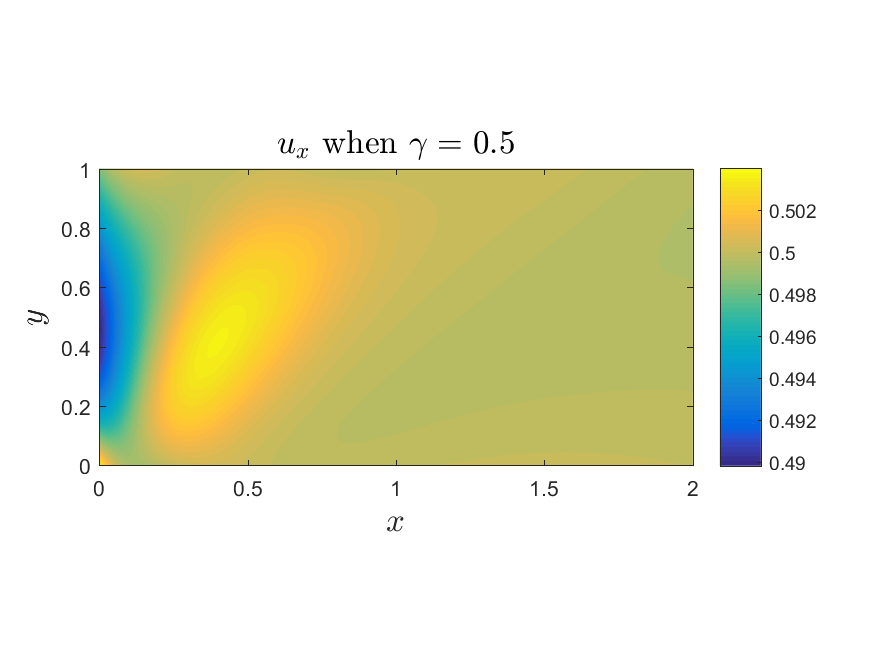}}
\end{minipage}
\hfill
 \begin{minipage}[]{0.2 \textwidth}
 \leftline{\small\textbf{(b)}}
\centerline{\includegraphics[height=3.5cm]{figure/gamma05_3x128_tau500_srelu_lr3_regxx16_T2_uxp_opt.png}}
\end{minipage}
\hfill
\begin{minipage}[]{0.2 \textwidth}
 \leftline{\small\textbf{(c)}}
\centerline{\includegraphics[height=3.5cm]{figure/gamma05_5x128_tau500_srelu_lr3_regxx16_T2_uxp_opt.png}}
\end{minipage}
\hfill
\begin{minipage}[]{0.2 \textwidth}
 \leftline{\small\textbf{(d)}}
\centerline{\includegraphics[height=3.5cm]{figure/gamma05_7x128_tau500_srelu_lr3_regxx16_T2_uxp_opt.png}}
\end{minipage}
 \begin{minipage}[]{0.2 \textwidth}
 \leftline{\small\textbf{(e)}}
\centerline{\includegraphics[height=3.5cm]{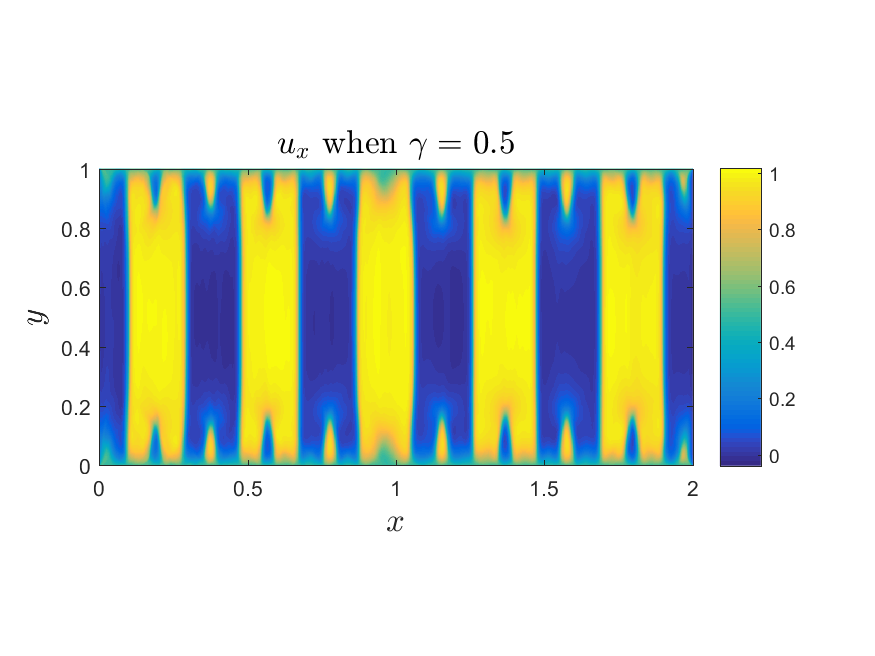}}
\end{minipage}
\hfill
\begin{minipage}[]{0.2 \textwidth}
 \leftline{\small\textbf{(f)}}
\centerline{\includegraphics[height=3.5cm]{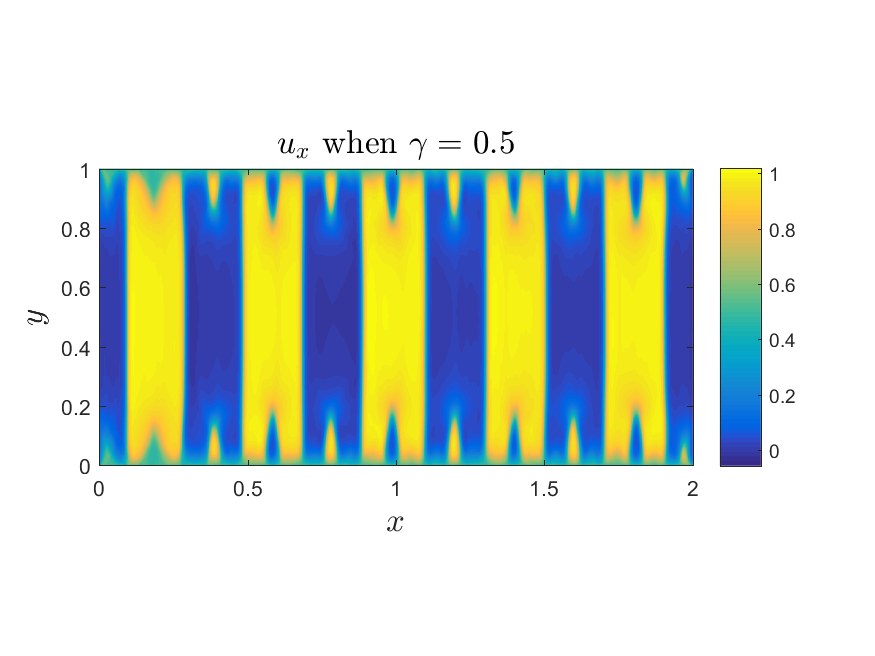}}
\end{minipage}
\hfill
\begin{minipage}[]{0.2 \textwidth}
 \leftline{\small\textbf{(g)}}
\centerline{\includegraphics[height=3.5cm]{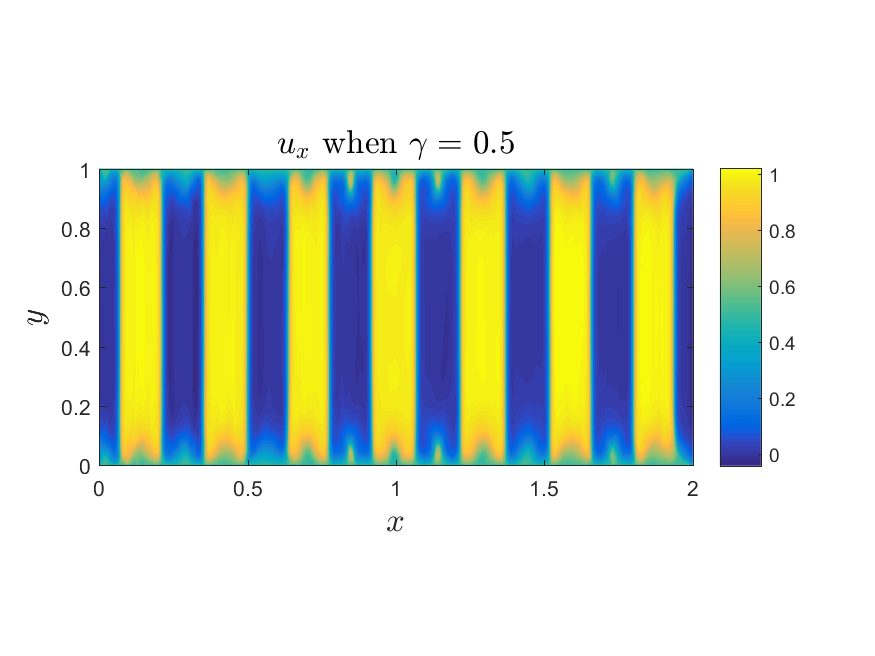}}
\end{minipage}
\hfill
\begin{minipage}[]{0.2 \textwidth}
 \leftline{\small\textbf{(h)}}
\centerline{\includegraphics[height=3.5cm]{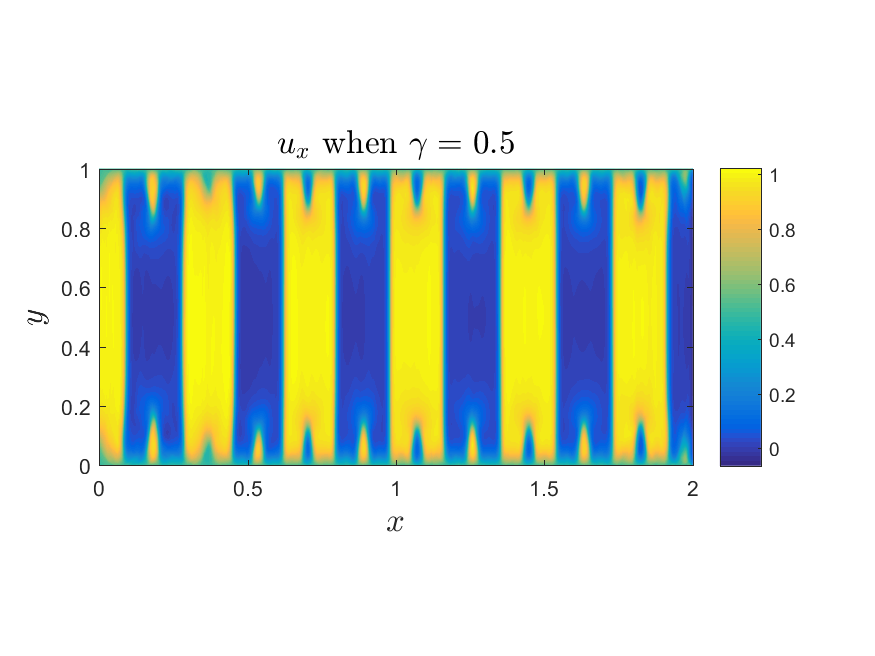}}
\end{minipage}
\begin{minipage}[]{0.2 \textwidth}
 \leftline{\small\textbf{(i)}}
\centerline{\includegraphics[height=3.5cm]{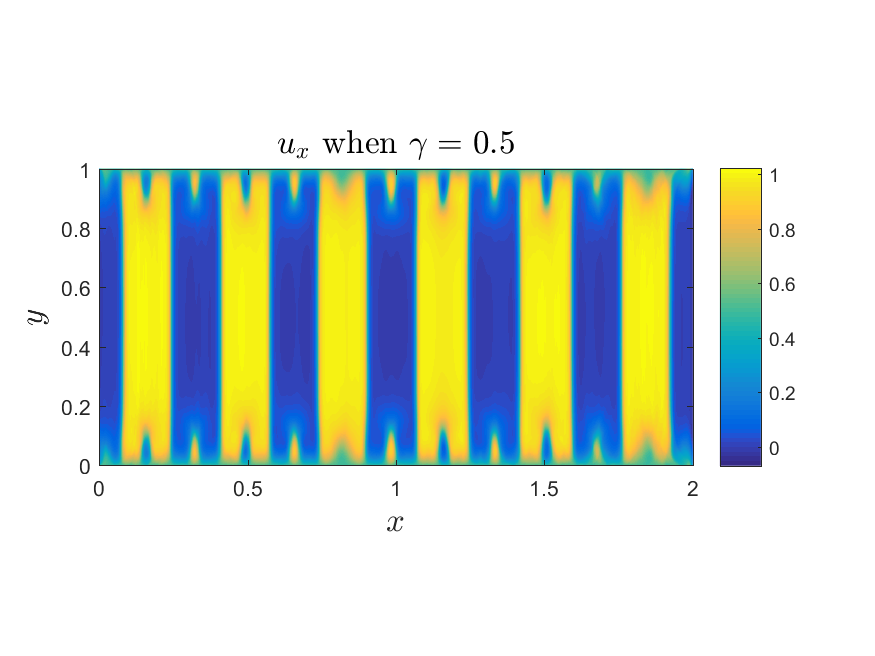}}
\end{minipage}
\hfill
\begin{minipage}[]{0.2 \textwidth}
 \leftline{\small\textbf{(j)}}
\centerline{\includegraphics[height=3.5cm]{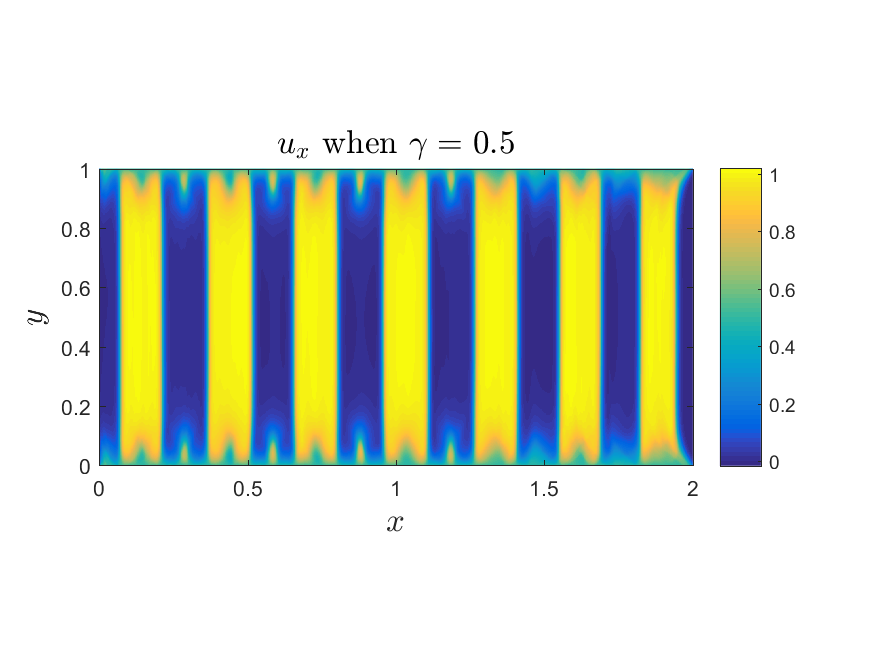}}
\end{minipage}
\hfill
\caption{\textbf{ 2d regularized problem with SmReLU activation function.} Here $\gamma=0.5$,  $iteration=300,000$, $\eta=10^{-3}$ and $\varepsilon=\frac{0.1}{16}$. Top: the energy with different  (a) NN: $1\times 128$; (b) NN: $3\times 128$; (c) NN: $5\times 128$; (d) NN: $7\times 128$; (e) NN: $9\times 128$; (f) NN: $11\times 128$; (g) NN: $13\times 128$;  (h) NN: $15\times 128$; (i) NN: $17\times 128$;  (j) NN: $19\times 128$.}
\label{2d_reg16_depth}
\end{figure}

Here, we fix the width of the DNN as 128 and $\varepsilon=\frac{1}{160}$. We explore how the depth affects the results, which are shown in Fig.~\ref{2d_reg16_depth}. Once again, one layer is insufficient as it does not capture the banded microstructure. The energy decreases with the increasing depth of the DNN after some oscillations.

The number of bands more or less increases with increasing depth, the maximum observed being 7 yellow bands. This is probably because the global minimum is probably reached, although this is difficult to prove. More bands would increase the interfacial energy of transition layers, which is proportional to $\eps$ times the total length of transition layers between the blue and yellow phases.

We also see a refinement mechanism in Fig.~\ref{2d_reg16_depth}(b4) where a yellow band near the middle is being split almost completely into two bands. The energy eventually plateaus out as the depth increases.

\begin{figure}[ht]
    \begin{minipage}[]{0.2 \textwidth}
     \centerline{\includegraphics[height=3.2cm]{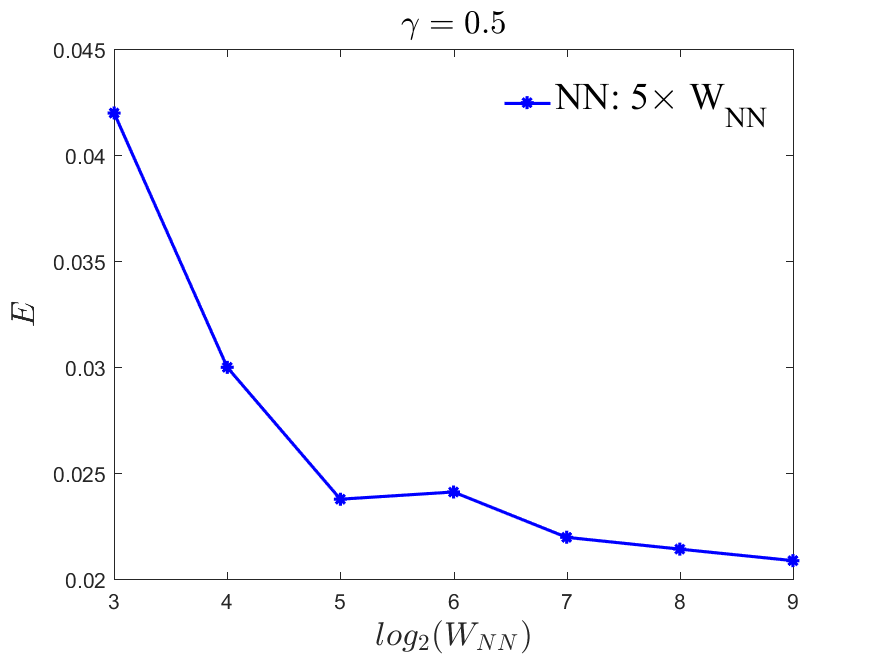}}
     \end{minipage}
     \hfill
   \begin{minipage}[]{0.2 \textwidth}
  \leftline{\small\textbf{(a)}}
\centerline{\includegraphics[height=3.5cm]{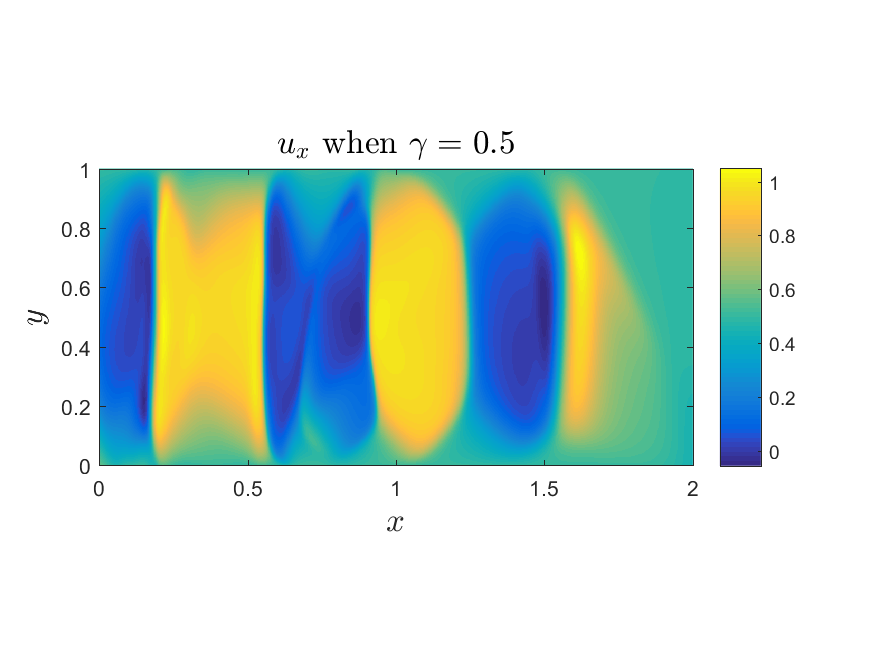}}
\end{minipage}
\hfill
 \begin{minipage}[]{0.2 \textwidth}
  \leftline{\small\textbf{(b)}}
\centerline{\includegraphics[height=3.5cm]{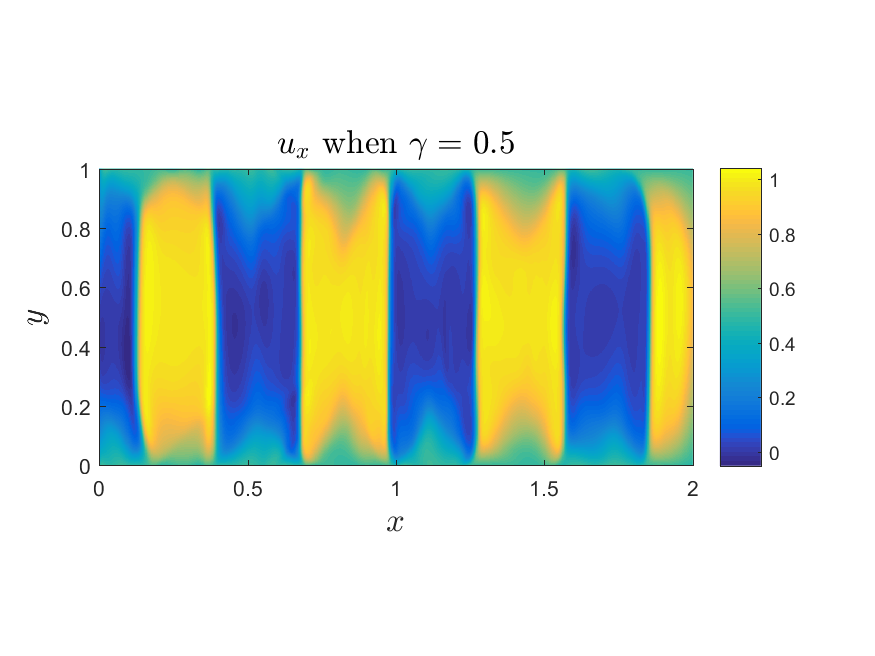}}
\end{minipage}
\hfill
\begin{minipage}[]{0.2 \textwidth}
  \leftline{\small\textbf{(c)}}
\centerline{\includegraphics[height=3.5cm]{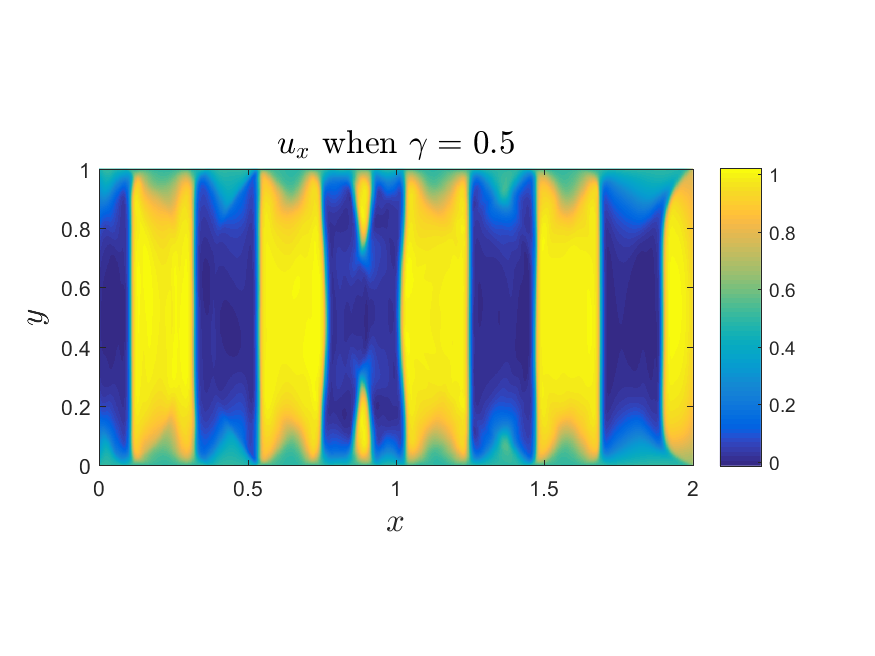}}
\end{minipage}
\begin{minipage}[]{0.2 \textwidth}
  \leftline{\small\textbf{(d)}}
\centerline{\includegraphics[height=3.5cm]{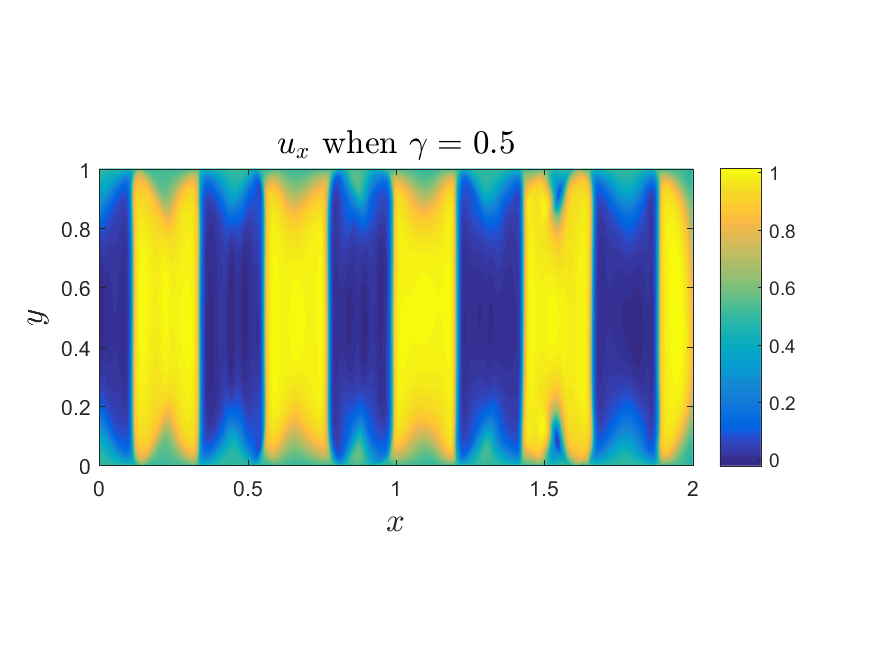}}
\end{minipage}
\hfill
 \begin{minipage}[]{0.2 \textwidth}
  \leftline{\small\textbf{(e)}}
\centerline{\includegraphics[height=3.5cm]{figure/gamma05_5x128_tau500_srelu_lr3_regxx16_T2_uxp_opt.png}}
\end{minipage}
\hfill
\begin{minipage}[]{0.2 \textwidth}
  \leftline{\small\textbf{(f)}}
\centerline{\includegraphics[height=3.5cm]{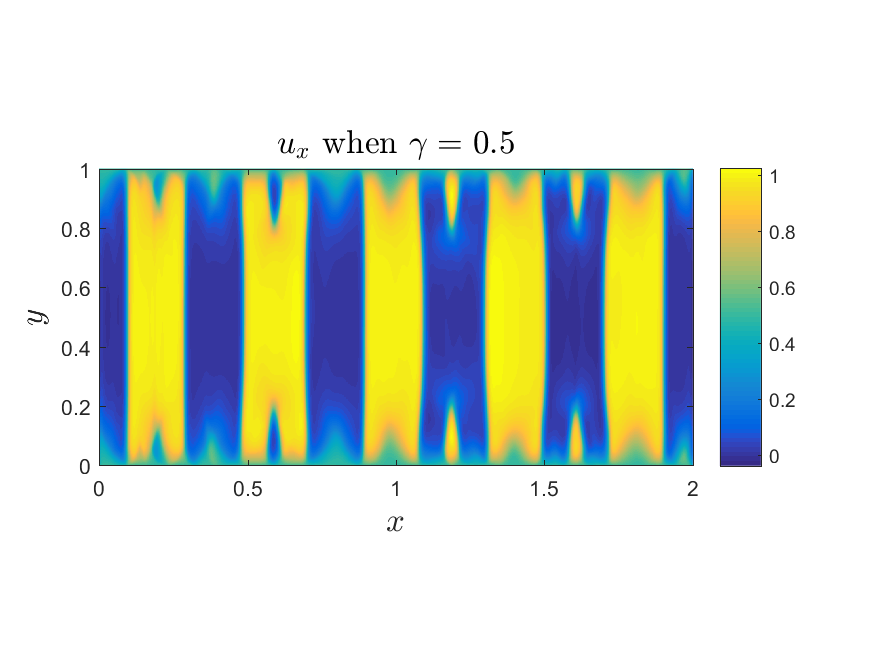}}
\end{minipage}
\hfill
\begin{minipage}[]{0.2 \textwidth}
  \leftline{\small\textbf{(g)}}
\centerline{\includegraphics[height=3.5cm]{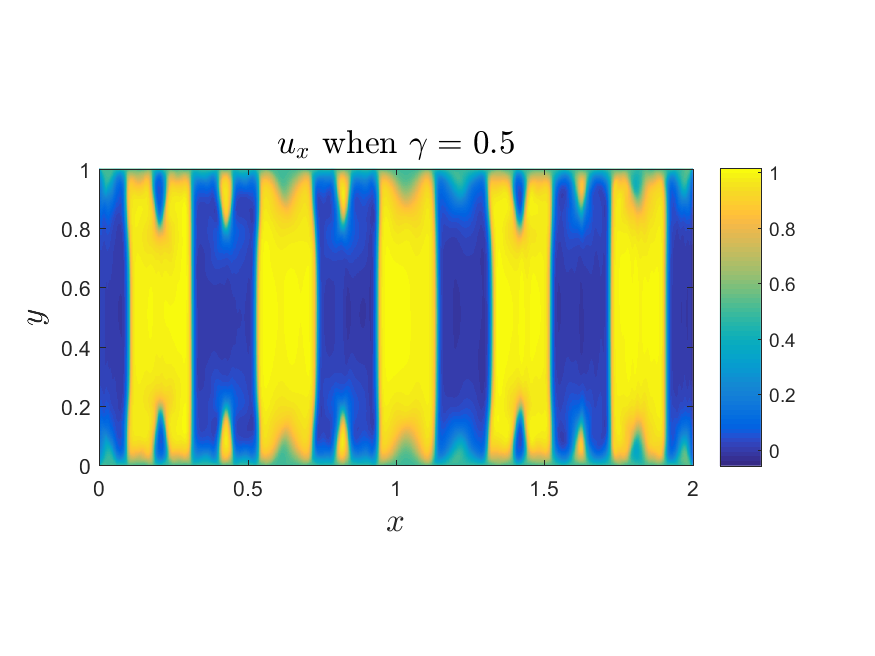}}
\end{minipage}
\caption{\textbf{ 2d regularized problem with SmReLU activation function.} Here $\gamma=0.5$, $iteration=300,000$, $\eta=10^{-3}$ and $\varepsilon=\frac{0.1}{16}$. Top left: the energy with different (a) NN: $5\times 8$; (b) NN: $5\times 16$; (c) NN: $5\times 32$; (d) NN: $5\times 64$; (e) NN: $5\times 128$; (f) NN: $5\times 256$; (f) NN: $5\times 512$.}
\label{2d_reg16_width}
\end{figure}

In order to explore how the width affects the results, we fix the depth of the DNN as 5 and $\varepsilon=\frac{1}{160}$.   The results are shown in Fig.~\ref{2d_reg16_width}. We can see the energy and the number of the yellow bars will decrease with increasing width of the DNN. For a fixed depth of 5 layers of the DNN,  increasing the width beyond $128$ does not result in more than 5 yellow bands in our solutions.

\subsubsection{Collocation Point Placement}
\begin{figure}[ht]
 \begin{minipage}[]{0.2 \textwidth}
\centerline{\includegraphics[height=3.5cm]{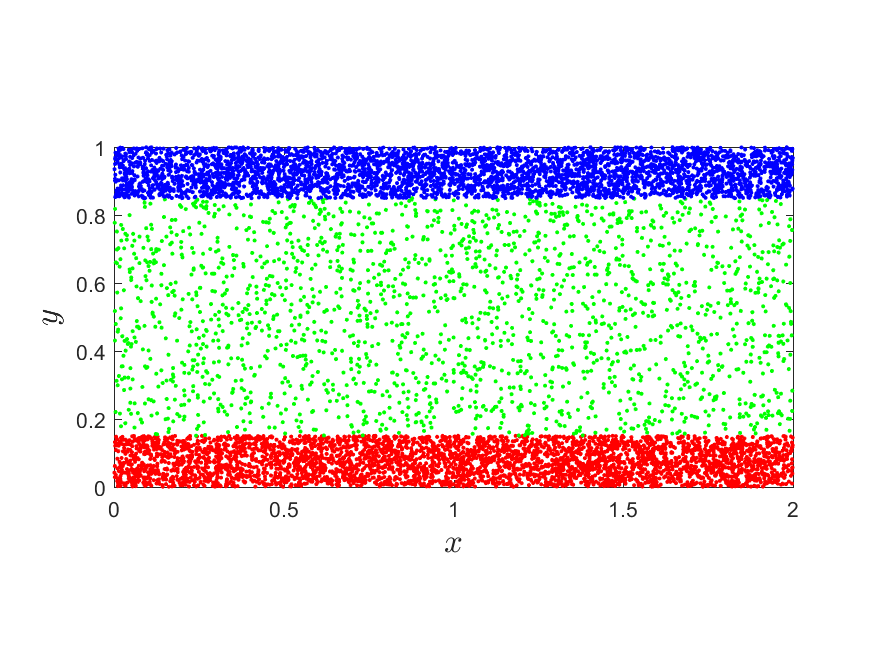}}
\end{minipage}
\hfill
\begin{minipage}[]{0.2 \textwidth}
\centerline{\includegraphics[height=3.2cm]{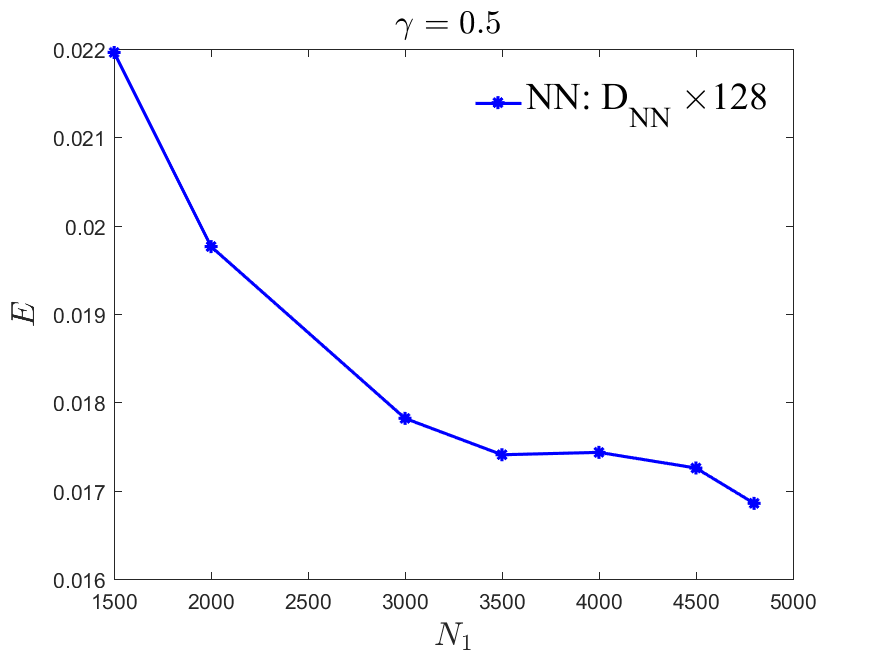}}
\end{minipage}
\hfill
 \begin{minipage}[]{0.2 \textwidth}
  \leftline{\small\textbf{(a)}}
\centerline{\includegraphics[height=3.5cm]{figure/gamma05_5x128_tau500_srelu_lr3_regxx16_T2_uxp_opt.png}}
\end{minipage}
\hfill
\begin{minipage}[]{0.2 \textwidth}
  \leftline{\small\textbf{(b)}}
\centerline{\includegraphics[height=3.5cm]{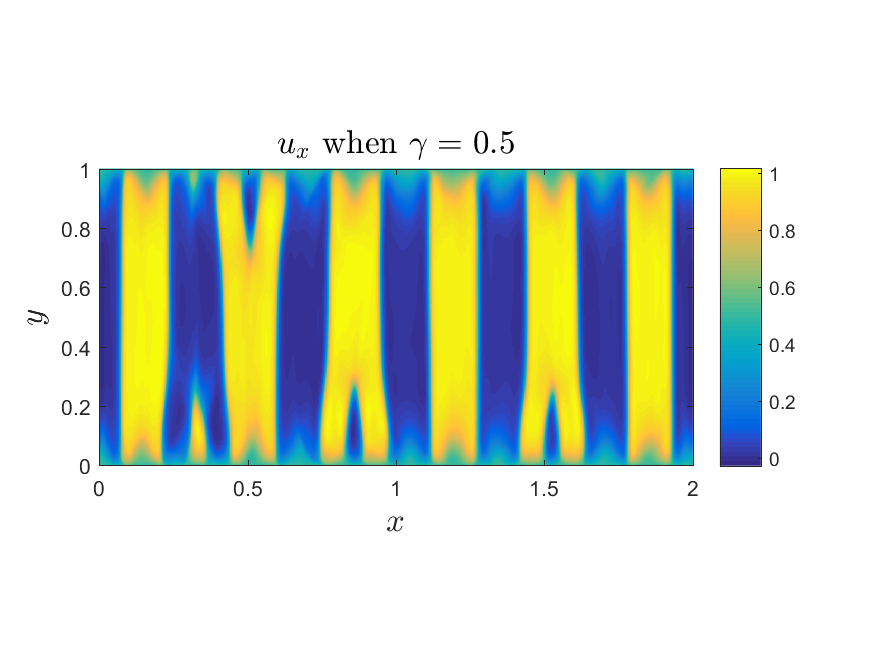}}
\end{minipage}
\begin{minipage}[]{0.2 \textwidth}
  \leftline{\small\textbf{(c)}}
\centerline{\includegraphics[height=3.5cm]{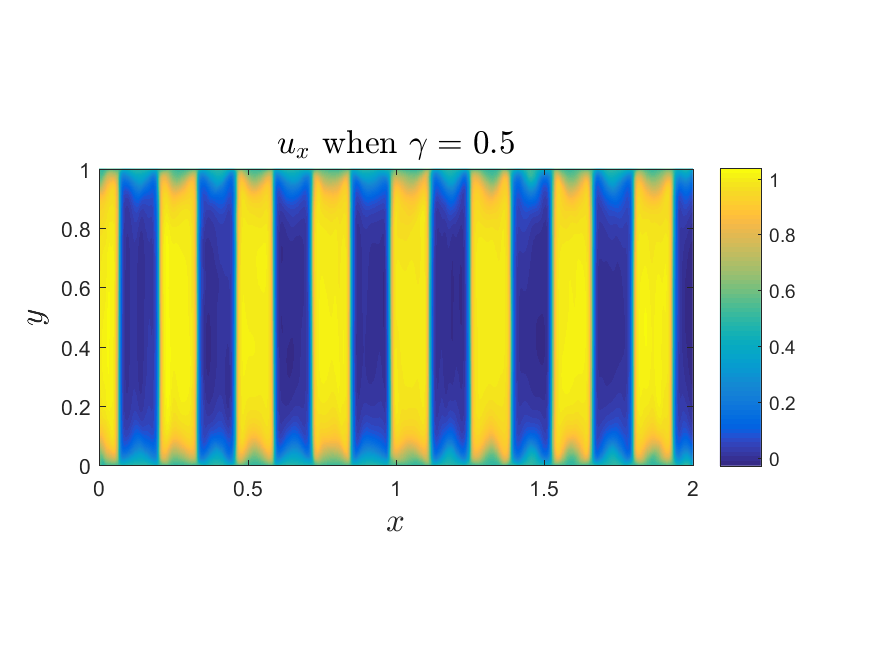}}
\end{minipage}
\hfill
\begin{minipage}[]{0.2 \textwidth}
  \leftline{\small\textbf{(d)}}
\centerline{\includegraphics[height=3.5cm]{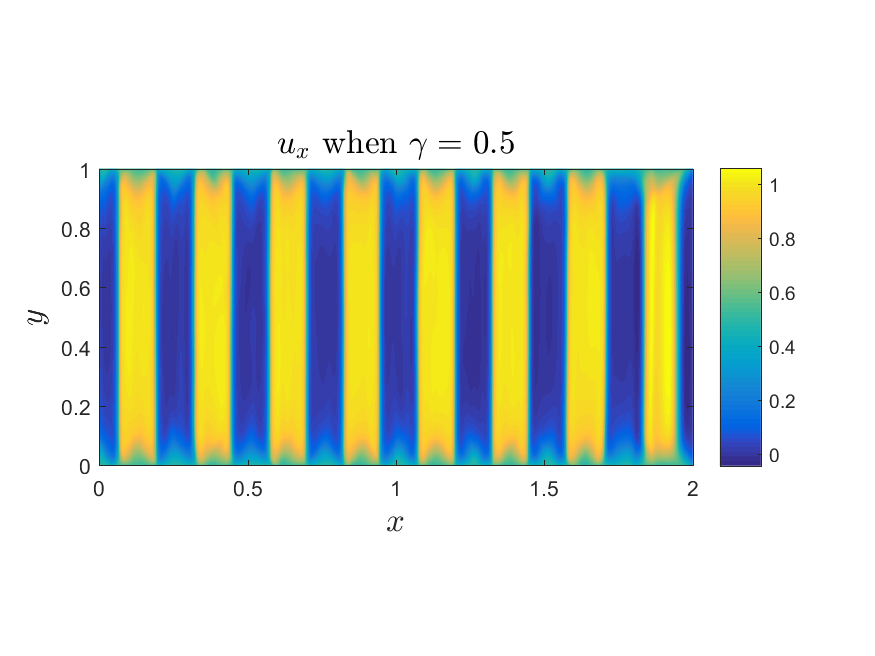}}
\end{minipage}
\hfill
\begin{minipage}[]{0.2 \textwidth}
  \leftline{\small\textbf{(e)}}
\centerline{\includegraphics[height=3.5cm]{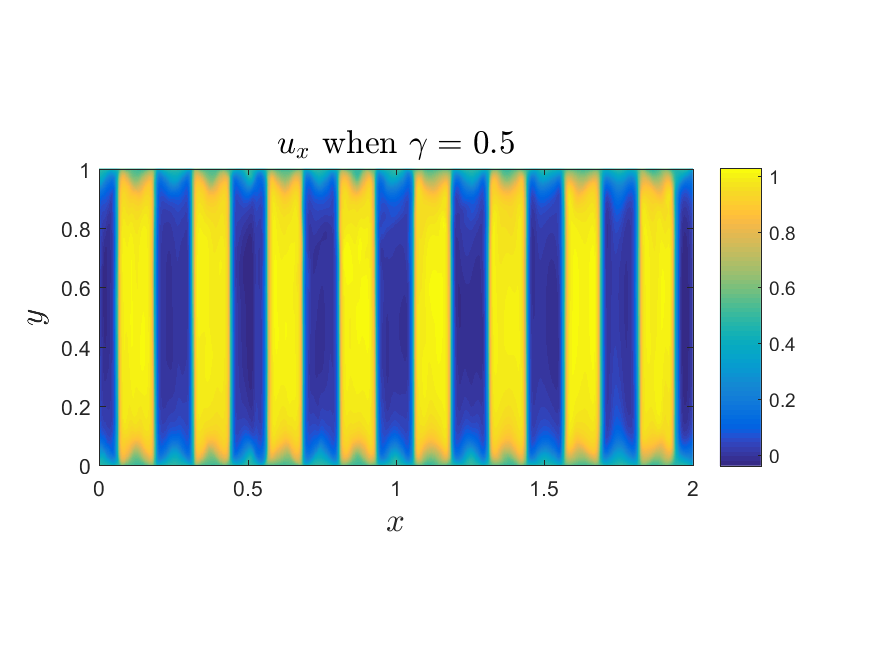}}
\end{minipage}
\hfill
\begin{minipage}[]{0.2 \textwidth}
  \leftline{\small\textbf{(f)}}
\centerline{\includegraphics[height=3.5cm]{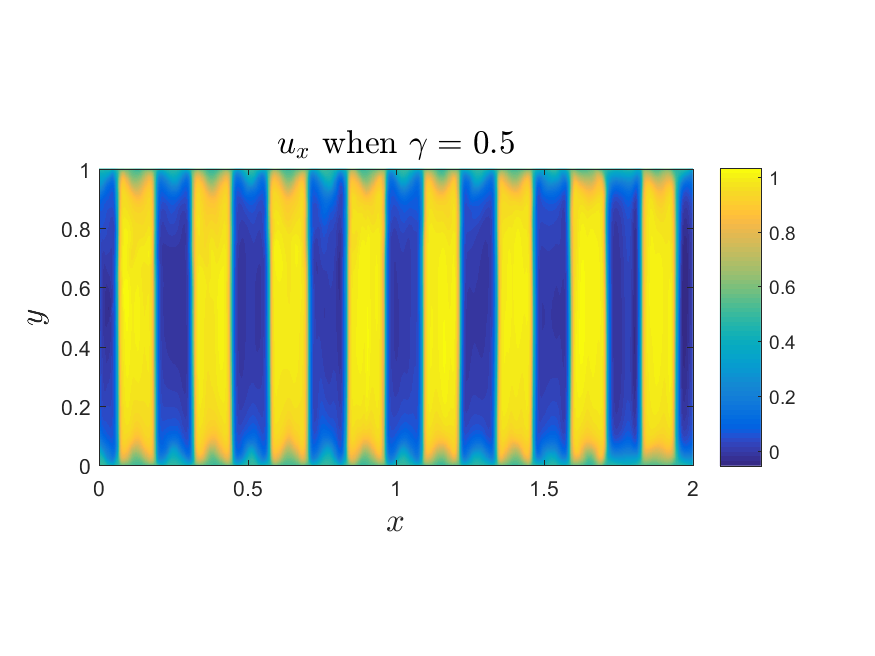}}
\end{minipage}
\caption{\textbf{Adapted collocation points }2d regularized problem with SmReLU activation function, $\gamma=0.5$, $iteration=300,000$, $\eta=10^{-3}$ and $\varepsilon=\frac{0.1}{16}$. First figure: the position of collocation point; second figure: the energy with different (a) $N_1=N_3=1500$ and $N_2=7000$; (b) $N_1=N_3=2000$ and $N_2=6000$; (c) $N_1=N_3=3000$ and $N_2=4000$; (d) $N_1=N_3=3500$ and $N_2=3000$; (e) $N_1=N_3=4000$ and $N_2=2000$; (f) $N_1=N_3=4500$ and $N_2=1000$.}
\label{2d_reg16_adapted}
\end{figure}

In computed minimizers we observe that the energy density near  the boundaries  is higher,  whereas we use a random choice of collocation points in the domain $\Omega$. In order to minimize  the energy contribution near the boundary more effectively, we select adaptively a higher proportion of  collocation points near the top and bottom parts of domain. We divide the domain $\Omega$ into three parts,  $\Omega_1=(0,2)\times(0,0.15)$, $\Omega_2=(0,2)\times(0.15,0.85)$ and $\Omega_3=(0,2)\times(0.85,1)$. And in each domain, we randomly choose $N_1$, $N_2$ and $N_3$ points, and let $N_1=N_3$. For details see Fig.~\ref{2d_reg16_adapted}.  When we increase the number $N_1=N_3$, keeping the sum of $N_1$, $N_2$ and $N_3$ fixed, the energy decreases.

\clearpage{}
\subsection{Curved Interfaces }
We consider the following energy
$$W(\nabla u)=\frac{1}{2}[ (\cos(\phi) u_x + \sin(\phi) u_y)   ^2(1-(\cos(\phi) u_x + \sin(\phi) u_y))^2+(-\sin(\phi) u_x + \cos(\phi) u_y )^2],$$
with Dirichlet boundary condition
\begin{equation}\label{2d_DB2}
u(x,y)=\gamma (\cos(\phi) x + \sin(\phi) y), \quad \forall (x,y)\in \partial \Omega.
\end{equation}
This corresponds to rotating one of the minima of $W$ from $(0,1)$ to $(\cos\phi,\sin\phi)$, where $\phi$ is a fixed parameter. Incompatibility with the boundary conditions now forces the twin boundaries (interfaces) to bend and curve, demonstrating the advantage of the mesh-free nature of the method.
It is easy to see that previously we have considered the case with $\phi=0$.

\begin{figure}[ht]
   \begin{minipage}[]{0.25 \textwidth}
 \leftline{\small\textbf{(a1)}}
\centerline{\includegraphics[height=5cm]{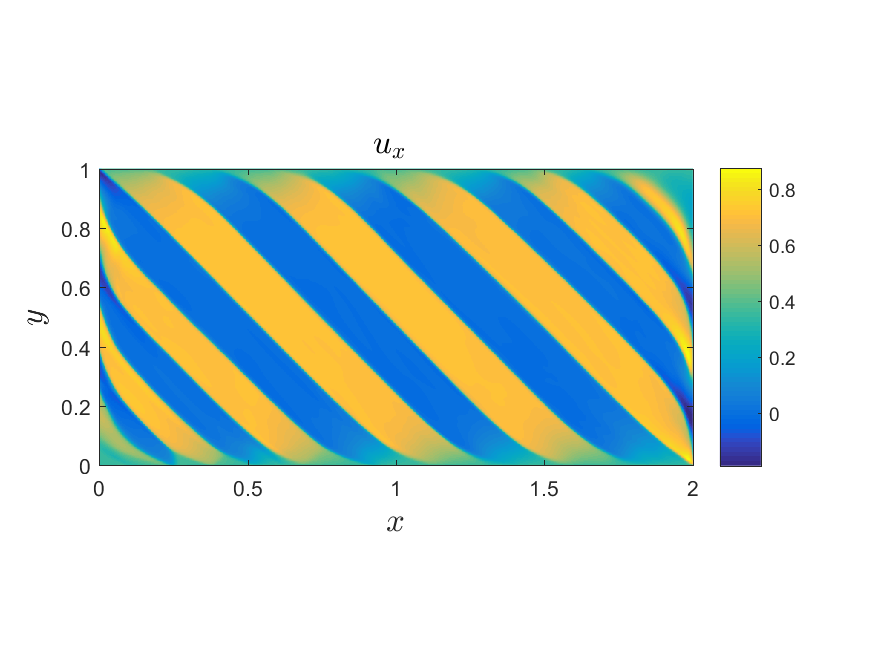}}
\end{minipage}
\hfill
 \begin{minipage}[]{0.25 \textwidth}
 \leftline{\small\textbf{(a2)}}
\centerline{\includegraphics[height=5cm]{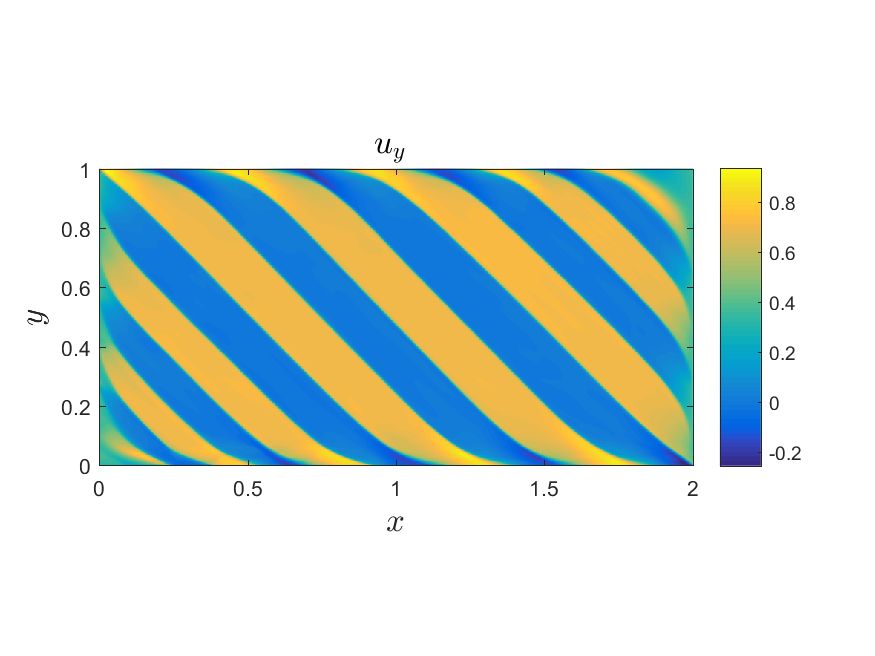}}
\end{minipage}
\hfill
\begin{minipage}[]{0.25 \textwidth}
 \leftline{\small\textbf{(a3)}}
\centerline{\includegraphics[height=5cm]{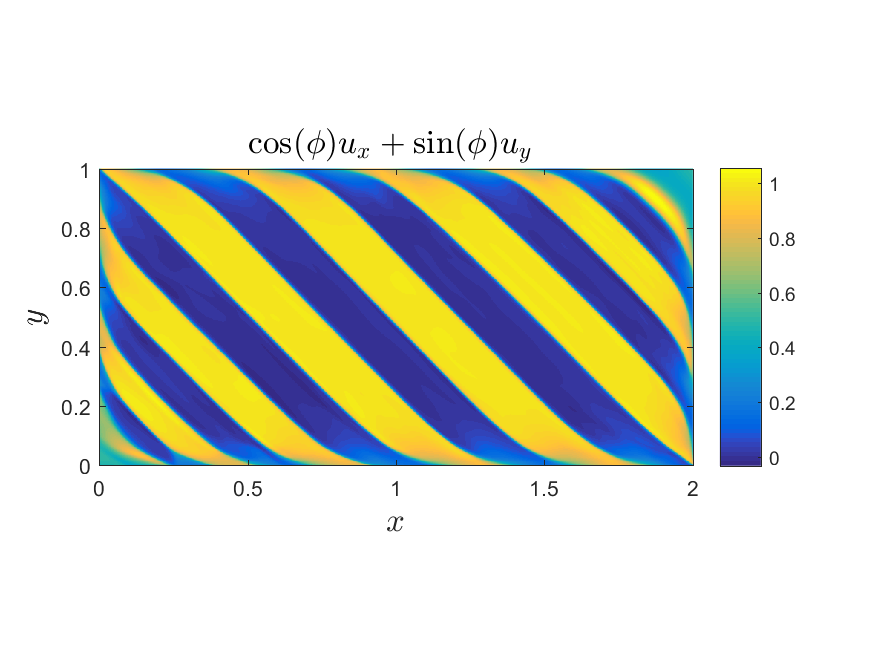}}
\end{minipage}
   \begin{minipage}[]{0.25 \textwidth}
 \leftline{\small\textbf{(b1)}}
\centerline{\includegraphics[height=5cm]{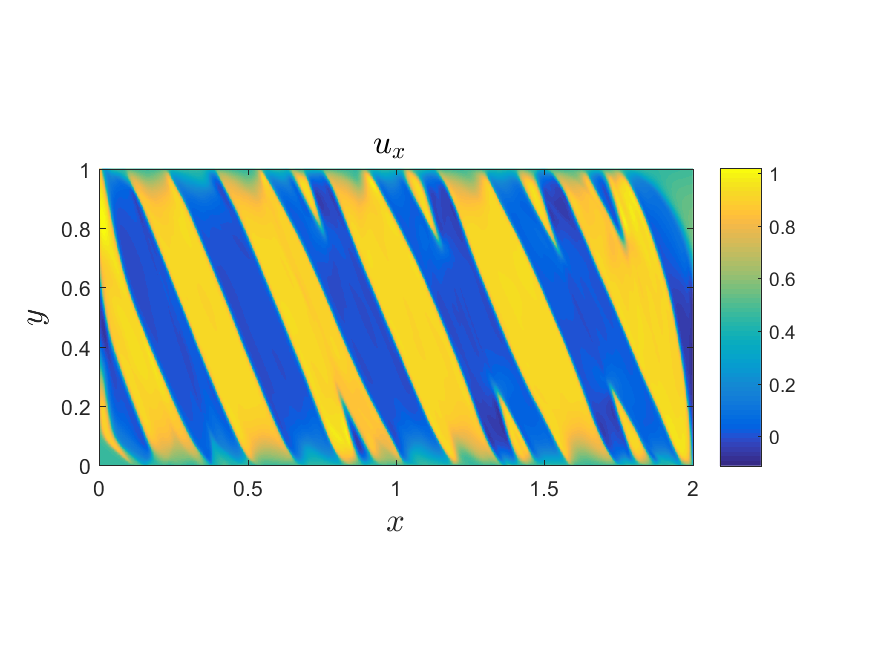}}
\end{minipage}
\hfill
 \begin{minipage}[]{0.25 \textwidth}
 \leftline{\small\textbf{(b2)}}
\centerline{\includegraphics[height=5cm]{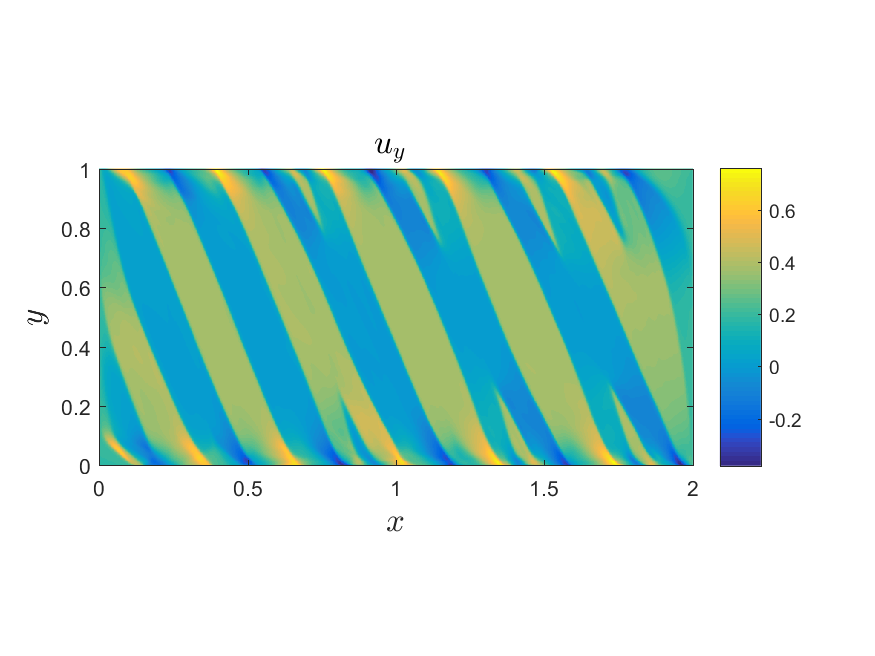}}
\end{minipage}
\hfill
\begin{minipage}[]{0.25 \textwidth}
 \leftline{\small\textbf{(b3)}}
\centerline{\includegraphics[height=5cm]{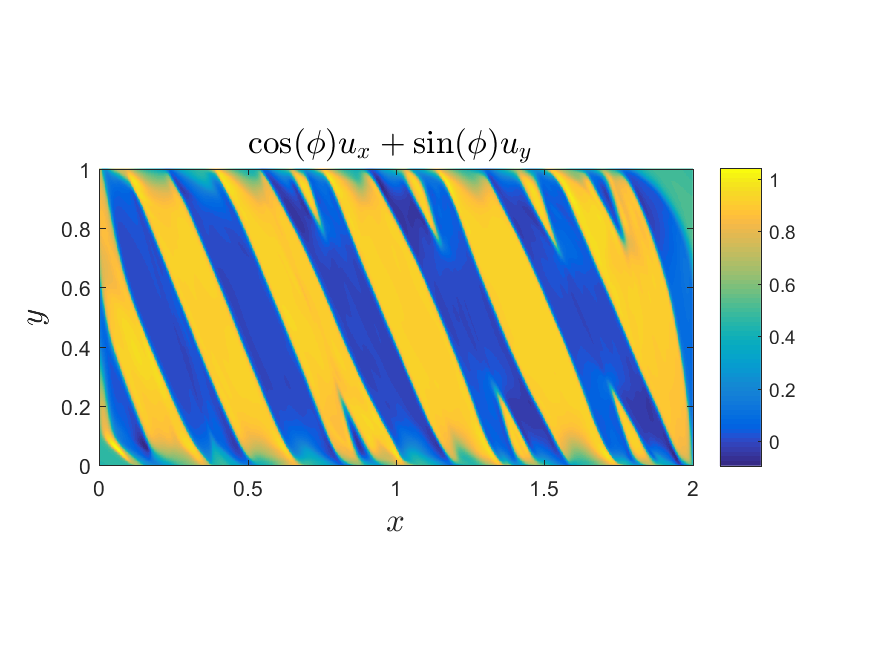}}
\end{minipage}
\caption{\textbf{ Rotated minimum 2d problem with SmReLU activation function.} Here $iteration=300,000$, $\eta=10^{-3}$ and $\varepsilon=\frac{0.1}{32}$. (a) $\gamma=0.5$ and $\phi=\frac{\pi}{4}$; (b) $\gamma=0.5$ and $\phi=\frac{\pi}{8}$. left: $u_x$; middle: $u_y$; right: $\cos(\phi) u_x+\sin(\phi)u_y$.}
\label{2d_reg_xy_new}
\end{figure}

The results are shown in Fig.~\ref{2d_reg_xy_new}.


 \begin{figure}[ht]
  \begin{minipage}[]{0.25 \textwidth}
 \leftline{\small\textbf{(a)}}
\centerline{\includegraphics[height=5cm]{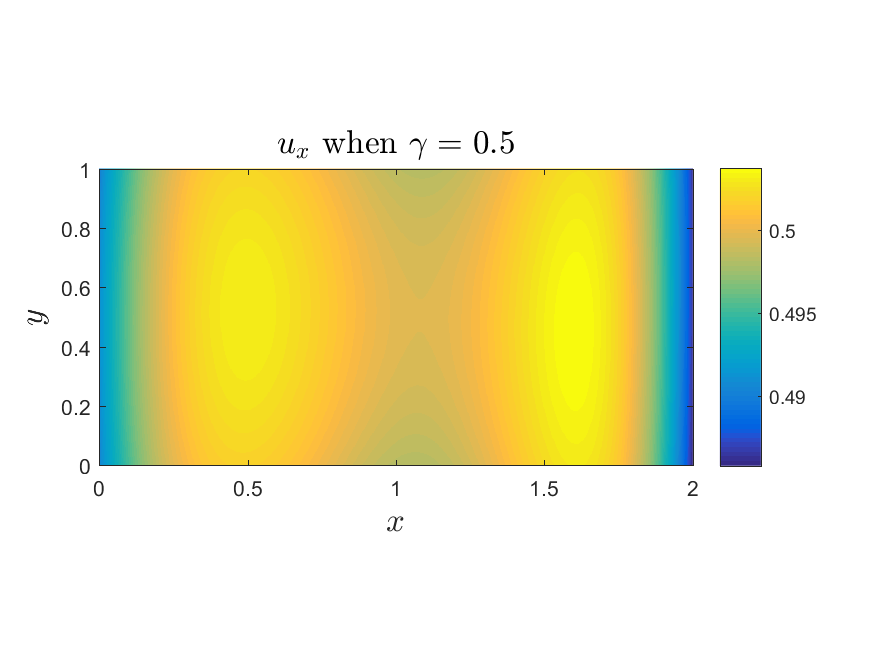}}
\end{minipage}
\hfill
  \begin{minipage}[]{0.25 \textwidth}
 \leftline{\small\textbf{(b)}}
\centerline{\includegraphics[height=5cm]{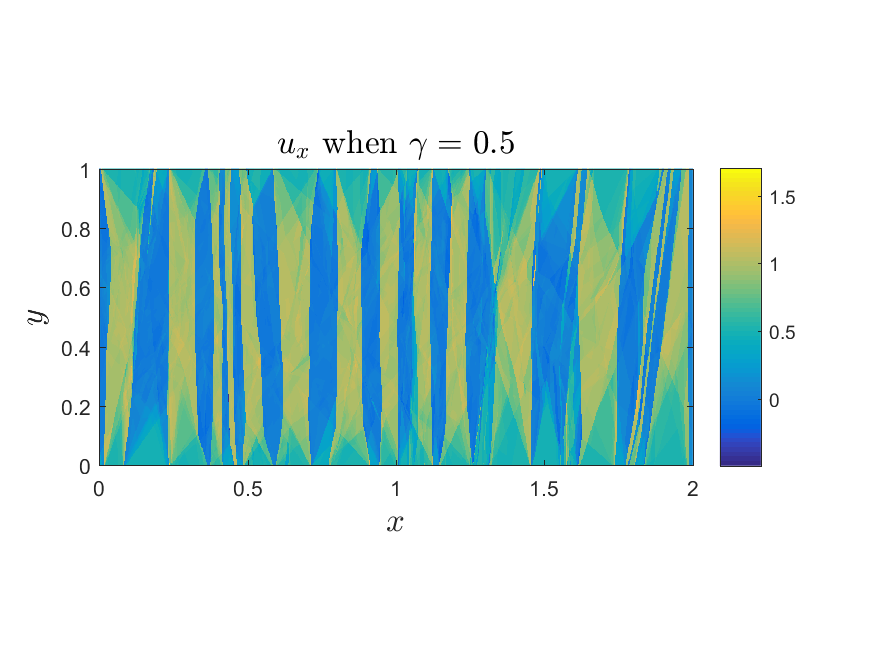}}
\end{minipage}
\hfill
 \begin{minipage}[]{0.25 \textwidth}
 \leftline{\small\textbf{(c)}}
\centerline{\includegraphics[height=5cm]{figure/gamma05_7x128_tau500_srelu_lr3_regxx32_T2_uxp_opt.png}}
\end{minipage}
\caption{\textbf{ Different activation functions in the 2d regularized problem. } Here $iteration=300,000$, $\eta=10^{-3}$ and $\varepsilon=\frac{0.1}{32}$. (a)Tanh activation function; (b) ReLU activation function; (c) SmReLU activation function.}
\label{2d_reg_activation}
\end{figure}

\clearpage
\section{Conclusions and Discussions}\label{sec:conclusion}
\noindent
This work makes use of the Deep Ritz Method for the first time in order to solve a nonconvex gradient-energy minimization problem from materials science. Such problems are challenging in two ways. First,  in general they do not possess a global minimum but only minimizing sequences, as well as a complex large set of local minima. Second, minimizers, including local ones, are characterized by complicated geometric microstructures, comprising multiple layers separated by twin interfaces, which are in effect free boundaries in the form of gradient discontinuities. These exhibit needle tapering and  tip-splitting topological transitions near  boundaries of the domain  \cite{healey,kohn,conti,dondl,james,hourosakis}.

A DNN represents the minimizer-candidate, which allows representation of the energy as a function of the weights and biases (the DNN parameters), and optimization with respect to them. Our results confirm the ability of the Deep Ritz method to spontaneously capture the above microstructure complexities of local or global minimizers.
A crucial new ingredient of our method is the activation function proposed here for the first time. To begin with, we observe that the ReLU activation function captures exact global minimizers in a simple unregularized 1D problem, due to its piecewise linear character. Other activation functions, such as Tanh and Sigmoid, struggle and are indeed unable to approximate these weak solutions with derivative jumps Fig.~\ref{ac_result}. For regularized problems, we introduce a new smoothened version of ReLU which we term SmReLU; see \eqref{SmReLU}.  It is quite successful in producing results that agree with careful FD computations  \cite{healey} and
captures the correct microstructure Fig.~\ref{twins} (b), (c),  while it outperforms other traditional activation functions, which fail at this task; see Fig.~\ref{2d_reg_activation}.

We find that it is essential for the DNN to have both sufficient width and depth to capture local minimizers. In general, increasing the number of layers (depth) increases the number of twin bands appearing in  our solution until a point where further depth increases does not improve the solution. In that sense, depth plays a role analogous to mesh size in FE. We find that for fixed depth, it is not necessary to increase width past a certain point, so here depth plays a more important role.
Equipped with our new SmReLU activation function, our method has no problem capturing both sharp gradient discontinuities across twin boundaries and smoothened transition layers. It also approximates curved interfaces successfully; see Fig.~\ref{2d_reg_xy_new}.
This is because our approach is mesh-free, which is a great advantage over FEM and FD methods, especially in the presence of interfacial discontinuities carrying surface energy, where problems can arise due to mesh misorientation compared to interfaces \cite{negri}.
Another advantage of this method is its simplicity, ease of programming, and the ability to naturally capture nonsmooth solutions with geometrically complex free interfaces in more than one dimension without special treatment such as discontinuous Galerkin FE for higher gradients in nonconvex elasticity \cite{grekas}.


\section*{Acknowledgements}
\noindent
The research of X. Chen is supported by the Ministry of Education, Singapore, under its Research Centre of Excellence award to the Institute for Functional Intelligent Materials (I-FIM, project No. EDUNC-33-18-279-V12).
The research of Z. Zhang is supported by Hong Kong RGC grant (projects 17300318 and 17307921), National Natural Science Foundation of China (project 12171406), and Seed Funding for Strategic Interdisciplinary Research Scheme 2021/22 (HKU).

\appendix

\bibliographystyle{siam}
\bibliography{ZWpaper_DeepLearningPDEDiscCoef}
\end{document}